\documentclass[aps,nofootinbib,preprint]{revtex4}
\usepackage{epsf,epsfig}

\usepackage{graphicx,subfigure}

\usepackage{amsmath}
\usepackage{amsfonts}
\usepackage{amssymb}
\usepackage{hyperref}
\usepackage{color}

\newcommand{\be}{\begin{equation}}
\newcommand{\ee}{\end{equation}}

\def\beq{\begin{equation}}
\def\eeq{\end{equation}}
\def\bea{\begin{eqnarray}}
\def\eea{\end{eqnarray}}
\def\ov{\overline}

\begin{document}
\title{The case for three-body decaying dark matter}

\vspace*{0.2cm}

\author{\vspace{0.5cm}
Hsin-Chia Cheng$\,^{a}$, Wei-Chih Huang$\,^{b}$, Ian Low$\,^{c,d}$, and Gabe Shaughnessy$\,^{c,d,e}$}
\affiliation{\vspace{0.5cm}
\mbox{$^{a}$Department of Physics, University of California, Davis, CA 95616}\\
\mbox{$^{b}$SISSA and INFN-sezione di Trieste, Via Bonomea 265,
34136 Trieste, Italy}
 \mbox{$^{c}$High Energy Physics Division, Argonne National Laboratory,
Argonne, IL 60439}\\
\mbox{$^{d}$Department of Physics and Astronomy, Northwestern University,
Evanston, IL 60208}\\
\mbox{$^{e}$Department of Physics, University of Wisconsin, Madison, WI 53706}
\vspace{0.8cm}
}

\begin{abstract}
\vspace*{0.5cm}
Fermi-LAT has confirmed the excess in cosmic positron fraction  observed by PAMELA, which could be explained by dark matter annihilating or decaying in the center of the galaxy. Most existing models postulate that the dark matter annihilates or decays into final states with two or four leptons, which would produce diffuse gamma ray emissions that are in tension with data measured by Fermi-LAT. We point out that the tension could be alleviated if the dark matter decays into three-body final states with a pair of leptons and a missing particle. Using the goldstino decay in a certain class of supersymmetric theories as a prime example, we demonstrate that simultaneous fits to the total $e^++e^-$ and the fractional $e^+/e^-$  fluxes from Fermi-LAT and PAMELA could be achieved for a 2 TeV parent particle and a 1 TeV missing particle, without being constrained by gamma-ray measurements. By studying different effective operators giving rise to the dark matter decay,  we show that this feature is generic for three-body decaying dark matter containing a missing particle. Constraints on the hadronic decay widths from the cosmic anti-proton spectra are also discussed.

\end{abstract}


\maketitle

\section{Introduction}
\label{sect:intro}

The intriguing positron excess in PAMELA~\cite{Adriani:2008zr} has been recently confirmed by Fermi-LAT~\cite{:2011rq} with the help of the Earth's magnetic field to distinguish positrons from electrons due to the lack of the magnetic field on board. Fermi-LAT also extends the positron excess range up to a higher energy scale, around $200$ GeV. In addition, measurements in the total $e^++e^-$ by Fermi-LAT also exhibit some interesting feature well beyond the 100 GeV region \cite{Abdo:2009zk}. These results imply the existence of new sources of primary positrons. The new source could be astrophysical, for example,  nearby pulsars~\cite{Hooper:2008kg}. The more interesting possibility is that they may be a sign of the dark matter (DM), due to either the  dark matter annihilation in the galaxy halo~\cite{DM annihilation}, or the decaying dark matter~\cite{DM decay}. In the case of the dark matter annihilation, it requires a very large boost factor due to the fact that both the resulting positron flux and the relic abundance are determined by the same annihilation channel. Some of the possible solutions include the Sommerfeld enhancement~\cite{Hisano:2003ec}~\cite{ArkaniHamed:2008qn}~\cite{Cirelli:2009uv} and Breit-Wigner Enhancement~\cite{Feldman:2008xs}~\cite{Ibe:2008ye}. On the other hand, the decaying dark matter has no such a limitation, but a mechanism responsible for a long lifetime has to be realized in this case.

The possibility that the excesses in the PAMELA positron fraction and Fermi-LAT $e^+ +e^-$ spectrum come from the dark matter annihilation or decay and the corresponding constraints from the antiproton and gamma ray spectra has been the subject of intensive investigations. Most of the studies in the literature focused on the two-body and four-body final states of the standard model (SM) particles~\cite{DM decay}, with the latter comes from an intermediate stage of two light ``portal'' particles~\cite{ArkaniHamed:2008qn}~\cite{Pospelov:2007mp}~\cite{Cholis:2008vb}~\cite{ArkaniHamed:2008qp}~\cite{Nomura:2008ru}. The general lesson from these studies is that final states have to be dominantly $\mu$'s or $\tau$'s, since a significant fraction of the electron final states would produce a sharp feature in the total $e^+ + e^-$ flux which is not seen in the Fermi-LAT data. However, these charged final states would produce diffuse gamma ray emissions through inverse Compton scatterings that are excluded by the Fermi-LAT data~\cite{ICS exclusion}~\cite{Cirelli:2009dv}~\cite{Papucci:2009gd}~\cite{Cirelli:2012ut}. There also exist constraints on decays into final state producing hadrons from the anti-proton measurements made by PAMELA and on prompt decays into photons from Fermi-LAT~\cite{Abdo:2010dk}, HESS~\cite{HESS dark matter}, and VERITAS~\cite{VERITAS dark matter}, especially for annihilating dark matter.

In a previous paper~\cite{Cheng:2010mw}, we proposed a novel scenario of the decaying dark matter where the final states include two standard model particles and a heavy missing particle.\footnote{Other possibilities of three-body DM decays were considered in Ref.~\cite{Pospelov:2008rn} ,~\cite{Demir:2009kc} ,~\cite{Kohri:2009yn} ,~\cite{Carone:2011ur} and by A.~Ibarra et al in Ref.~\cite{DM decay}.} In a certain class of supersymmetric theories where supersymmetry (SUSY) is spontaneously broken in multiple sequestered sectors, there will be a goldstino in each SUSY breaking sector. Only one linear combination of them is eaten and becomes the longitudinal component of the gravitino. The others are dubbed as goldstini in Ref.~\cite{Cheung:2010mc}.
If the lightest supersymmetric particle (LSP) is the gravitino and the next lightest supersymmetric particle (NLSP) is a goldstino, and  $R$-parity is conserved, the goldstino decays through dimension-8 operators to three-body final states containing the gravitino plus a pair of standard model (SM) particles. As a result, the goldstino could have a long lifetime and be cosmologically stable. For suitable SUSY breaking scales in the hidden sectors~\cite{Cheng:2010mw}, the lifetime can naturally be in the  range of $10^{26}-10^{27}$ sec, which is necessary for the decaying dark matter interpretation of the observed $e^+/e^-$ excess.  Unlike the gravitino, the goldstino interactions with the SM fields are not universal and
it is easy to come up with scenarios where the goldstino decays dominantly to leptons \cite{Cheng:2010mw}, thereby producing the observed excesses. A distinct feature of this scenario is that the decay of the dark matter is three-body, with the gravitino escaping detection and carrying off part of the energy. The SM particle pair in the final states has a smooth and soft injection energy spectrum, and consequently, a good fit to both the PAMELA $e^+/e^-$ and Fermi-LAT $e^++e^-$ data can be achieved with a universal coupling to all three generations of leptons, which is a welcoming feature from the model-building point of view. This is in contrast with the case of two-body or four-body final states where decays into $e^+e^-$ pair are disfavored.

In a more general context, one can imagine other models where the dark matter decays into three-body final states containing a missing particle.  Given the very different energy spectra from the well-studied cases of two-body and four-body decays, the scenario of three-body decays obviously deserves a detailed investigation to map out the parameter space allowed by various cosmic ray measurements such as  the antiproton and gamma ray data. This is the main purpose of this work. While we use the goldstini scenario as the prime illustrative example, more general setup can be included by studying various  effective operators giving rise to the decay of the dark matter. We will see that, for the mass range that could fit both the $e^+/e^-$ and $e^++e^-$ excesses, the injection energy spectra are quite insensitive to the type of operators mediating the decay as well as the spin of the dark matter.

This work is organized as follows. In the next section, we first perform fits to PAMELA and Fermi-LAT $e^+/e^-$  and Fermi-LAT $e^++e^-$ data for a wide range of the dark matter and missing particle masses, using the goldstino decay as an example. Then, we consider different types of higher-dimensional operators that lead to three-body decays and compare the spectra of the decay products from different decay operators. We also compare their fits to the PAMELA and Fermi-LAT $e^+/e^-$ data. There are small variations but they in general give similar results. In section~\ref{sect:constraints}, we study astrophysical constraints on three-body decaying dark matter using the gamma-ray and anti-proton data. Our conclusions are drawn in section~\ref{sect:concl}.

\section{Three-Body Dark Matter Decay for the positron Excess}
\label{sect:positron}

In the scenario of goldstino dark matter~\cite{Cheng:2010mw}, the goldstino decays to the gravitino through a three-body process. It was shown in Ref.~\cite{Cheng:2010mw} that the leptons pair produced in the decay has a smooth spectrum and can give a good fit to the PAMELA positron excess and the Fermi-LAT $e^+ +e^-$ spectrum, thereby providing a possible explanation of the observed anomalies. In this section we perform a general analysis of the three-body decaying dark matter explanation of the positron excesses in PAMELA and Fermi-LAT data by including the new Fermi-LAT result of the positron excess up to $\sim 200$~GeV in our fitting. In particular, we vary both the dark matter and missing particle masses to obtain a best fit and go beyond the goldstino scenario by studying various effective operators mediating the three-body decay.

\subsection{Goldstini}
\label{sect:gold}

Goldstini arise when there are multiple sequestered sectors which break SUSY. For simplicity, let us consider that SUSY is spontaneously broken in two hidden sectors, then there is a goldstino in each sector. If the superpartners of the SM particles receive SUSY breaking masses from both sectors, there will be couplings of the SM particles and their superpartners to each goldstino. If the SM superpartners are heavier than the goldstini, upon integrating out the superpartners, we will obtain dimension-8 operators between a pair of goldstini and a pair of SM particles.\footnote{The operators involving the Higgs fields actually have a lower dimension. However, it is suppressed by the same SUSY breaking scale with the mass dimension made up by the Higgs mass parameters in the numerator.} One linear combination of the goldstini is eaten and becomes the longitudinal mode of the gravitino. The other  obtains a mass due to supergravity effects \cite{Cheung:2010mc}. If there is a hierarchy in the SUSY breaking scales of the two hidden sectors, the eaten goldstino mostly comes from the sector with a larger SUSY breaking scale. The uneaten goldstino, which we assume to be the dark matter, is made mostly of the goldstino in the  sector with a smaller SUSY breaking scale. After going to the mass eigenstates, the dimension-8 operators  contains interactions which allow the uneaten goldstino decays to a gravitino and a pair of SM particles. If $R$-parity is conserved, the lifetime is naturally longer than the age of the universe and, with suitable choices of SUSY breaking scales of the hidden sectors, can be the required time scale to explain the $e^+/e^-$ excess observed by PAMELA and Fermi-LAT.

The goldstino decay operators were derived in the previous paper~\cite{Cheng:2010mw}. We have for SM fermions,
\bea
\label{eq:Leff}
{\cal L}_{2f}^{(1)} =  -\frac1{f_{eff}^2}\left(\frac{\widetilde{m}_1^2 \tan \theta - \widetilde{m}_2^2 \cot \theta}{m_{\widetilde{q}}^2}\right)
 \partial_\mu (\ov{\zeta}\ov{q}) \partial^\mu (\widetilde{G}_L q) + {\rm h.\, c.} \ ,
\eea
where $\zeta$ is the goldstion, $\widetilde{G}_L$ is the logitudinal mode of the gravitino, $q$ is the SM fermion, $f_{eff}= \sqrt{f_1^2+f_2^2}$ is the effective total SUSY breaking scale with $f_{1,2}$ being the SUSY breaking $F$-terms of the two SUSY breaking sectors, $\tan\theta = f_2/f_1$, $\widetilde{m}_{1,2}^2$ are the soft SUSY breaking mass of the superpartner of $q$  coming from the two SUSY breaking sectors respectively, and $m_{\widetilde{q}}^2=\widetilde{m}_1^2+\widetilde{m}_2^2$.
For gauge bosons,
\bea
\label{eq:gtoff}
{\cal L}_{2\gamma}^{(1)} &=&  \frac{-i}{ f_{eff}^2} \left( \frac{\widetilde{m}_1\tan\theta -\widetilde{m}_2 \cot\theta }{m_{{\lambda}} }\right)\ov{\widetilde{G}}_{L}\, \ov{F}\,\sigma \cdot \partial  \left(F\, \zeta \right)  \ ,
\eea
where $F(\ov{F})\equiv F_{\mu\nu} \sigma^{\mu\nu} (F_{\mu\nu} \ov{\sigma}^{\mu\nu})$ is the gauge field strength tensor, $\widetilde{m}_{1,2}$ are the SUSY breaking gaugino masses coming from the two SUSY breaking sectors, and $m_\lambda=\widetilde{m}_1+\widetilde{m}_2$.
For the Higgs field, there are terms from both K\"ahler potential and superpotential,
\bea
{\cal L}_{2h}^{(0)} &=&-\frac{1}{ \mu f_{\rm eff}^2}
\widetilde{G}_L \zeta \left[\left(m_{H_u}^2 + |\mu|^2 \right)
\phi_u^\dagger- B\mu \phi_d\right] \left[\delta m_d^2 \phi_d^\dagger-
\delta B\mu \phi_u \right] \nonumber \\
&&  + u \leftrightarrow d  + \rm{h.c.} \, , \\
{\cal L}_{2h}^{(1)} &=& \frac{1}{\mu^2 f_{\rm eff}^2} \left[\partial_\mu \left\{\left(\left(m_{H_u}^2
+ |\mu|^2 \right) \phi_u- B\mu \phi_d^\dagger\right) \ov{\widetilde{G}}_L
\right\}\, i\ov{\sigma}^\mu\, \left(\delta m_{u}^2 \phi_u^\dagger - \delta B\mu \phi_d \right) \zeta
\right. \nonumber \\
& & \left. + \partial_\mu \left\{\left(\delta m_{u}^2 \phi_u - \delta
B\mu \phi_d^\dagger \right) \ov{\zeta}
\right\} \, i\ov{\sigma}^\mu\,  \left(\left(m_{H_u}^2
+ |\mu|^2 \right) \phi_u^\dagger- B\mu \phi_d\right) \ov{\widetilde{G}}_L\right]  \nonumber \\
&&  + u \leftrightarrow d  + \rm{h.c.} \ ,
\eea
where
\bea
m_{H_\alpha}^2 &=& \sum_i m_{i\alpha}^2 \ , \quad \delta m_\alpha^2 =m_{1 \alpha}^2 \tan \theta - m_{2\alpha}^2 \cot \theta \ , \quad \alpha= u, d \ , \\
B &=& \sum_i B_i\  \ \ , \  \ \quad \delta B = B_1 \tan \theta - B_2 \cot \theta \ ,
\eea
with $i=1,\,2$ indicating the SUSY breaking sector where the soft SUSY breaking parameters come from. The operators with the Higgs fields can actually induce two-body decay $\zeta \to \widetilde{G}_L h$ after substituting the Higgs vacuum expectation values. The two-body decay mode is suppressed by a factor $(v/m_\zeta)^2$. Compared with the phase space suppression of the three-body decay, the two-body decay rate is expected to be of the same order as the three-body decay rate but may be somewhat larger depending on the model parameters.

Unlike the gravitino, couplings of the goldstino to SM fields are not universal, but rather depend on the fractions of the soft SUSY breaking parameters of the corresponding superpartners coming from the two SUSY breaking sectors. To explain the cosmic $e^+/e^-$ excess while satisfying other cosmic ray constraints, the decays should dominantly go to leptons. The constraints on the branching fractions to other particles will be studied in Sec.~\ref{sect:constraints}.

To the leading order, the goldstino acquires a mass which is twice that of gravitino due to supergravity effects~\cite{Cheung:2010mc}.  However, the relation can be modified significantly beyond the leading order~\cite{Craig:2010yf}. In this subsection, we extend the analysis to explore a wider range of the goldstino mass $m_\zeta$ and the gravitino masse $m_{\widetilde{G}_L}$ by relaxing the mass relation of $m_\zeta=2m_{\widetilde{G}_L}$. We follow exactly the same procedure as in Ref.~\cite{Cheng:2010mw}.
To recap, we make use of the Bessel function method of Ref.~\cite{Delahaye:2007fr} to obtain the positron flux detected on the Earth
due to the DM decay by assuming the MED model parameters. For the background fluxes we employ the
``model" presented by the Fermi-LAT collaboration~\cite{Grasso:2009ma}, which can be analytically parameterized
as shown in Ref.~\cite{Ibarra:2009dr}. The Moore profile in Ref.~\cite{Diemand:2004wh} is used for the DM halo.
With the assumption that the dark matter density is $\rho_\odot = 0.3$~GeV/cm$^3$,
we perform combined fits to both PAMELA and Fermi-LAT data, including the new Fermi-LAT positron fraction data~\cite{:2011rq}, by varying the decaying
DM lifetime and the overall normalization of the primary $e^-$ component of the background flux, as described in Ref.~\cite{Ibarra:2009dr}. Because the $e^+/e^-$ flux with energies below $10$ GeV measured at the top of the atmosphere is significantly influenced by the solar modulation effect, we only include data points above $10$ GeV from PAMELA for the total $\chi^2$. We assume universal couplings to all three lepton flavors which give a better fit than couplings to any single flavor.

\begin{figure}[t]
\includegraphics[trim =55mm 2mm 69mm 2mm, clip,
width=0.445\textwidth]{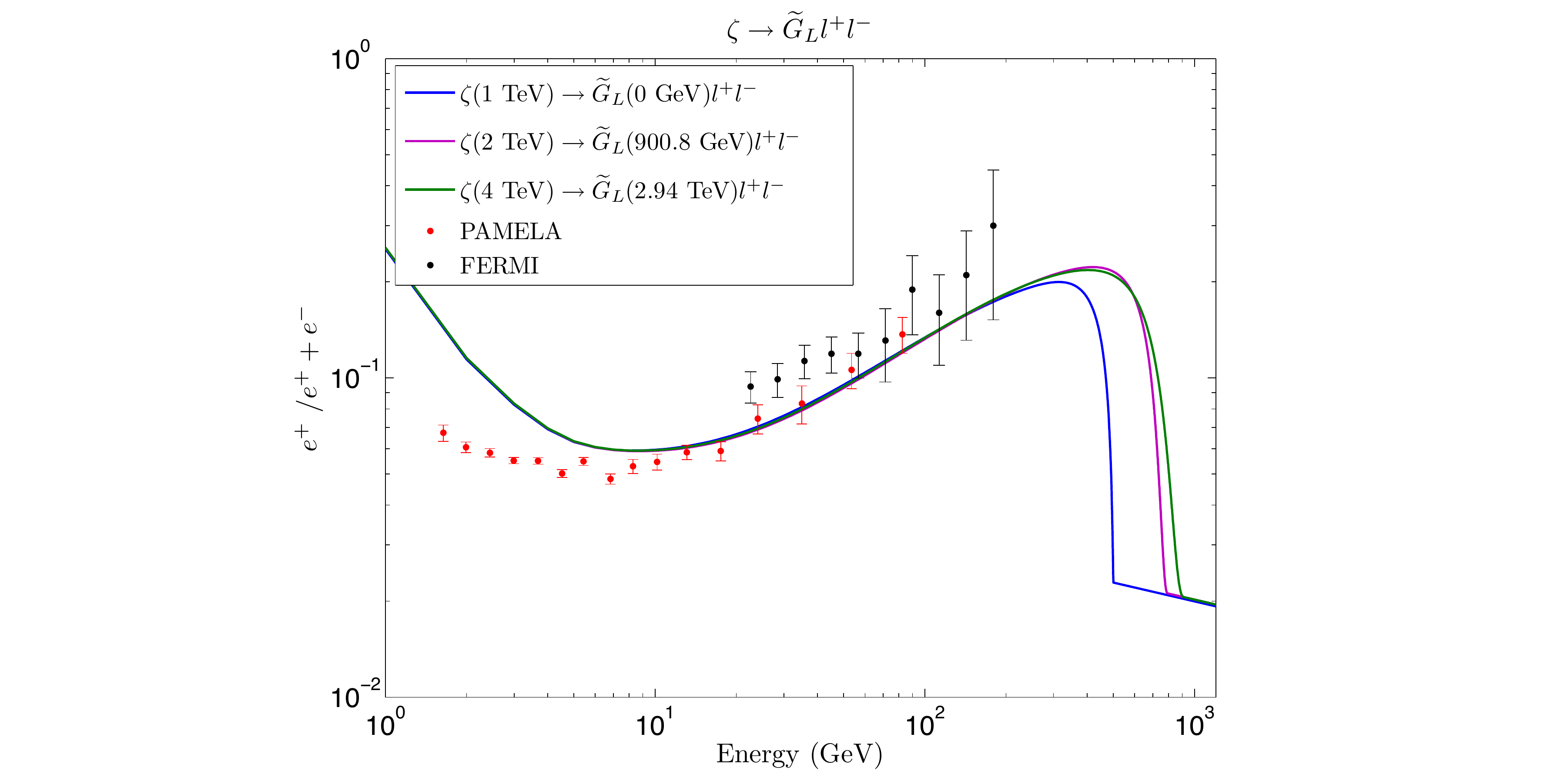}
\includegraphics[trim =55mm 2mm 69mm 2mm, clip,
width=0.445\textwidth]{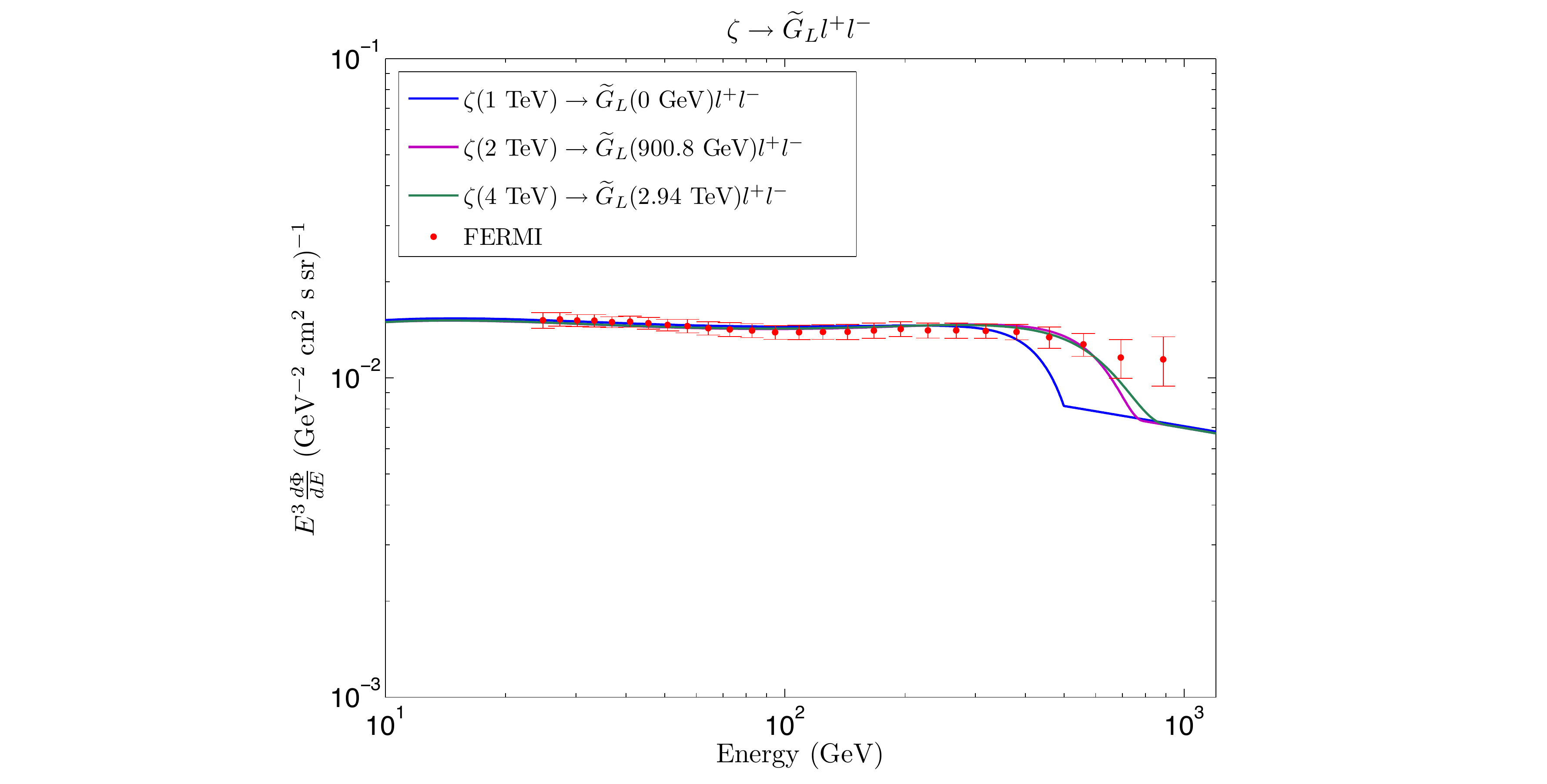}
\caption{{\em The best fits to the $e^+/(e^-+e^+)$ ratios observed by PAMELA and Fermi-LAT, and the Fermi-LAT $e^- +e^+$ spectrum for several goldstino masses.} \label{fig:sample_fits}}
\end{figure}

\begin{figure}[t]
\subfigure[]{
\includegraphics[trim =55mm 2mm 69mm 2mm, clip, width=0.445\textwidth]{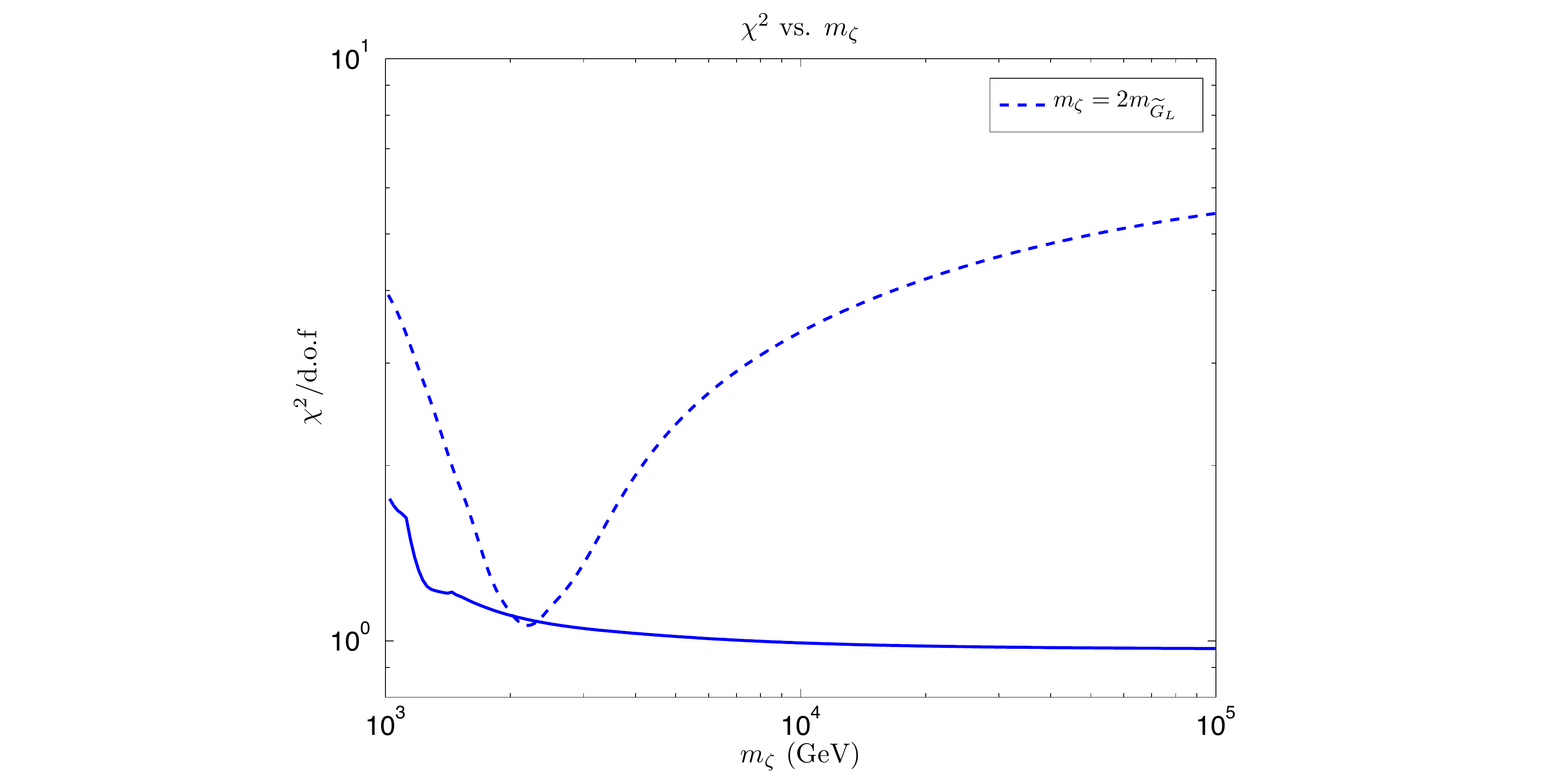}
            \label{fig:Mmscan}
            }
\subfigure[]{
\includegraphics[trim =55mm 2mm 69mm 2mm, clip, width=0.445\textwidth]{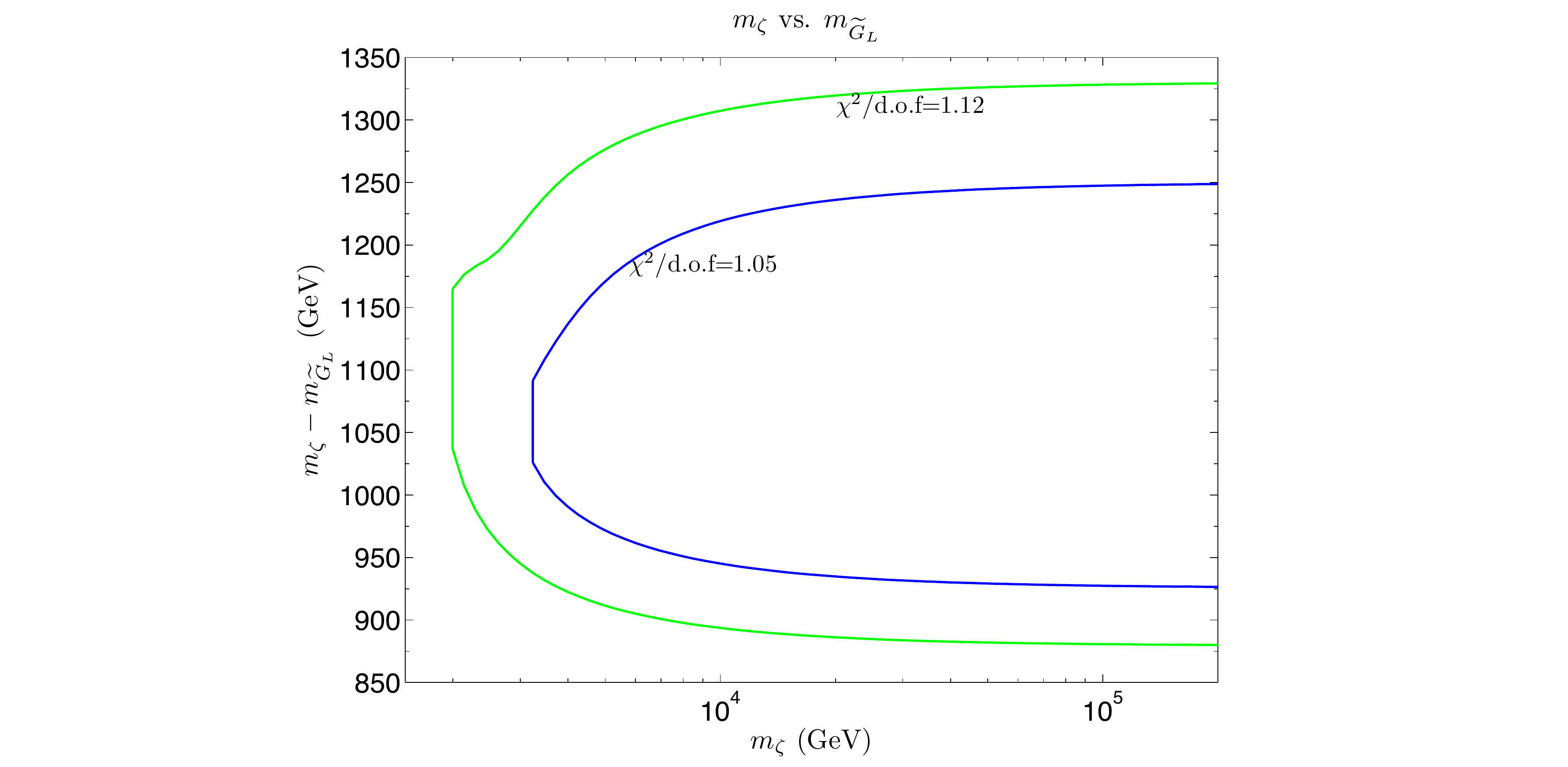}
            \label{fig:Mmscan2}
            }
\caption{\label{fig:fig2} \em  These plots display relations between $\chi^2$/d.o.f. of best fits to both PAMELA and Fermi-LAT data and the goldstino mass $m_\zeta$. On the left plot, we vary $m_\zeta$ and $m_{\widetilde{G}_L}$ independently (solid line) as well as dependently by assuming the mass relation: $m_\zeta=2m_{\widetilde{G}_L}$. The right plot shows  contours on the $m_\zeta$ and $m_\zeta-m_{\widetilde{G}_L}$ plane.
}
\end{figure}

The fits to the PAMELA and Fermi-LAT data for several sample goldstino masses are shown in Fig.~\ref{fig:sample_fits}. The $\chi^2$ per degree of freedom (d.o.f) of the fits as functions of the goldstino mass $m_\zeta$ and the gravitino mass $m_{\widetilde{G}_L}$ are shown in Fig.~\ref{fig:fig2}. The solid line in Fig.~\ref{fig:Mmscan} is the best fit $\chi^2$/d.o.f. for a given $m_\zeta$ by varying $m_{\widetilde{G}_L}$ freely, only subject to $m_{\widetilde{G}_L}\le m_\zeta$. We see that $\chi^2$/d.o.f. asymptotically approaches $\alt 1$ when $m_\zeta \agt 2$ TeV.  The dashed line, however, assumes  $m_\zeta=2m_{\widetilde{G}_L}$, in which case the minimum of $\chi^2$ occurs around $m_\zeta \sim 2$ TeV. Fig.~\ref{fig:Mmscan2} shows
 the contour plot of $m_\zeta$ versus $m_\zeta- m_{\widetilde{G}_L}$, from which we see that $\chi^2 \sim 1$ occurrs when $m_\zeta-m_{\widetilde{G}_L} \sim 1$ TeV and $m_\zeta \gtrsim 2$ TeV. Consequently, we will use $m_\zeta=2$ TeV as a benchmark value for later discussion. 

The underlying reason for these features is the hardening feature around $300\sim500$ GeV in the $e^+$ $+$ $e^-$ flux from Fermi-LAT, which dictates $m_\zeta-m_{\widetilde{G}_L}$ to be around $1$ TeV. This number comes about because we assume universal couplings to all three lepton flavors and the resulting softer energy spectra due to the $\mu$ and $\tau$ components drives the value of $m_\zeta-m_{\widetilde{G}_L}$ to be larger than twice the energy scale at which the hardening spectrum appears: $m_\zeta-m_{\widetilde{G}_L} \gtrsim  600 - 1000$ GeV. On the other hand,  the goldstino mass $m_\zeta$, which sets the overall energy scale, need to be heavier than 2 TeV since a smaller mass will produce  a feature in the total $e^++e^-$ flux at around $500 - 1000$ GeV, which worsens the fit with the Fermi-LAT data. (See Fig.~\ref{fig:sample_fits}.)

\begin{figure}[htbp]
\begin{center}
\includegraphics[scale=0.34]{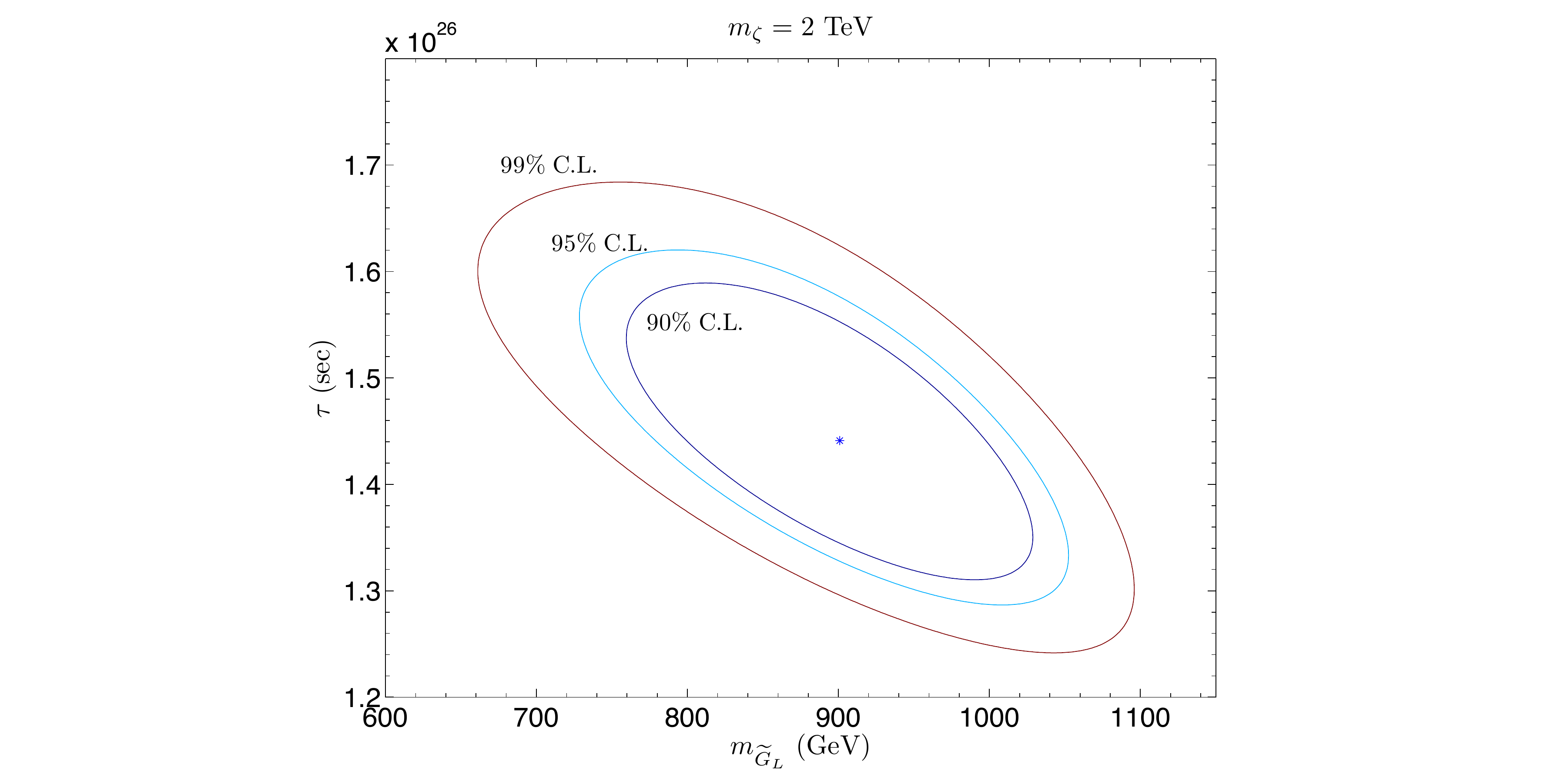}
\caption{\em The contour plot represents $90\%$, $95\%$ and $99\%$ C.L. for fits to PAMELA and Fermi-LAT data by varying the $m_{\widetilde{G}_L}$ and the lifetime $\tau$ of $\zeta$.
\label{fig:contoursplot}
}
\label{default}
\end{center}
\end{figure}
In addition to varying masses of the goldstino and gravitino, we also consider the best fit lifetime $\tau$ of the decaying goldstino, which is determined by the strength of the excess in the PAMELA $e^+/e^-$ flux in the energy regime of $10 - 200$ GeV. We present a contour plot corresponding to $90\%$, $95\%$ and $99\%$ C.L. of fits to PAMELA and Fermi-LAT data in the $m_{\widetilde{G}_L}$ versus $\tau$ plane in Fig.~\ref{fig:contoursplot}. It shows that for $m_{\zeta}=2$ TeV, the best fit corresponds to $m_{\widetilde{G}_L} \sim 900$ GeV. As $m_{\widetilde{G}_L}$ increases (decreases), it results in the softer (harder) injection spectrum, which can be compensated by a shorter (longer) lifetime.

 Before switching to three-body kinematics, we would like to comment that for $m_{\zeta}\geq 2$ TeV, the superpartners of the SM particles need to be heavier than a few TeV which implies some fine-tuning of the electroweak breaking scale. However, the absence of the SUSY signal at the LHC so far indicates that the superpartners may indeed be heavier than expected if SUSY exists at the TeV scale. The hint of a Higgs boson around 125 GeV is also consistent with a heavy SUSY spectrum.

\subsection{Three-Body Dark Matter Decay Kinematics}
\label{sect:kinematics}

If a dark matter particle decays to two-body final states, the energies of the decay products in the rest frame of the DM particle are fixed.
In contrast, the three-body decay will produce softer and broader spectra as we saw in the goldstino example. One can imagine there are other models where DM particles have similar three-body decays into a missing particle and two SM particles. The spectra of the decay products depend on the properties of DM particles and their interactions, but do share some generic features. By parametrizing different three-body decay mechanisms by effective higher dimensional operators, we will examine the similarities and differences of various three-body decay spectra in this subsection.

We start by comparing the injection energy spectrum of the SM fermions from decays via dimension-six four-fermi interactions with that from the benchmark scenario of goldstino decays, which occur through the dimension-eight operator in Eq.~(\ref{eq:Leff}). More specifically, we consider  the following Dirac structures~\cite{Carone:2011ur}: (pseudo-)scalar, (pseudo-)vector and tensor interactions, which are written as
\bea
&&
\frac{\lambda_{1}}{\Lambda^2}   \ov{\Psi}_{X} \Psi_{DM} \ov{\Psi}_q \Psi_q + {\rm h.\, c.},   \notag\\
&&
\frac{\lambda_{2}}{\Lambda^2}   \ov{\Psi}_{X}\gamma^5 \Psi_{DM} \ov{\Psi}_q \gamma^5\Psi_q + {\rm h.\, c.},   \notag\\
&&
\frac{\lambda_{3}}{\Lambda^2}   \ov{\Psi}_{X} \gamma^{\mu} \Psi_{DM} \ov{\Psi}_q \gamma_{\mu} \Psi_q + {\rm h.\, c.},   \\
&&
\frac{\lambda_{4}}{\Lambda^2}   \ov{\Psi}_{X} \gamma^{\mu} \gamma^5 \Psi_{DM} \ov{\Psi}_q \gamma_{\mu} \gamma^5 \Psi_q + {\rm h.\, c.},   \notag\\
&&
\frac{\lambda_{5}}{\Lambda^2}   \ov{\Psi}_{X} \sigma^{\mu\nu} \Psi_{DM} \ov{\Psi}_q \sigma_{\mu\nu} \Psi_q + {\rm h.\, c.},   \notag
\eea
where we assume both the dark matter (DM) and the missing particle in the decay product ($X$) are Dirac fermions, while $q$ refers to the SM fermions. The decay process induced by these operators is
\be
{\rm DM}\rightarrow X + q+\ov{q} \ .
\ee

\begin{figure}[t]
\subfigure[]{
\includegraphics[trim =55mm 2mm 69mm 2mm, clip, width=0.445\textwidth]{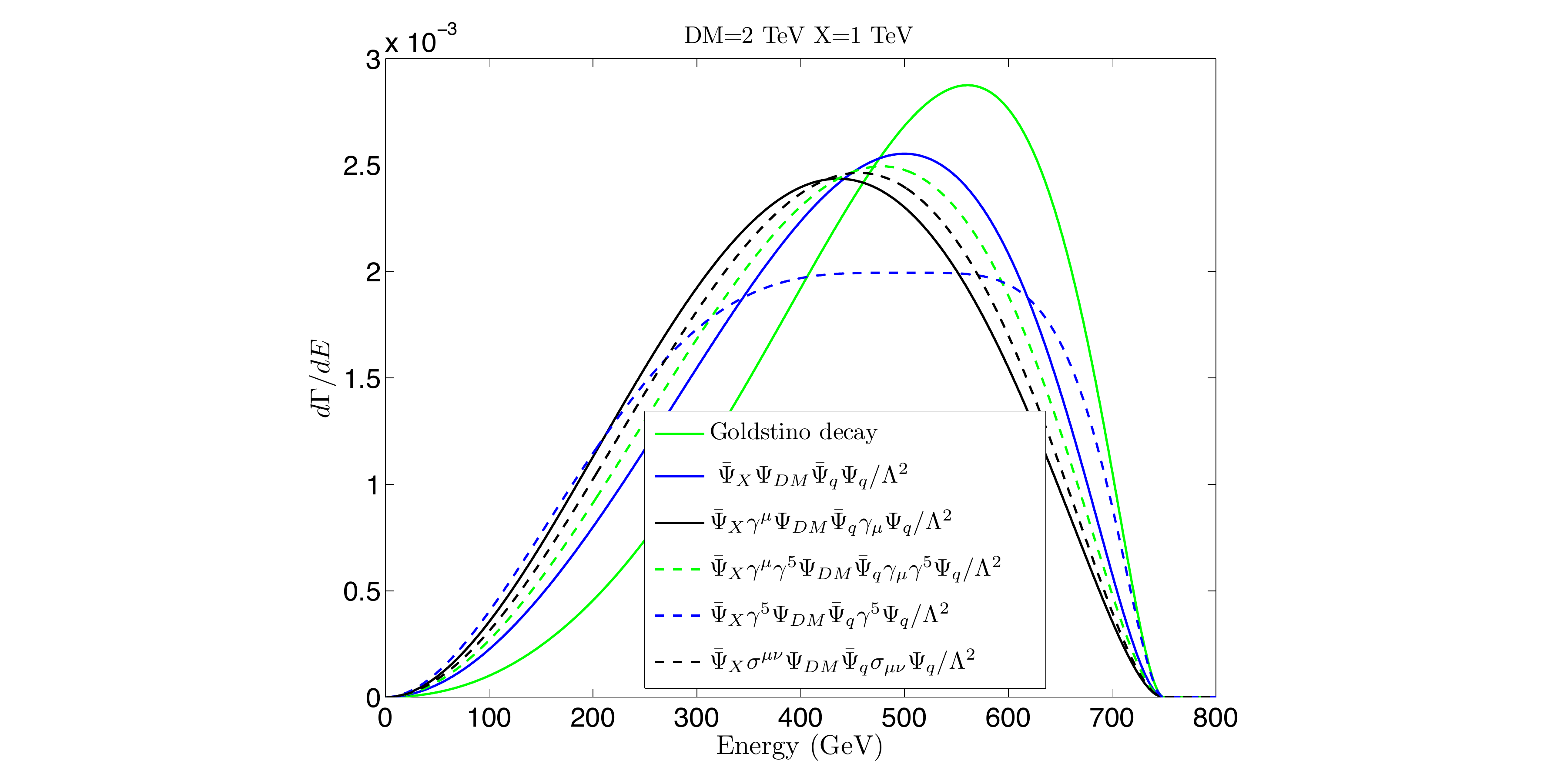}
           \label{fig:Edistf}
              }
\subfigure[]{
\includegraphics[trim =55mm 2mm 69mm 2mm, clip, width=0.445\textwidth]{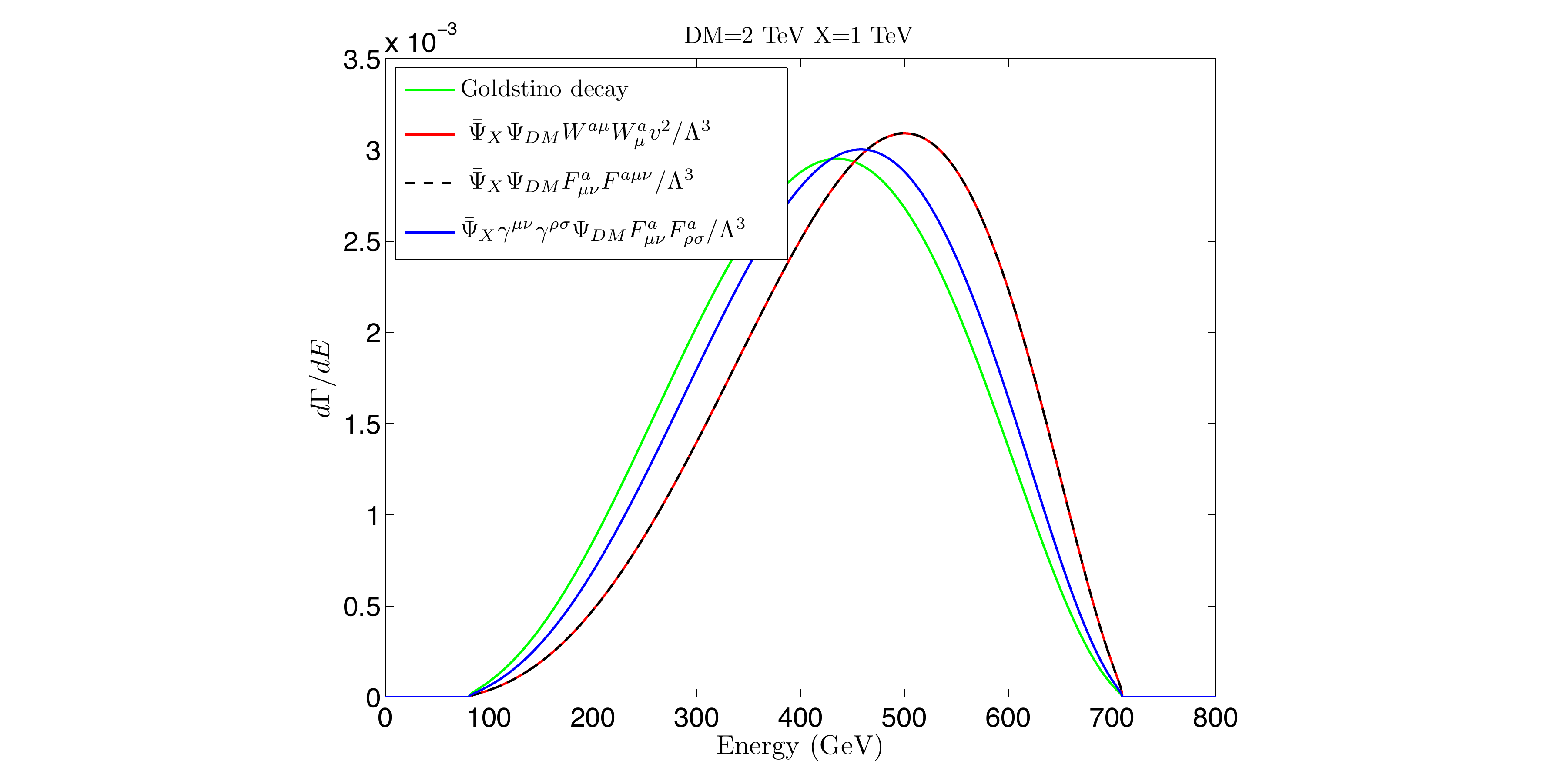}
            \label{fig:EdistW}
            }
\caption{\em The comparison among different DM decay mechanisms into fermions and $WW$, by normalizing the total decay width $\Gamma=1$. We assume the mass of the dark matter is $2$ TeV and $1$ TeV for X and $\widetilde{G}_L$.}
\end{figure}

Fig.~\ref{fig:Edistf} shows the comparison of energy spectra of SM fermions from the three-body decaying dark matter, where the total width is normalized to unity and the masses are fixed to $m_{\rm DM}=2 m_X = 2$ TeV. The goldstino decay results in the hardest spectrum due to fact that goldstino is derivatively coupled, which gives more weight to the higher energy modes. However, one sees that all spectra exhibit similar qualitative features, peaking in the $400 - 600$ GeV region. Therefore, we expect all of them to give rise to similar fits to the Fermi-LAT and PAMELA data points.

In addition to decays to SM fermions, we also compare various decay operators into SM gauge bosons, as they will be constrained by the anti-proton data from PAMELA as well as the gamma ray data from Fermi-LAT.  Here we list three types of different operators to be compared with the benchmark scenario of goldstino decay in Eq.~(\ref{eq:gtoff}):
\bea
&&
\frac{\lambda_{h'}v^2}{\Lambda^3}   \ov{\Psi}_{X} \Psi_{DM} W^a_{\mu}W^{a\mu} + {\rm h.\, c.},   \notag\\
&&
\frac{\lambda_{s'}}{\Lambda^3}   \ov{\Psi}_{X} \Psi_{DM} F^a_{\mu\nu}F^{a\mu\nu} + {\rm h.\, c.},   \\
&&
\frac{\lambda_{v'}}{\Lambda^3}  \ov{\Psi}_{X} \gamma^{\mu\nu}\gamma^{\rho\sigma}\Psi_{DM} F^a_{\mu\nu}F^a_{\rho\sigma}  + {\rm h.\, c.},   \notag
\eea
where $F^a_{\mu\nu}$ can be any of the $SU(3)$, $SU(2)$, and $U(1)$ SM gauge bosons.
The suppression of $v^2$ in front of the first operator signals its origin from the gauge invariant operator $({\lambda_{h'}}/{\Lambda^3})   \ov{\Psi}_{X} \Psi_{DM} D^{\mu}H^\dagger D_{\mu} H$. We see in Fig.~\ref{fig:EdistW} that, again, various operators have very similar qualitative features, even more so than the energy spectra of the SM fermions.

For the higher dimensional operators involving a pair of Higgs fields, if the Higgs fields are not both derivatively coupled, substituting in the Higgs VEV will induce a two-body decay $DM \to h\, X$. The Higgs boson will have a delta-function spectrum independent of the exact operators. We shall see that this two-body decay gives comparable constraints  to those from the corresponding three-body decay in section \ref{sect:constraints}.
\begin{figure}[t]
\subfigure[]{
\includegraphics[trim =55mm 2mm 69mm 2mm, clip, width=0.445\textwidth]{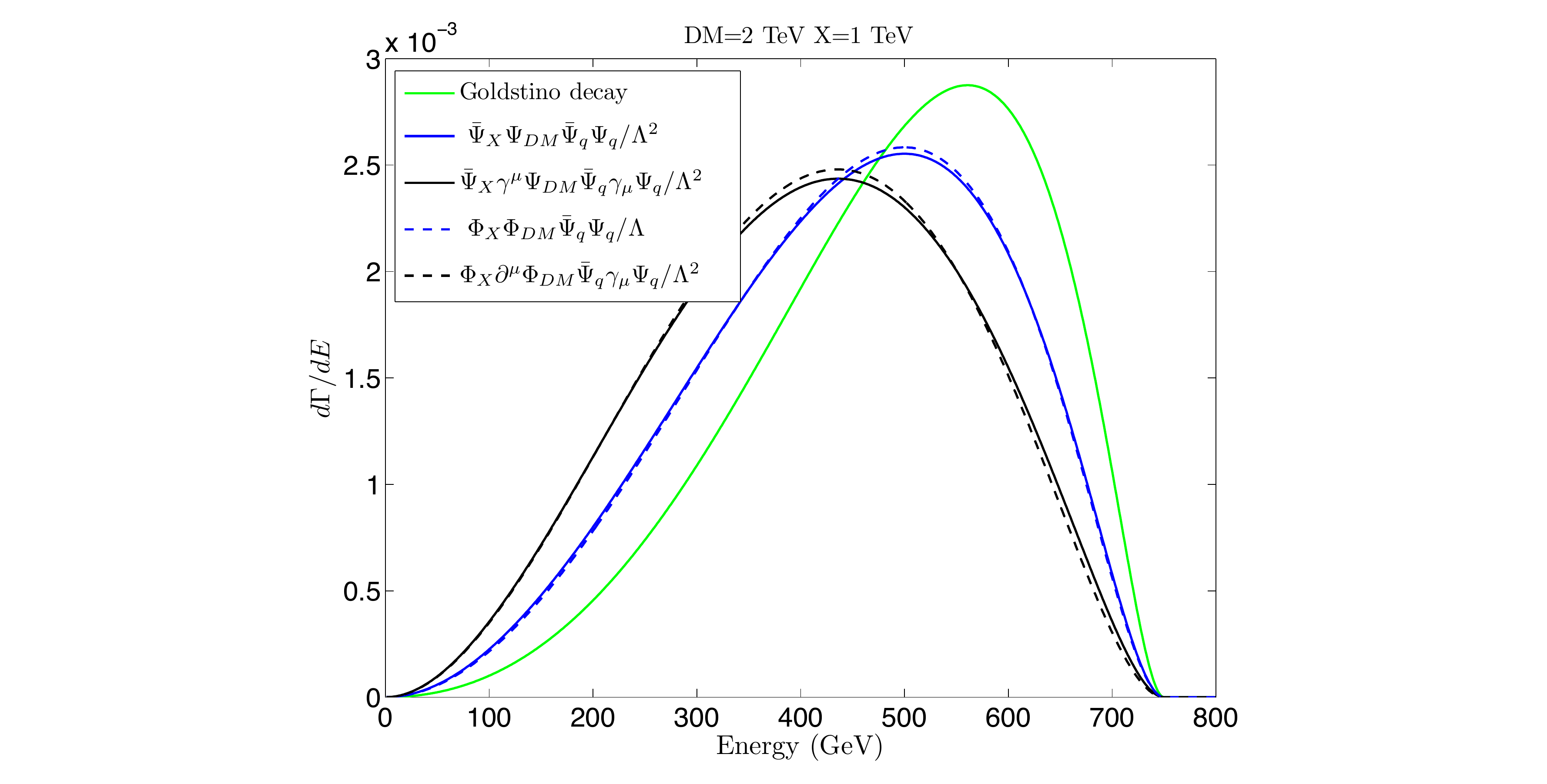}
            \label{fig:fermionscalar1}
           }
\subfigure[]{
\includegraphics[trim =55mm 2mm 69mm 2mm, clip, width=0.445\textwidth]{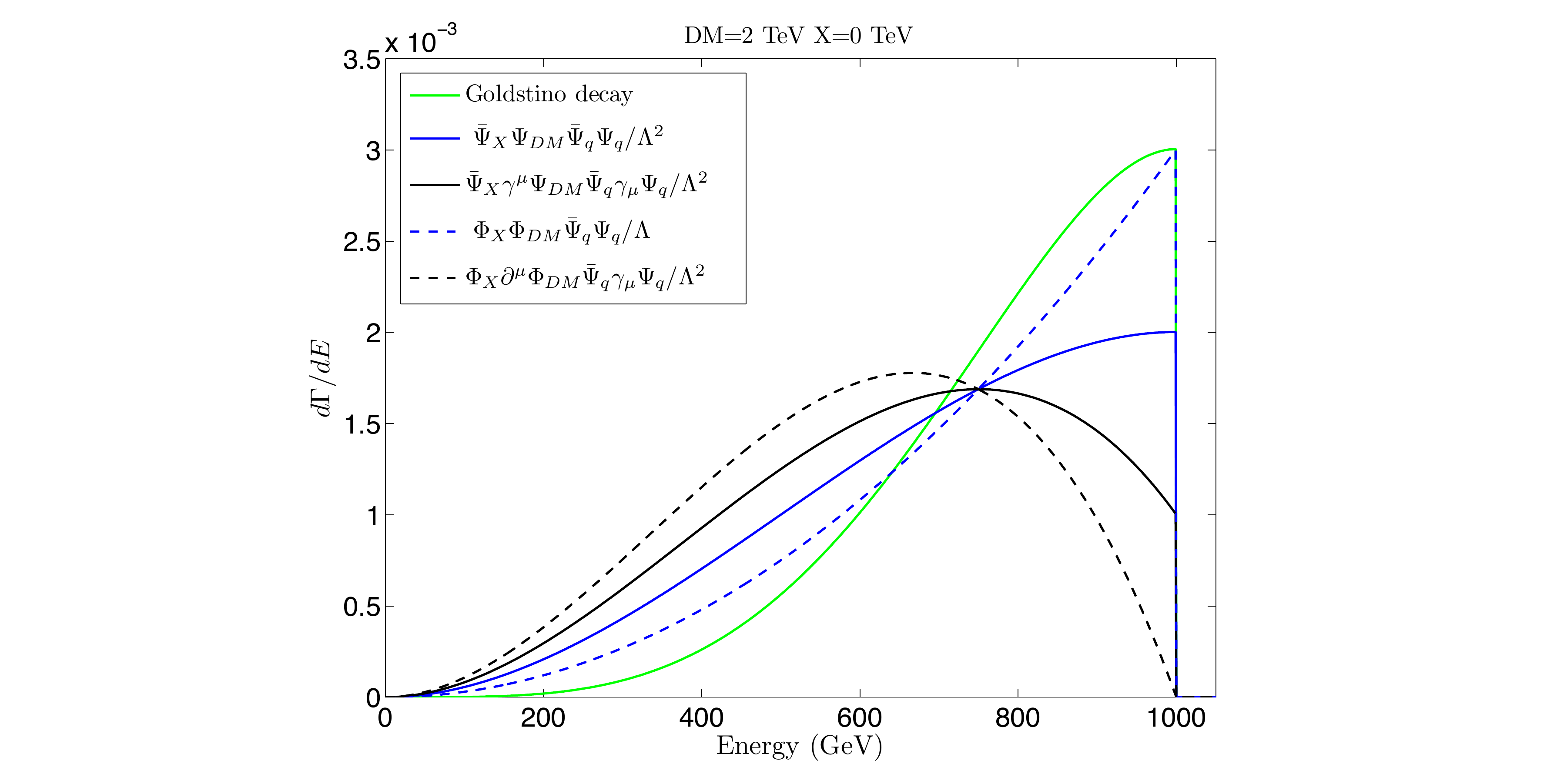}
           \label{fig:fermionscalar0}
            }
\caption{\em \label{fermionscalar} The comparison between spin-$0$  and spin-$1/2$ dark matter with several decay mechanisms. We assume $m_{\rm DM}=2m_X= 2$ TeV.}
\end{figure}

We conclude this subsection by comparing energy spectra of three-body decaying scalar dark matter with those of fermionic dark matter. From Fig.~\ref{fig:fermionscalar1} it is notable that, for $m_X={m_{DM}}/{2}$, spectra of scalar dark matter follows closely the counterparts of fermionic dark matter. However this behavior disappears in the limit of massless missing particle mass, $m_X=0$, as shown in Fig.~\ref{fig:fermionscalar0}. Therefore, in terms of fits to PAMELA and Fermi-LAT, we expect to have similar $\chi^2$/d.o.f. for both scalar and fermionic dark matter.

\subsection{$e^+$ and $e^-$}
\label{sect:edata}

Having established that the injection energy spectra of three-body decaying dark matter do not depend sensitively on the dynamics giving rise to the decay in the mass range of interest, we now illustrate how three-body decaying dark matter could give better fits to both  PAMELA and Fermi-LAT $e^+/e^-$ data than the conventional model of two-bdoy decaying dark matter.

\begin{figure}[t]
\begin{center}
\includegraphics[trim =55mm 2mm 69mm 2mm, clip, width=0.445\textwidth]{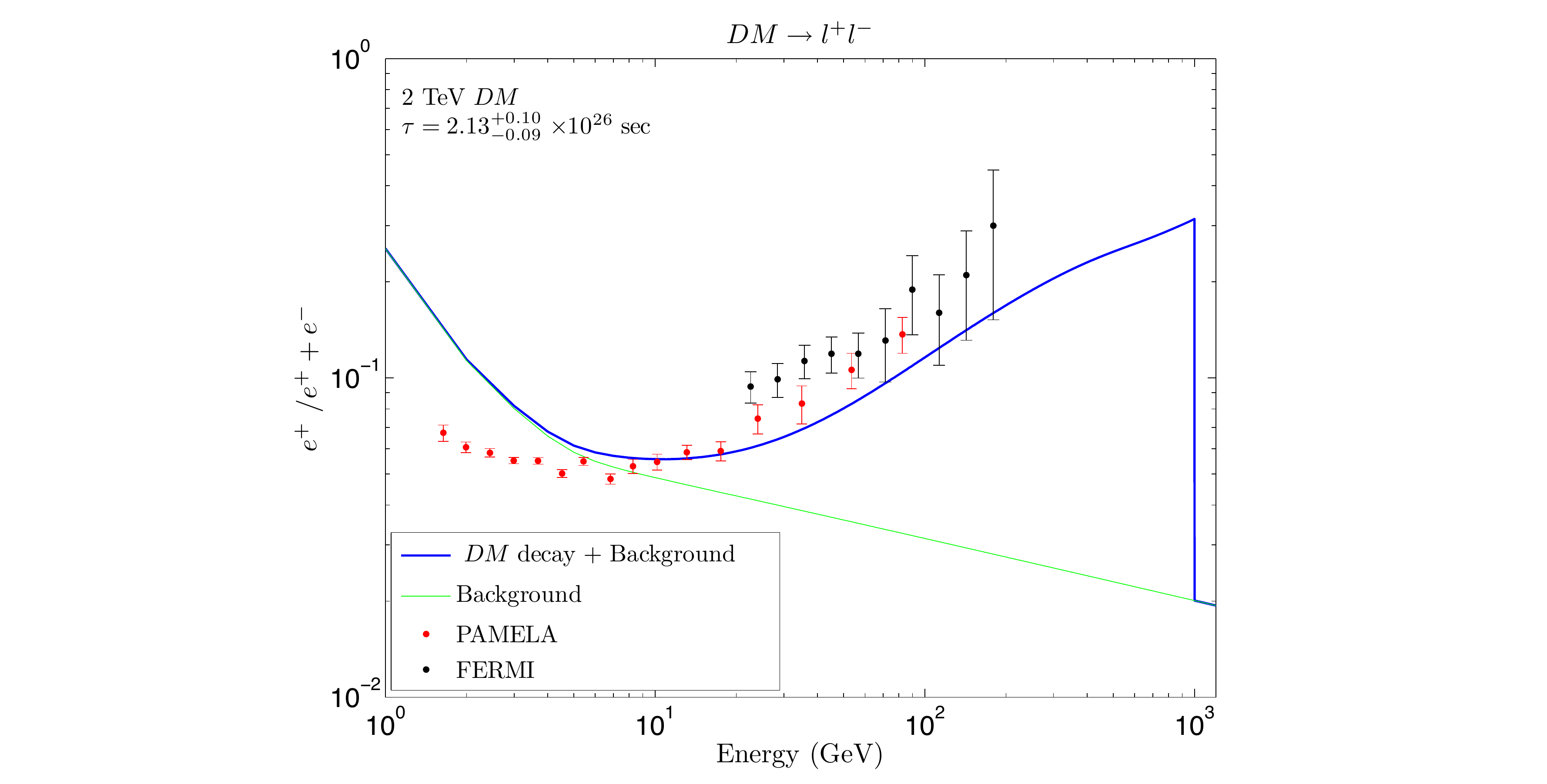}
\includegraphics[trim =55mm 2mm 69mm 2mm, clip, width=0.445\textwidth]{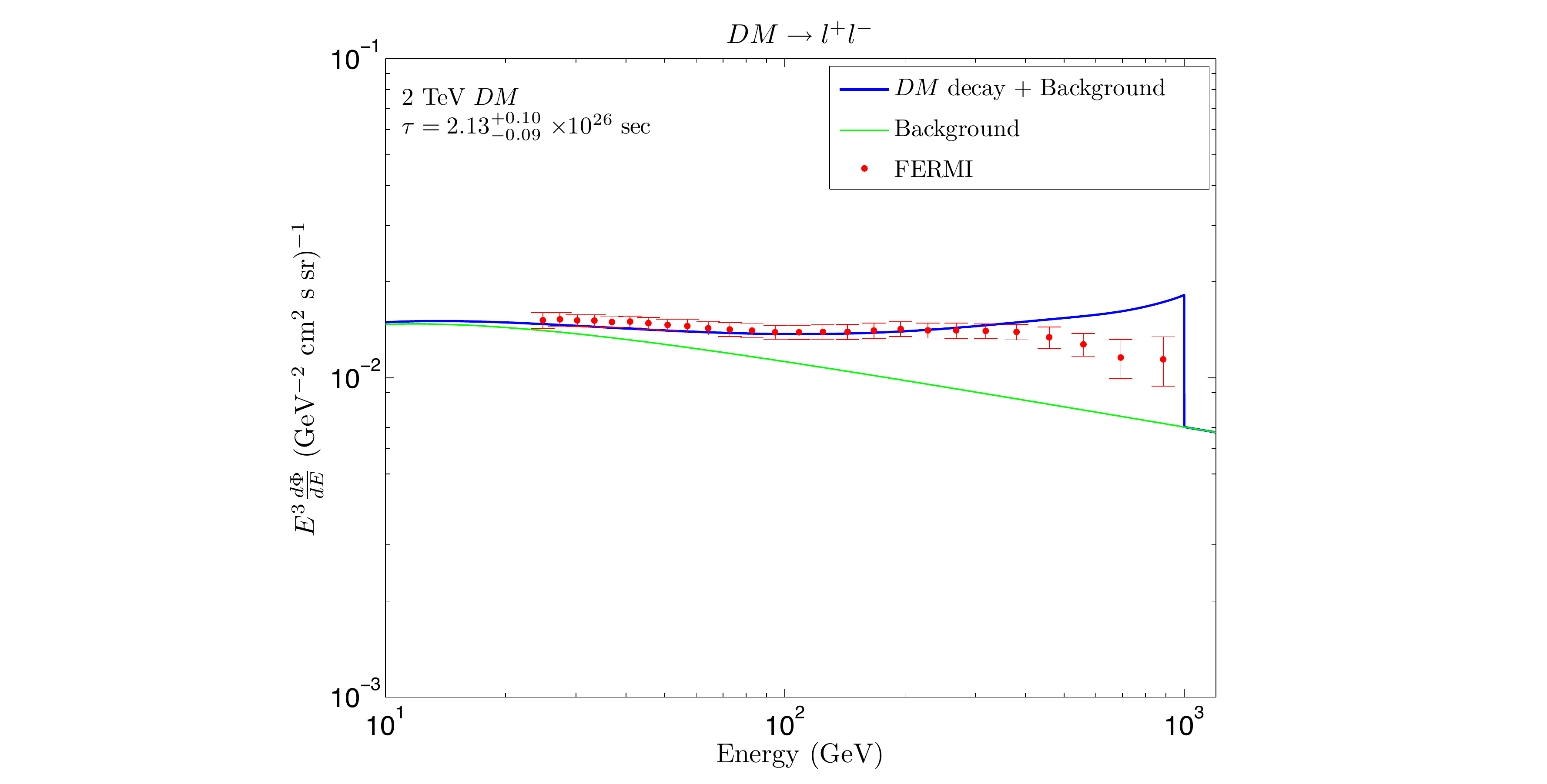}\\
\includegraphics[trim =55mm 2mm 69mm 2mm, clip, width=0.445\textwidth]{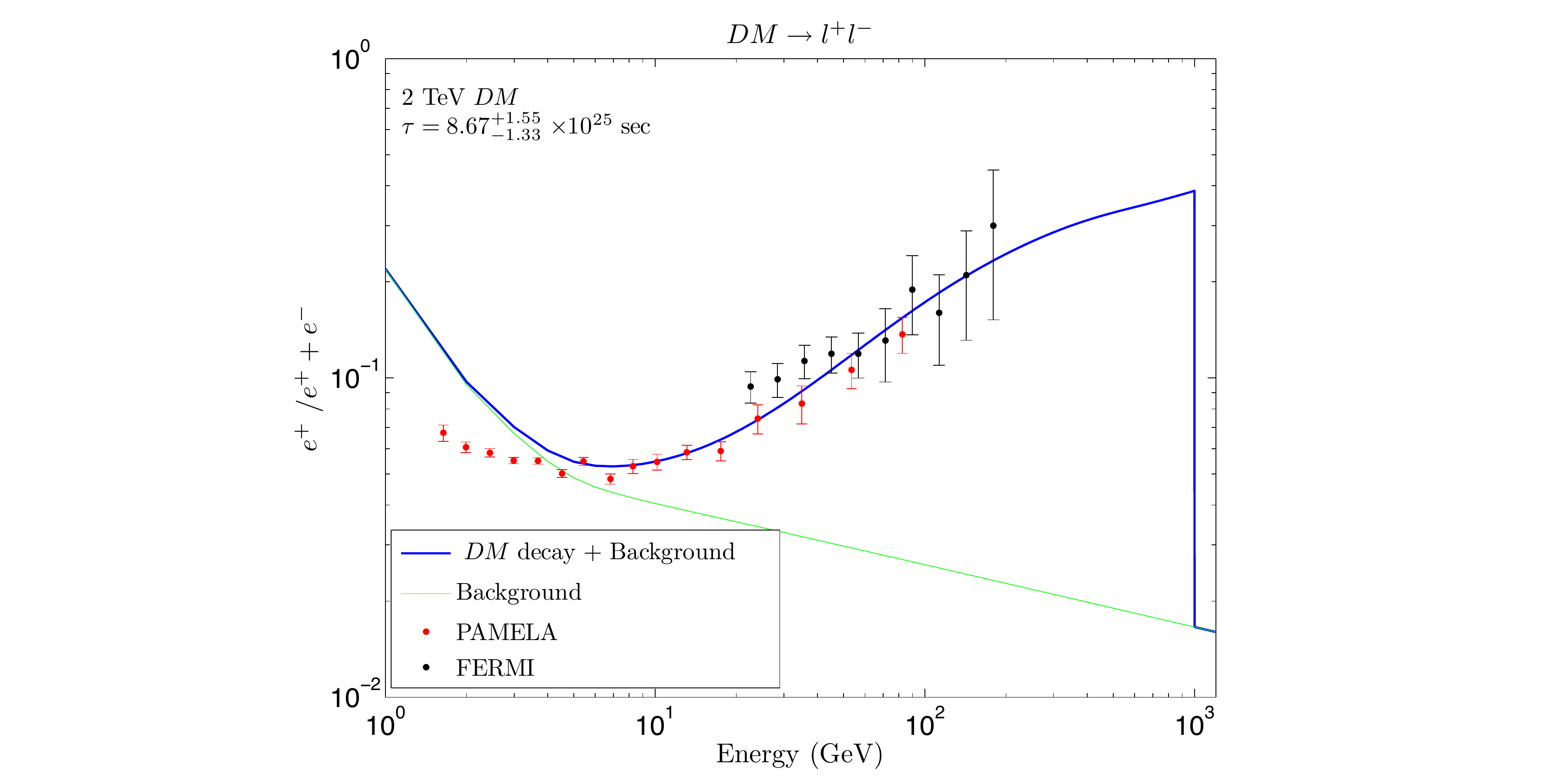}
\includegraphics[trim =55mm 2mm 69mm 2mm, clip, width=0.445\textwidth]{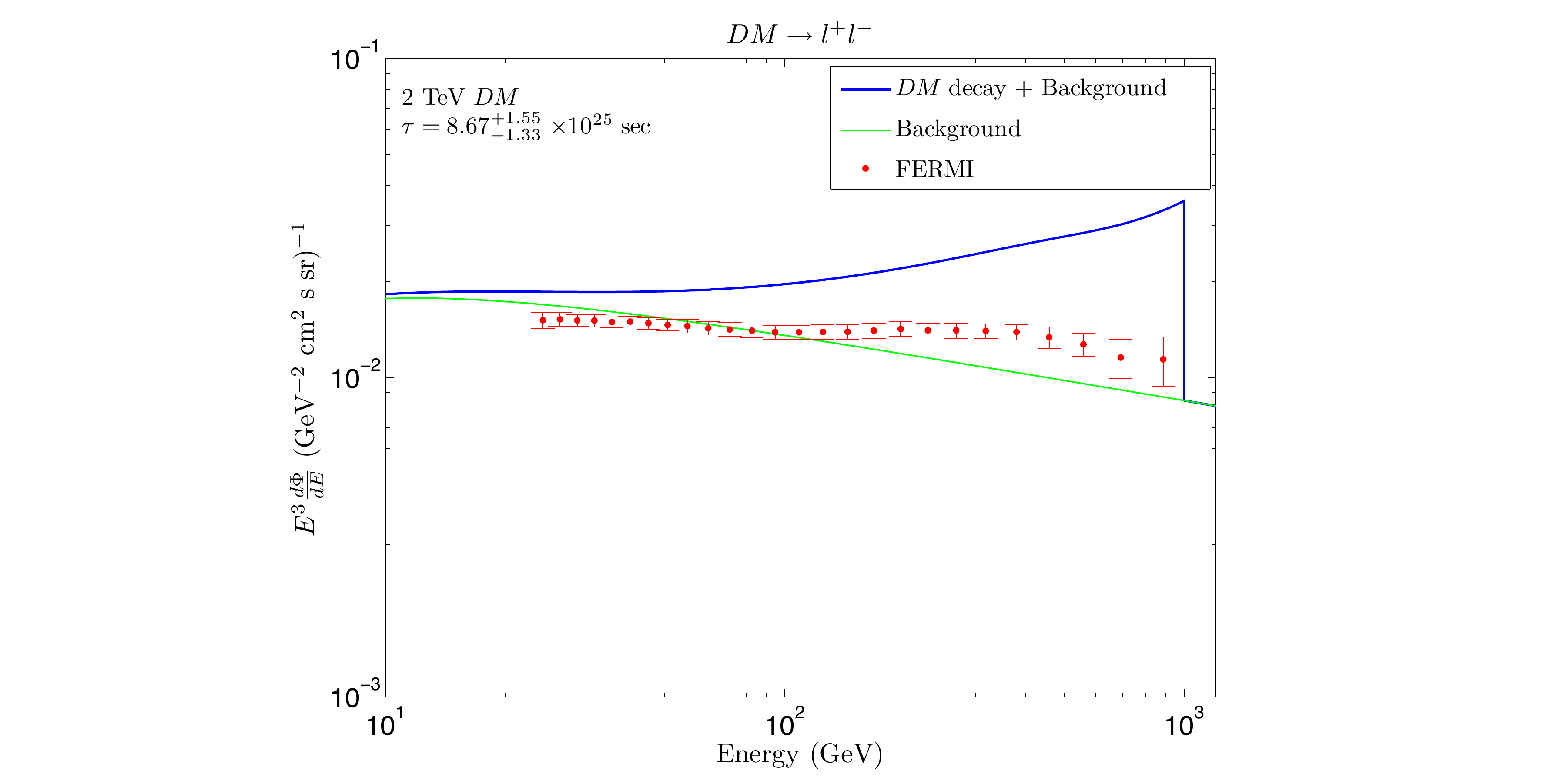}
\caption{\em Fits to PAMELA and Fermi-LAT using two-body decaying dark matter. For the upper panel we include data on both positron fraction and total  $e^- + e^+$ flux in the fit, while in the lower panel we only include the positron fraction data in computing the $\chi^2$. }
\label{e+e- twobody}
\end{center}
\end{figure}

\begin{figure}[t]
\begin{center}
\includegraphics[trim =55mm 2mm 69mm 2mm, clip, width=0.445\textwidth]{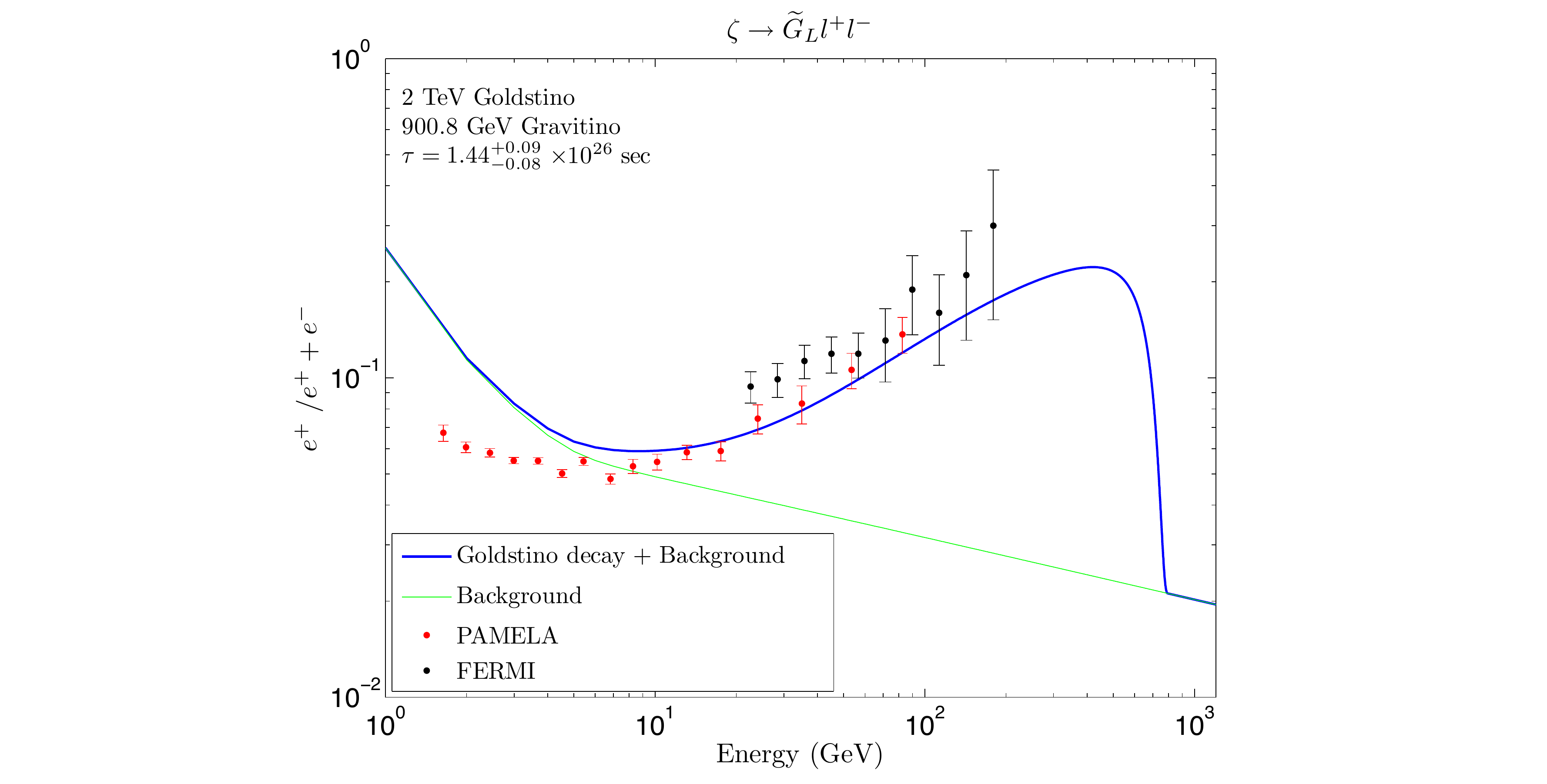}
\includegraphics[trim =55mm 2mm 69mm 2mm, clip, width=0.445\textwidth]{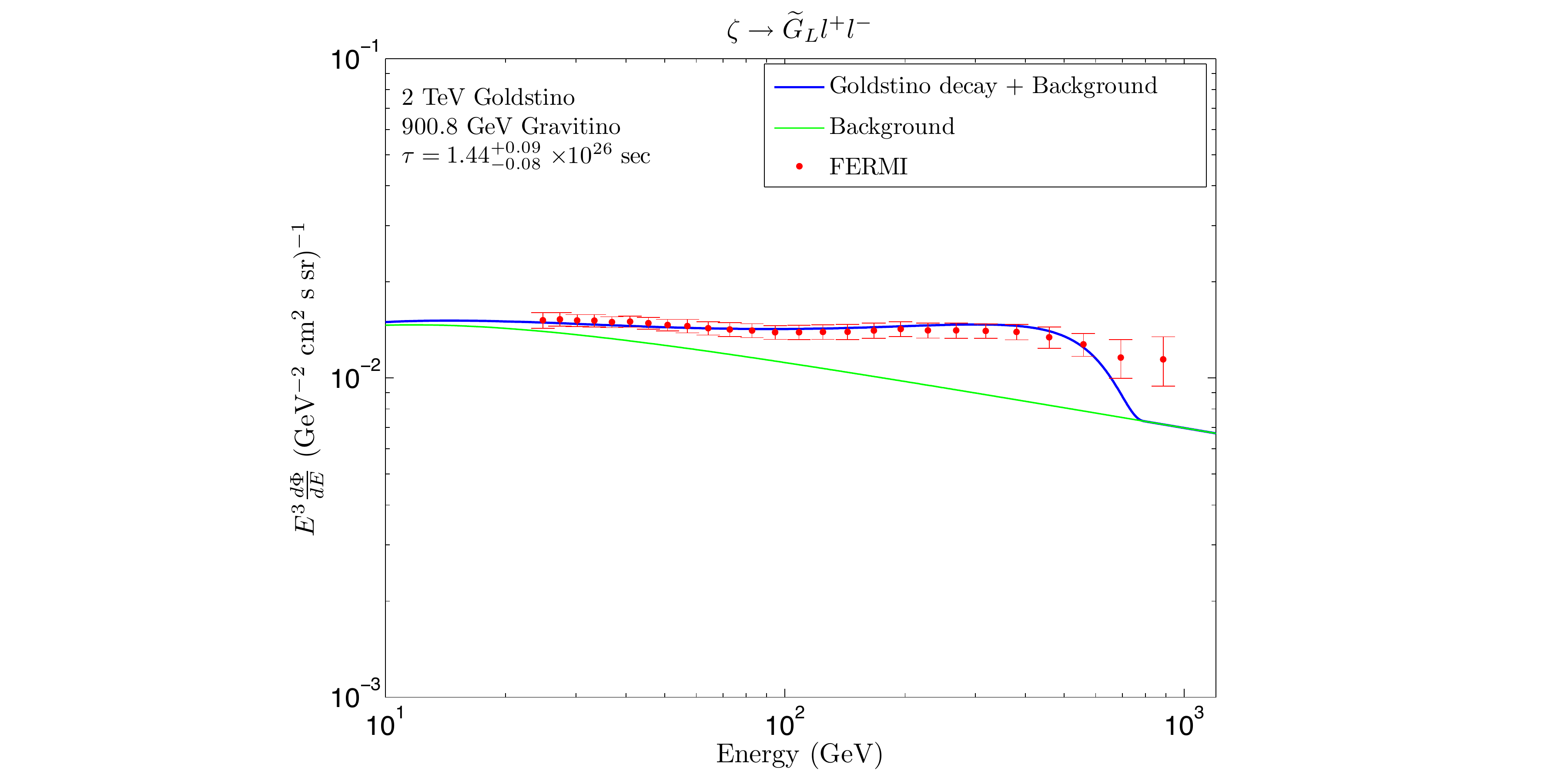}\\
\includegraphics[trim =55mm 2mm 69mm 2mm, clip, width=0.445\textwidth]{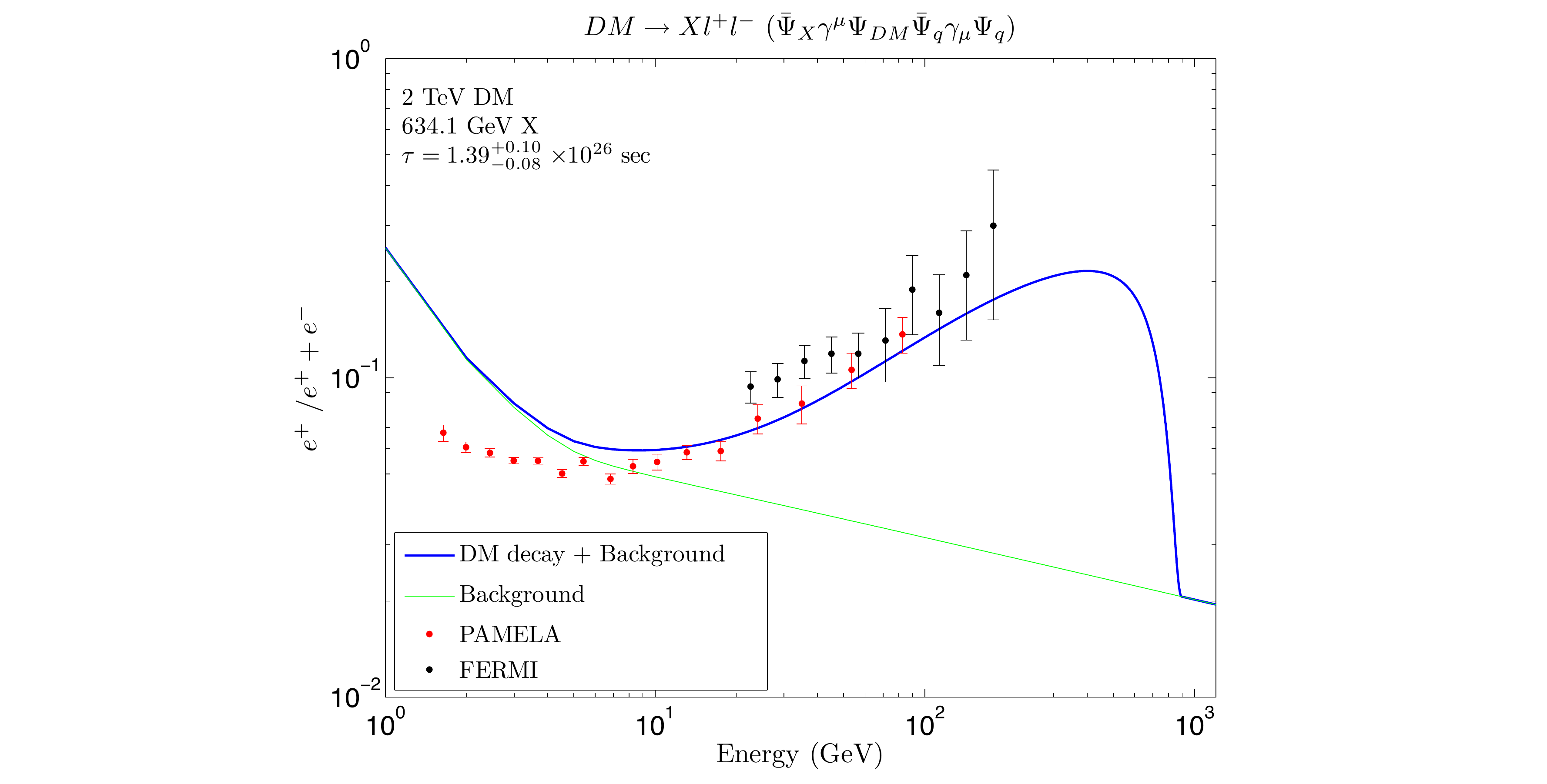}
\includegraphics[trim =55mm 2mm 69mm 2mm, clip, width=0.445\textwidth]{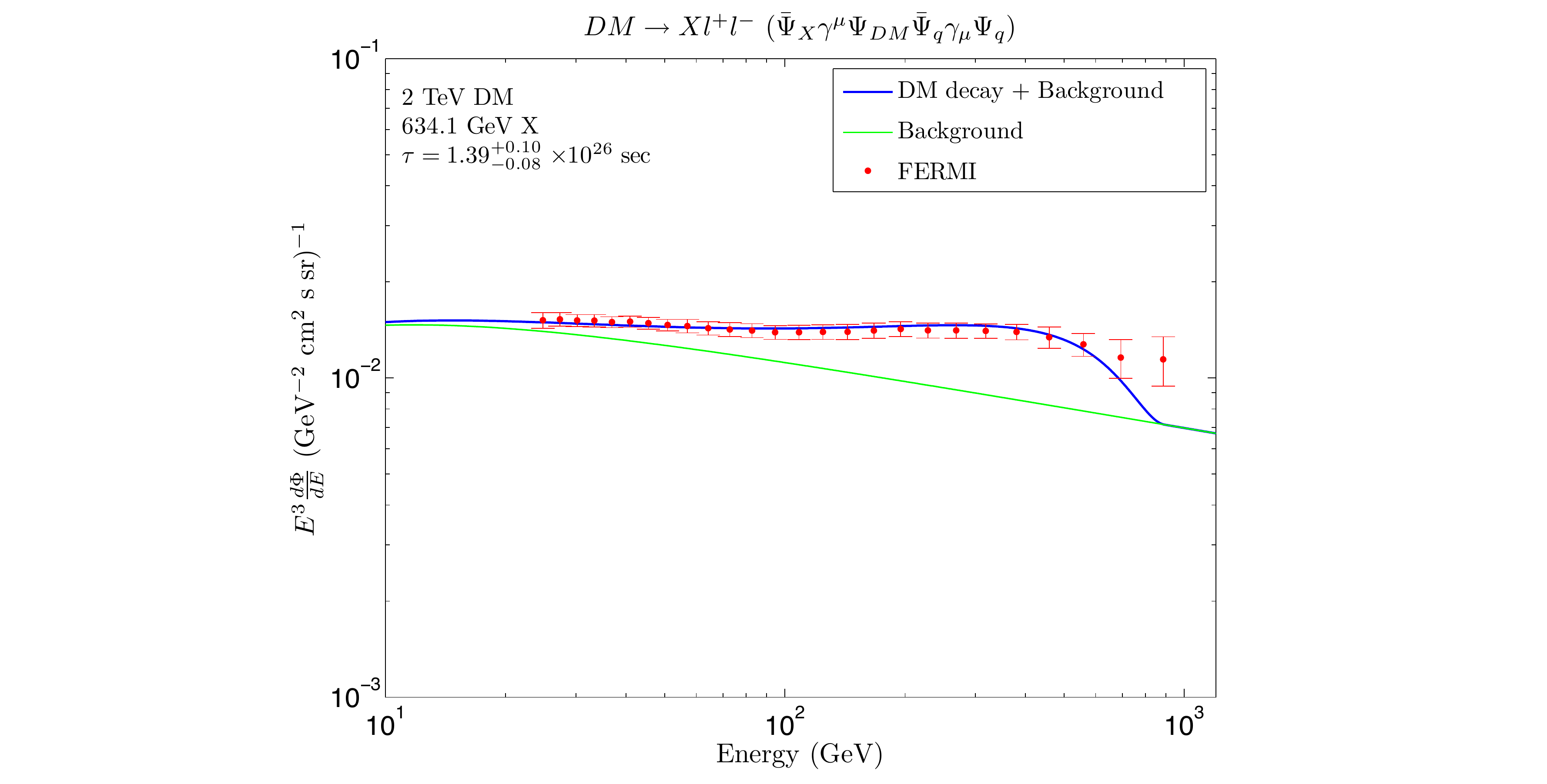} \\
\includegraphics[trim =55mm 2mm 69mm 2mm, clip, width=0.445\textwidth]{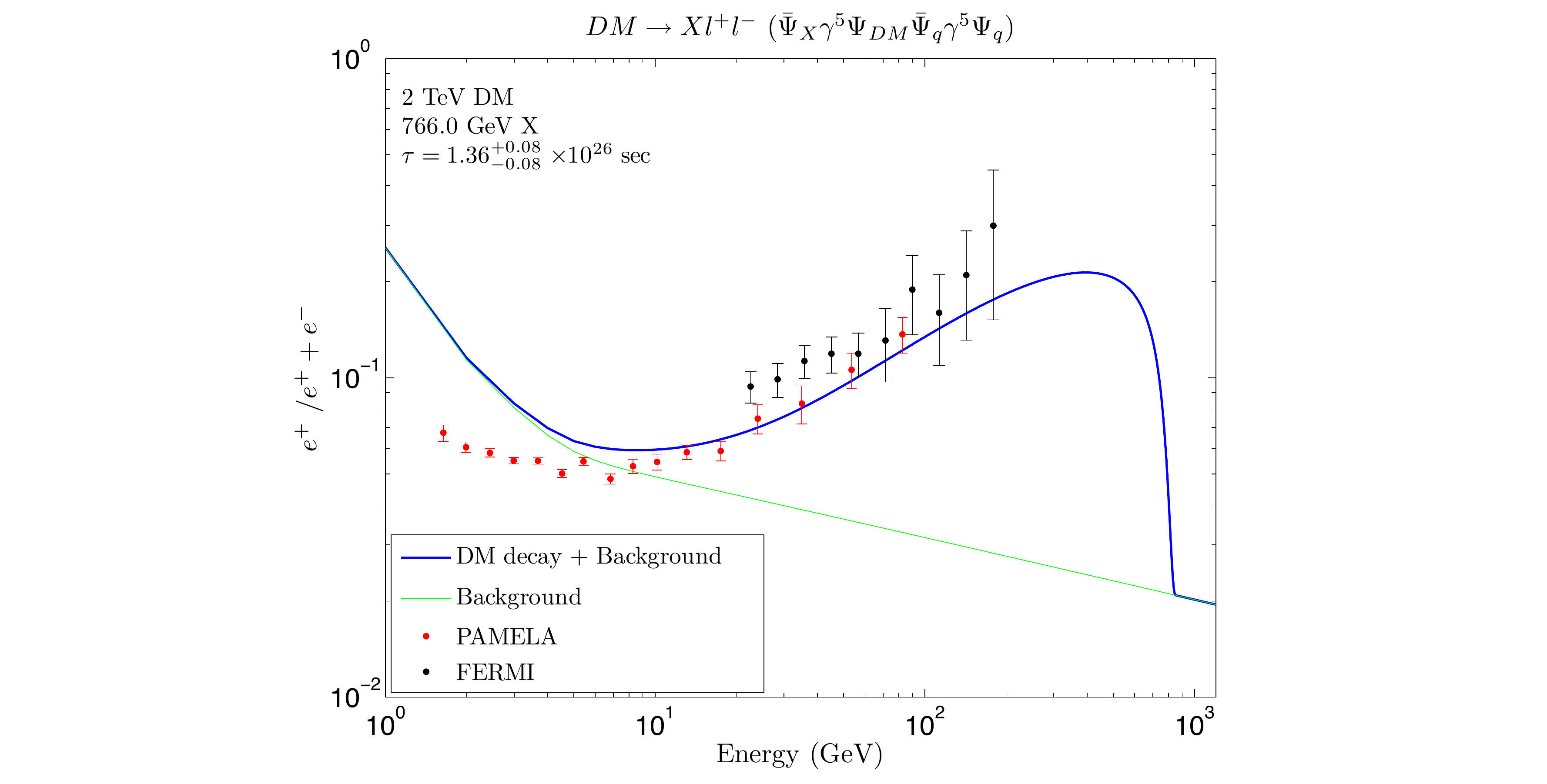}
\includegraphics[trim =55mm 2mm 69mm 3mm, clip, width=0.445\textwidth]{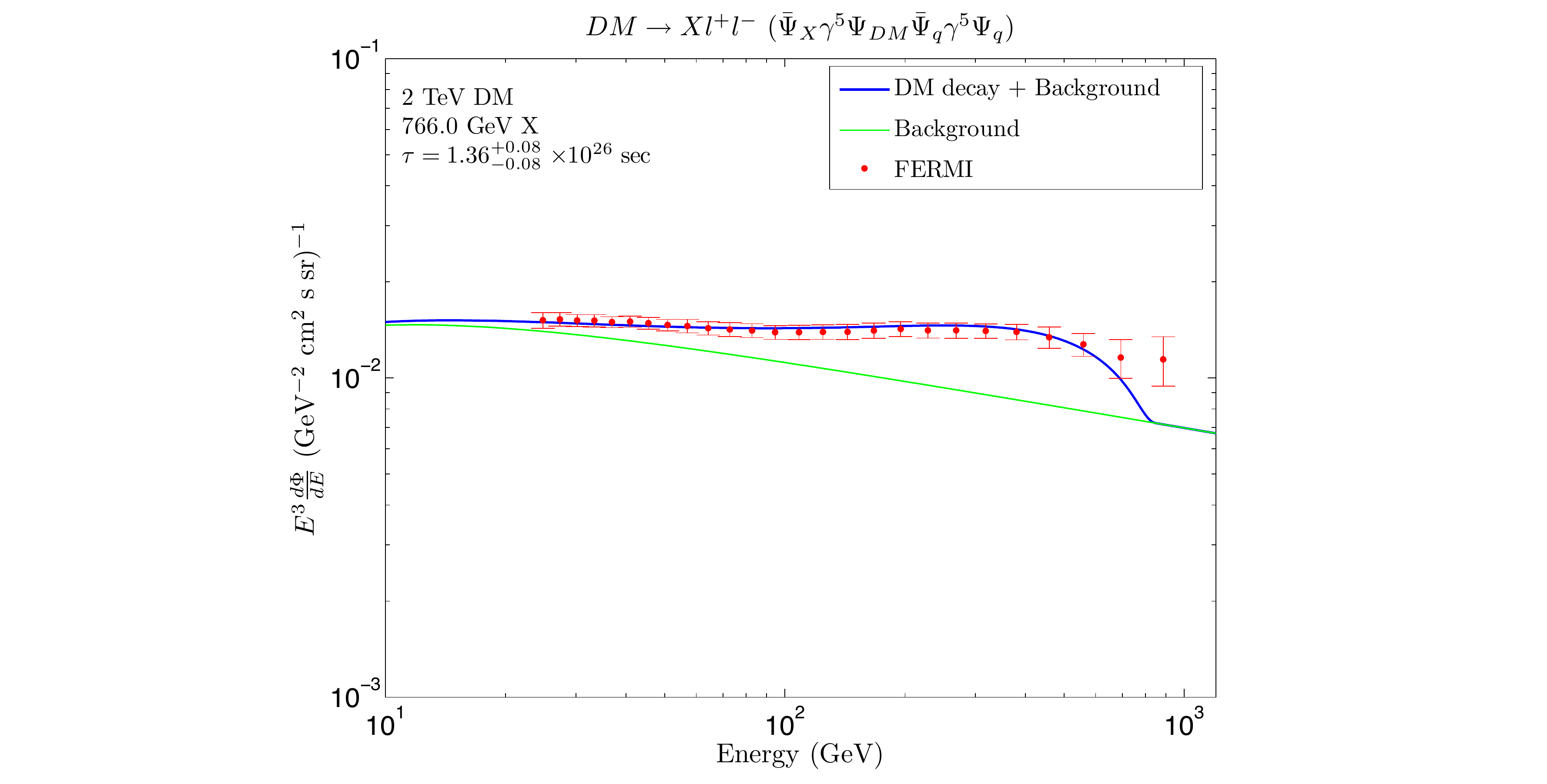} \\
\caption{\em The fits to PAMELA and Fermi-LAT between different operators. Three-body decays have the similar $\chi^2$ and lifetime. $m_{\widetilde{G}_L}$ controls the hardness of injection energy spectra; therefore, goldstino has the largest $m_{\widetilde{G}_L}$ to balance the hardest spectrum.}
\label{e+e- threebody}
\end{center}
\end{figure}

We first examine the fits for two-body decays of dark matter to a pair of leptons with universal couplings to all three generations.
As explained before, the observed excess of positron fraction from both PAMELA and Fermi-LAT can be accounted for by varying the lifetime of the decaying dark matter as long as the injection energy spectrum is not too soft. However, in order to explain the hardening feature in the total $e^- + e^+$ flux observed by Fermi-LAT, the injection spectrum must peak around $\cal O$$(400)$ GeV. As a consequence, the two-body decay has difficulty fitting both the positron fraction and total flux,  since the injection energy is a delta-function peak at half the mass of the dark matter in the centre-of-mass frame, which is demonstrated in the upper panels of Fig.~\ref{e+e- twobody}. Note that one could use only the PAMELA and Fermi-LAT $e^+/(e^- + e^+)$ data in computing the $\chi^2$/d.o.f., then the positron excess can be well described with a shorter lifetime compared to those of three-body decay cases. But the resulting total $(e^- + e^+)$ flux is in severe tension with the Fermi-LAT data, as can be seen in the lower panel of Fig.~\ref{e+e- twobody}.

The best fits for several three-body decay mechanisms are displayed in Fig.~\ref{e+e- threebody}, again assuming universal couplings to all three lepton flavors. We see the overall fits to both PAMELA and the Fermi-LAT are better than the case of two-body decay. Among the different decay mechanisms, the goldstino case has largest missing particle mass, which softens the harder energy spectrum resulting from its derivative couplings. Similarly, the pseudo-scalar operator needs a larger $m_{X}$ than the vector one due to a harder injection spectrum, but  a shorter lifetime, driven mostly by the excess of $e^+/(e^- + e^+)$, to compensate for the flatness in the spectrum shown in Fig.~\ref{fig:Edistf}.

\begin{figure}[t]
\includegraphics[trim =55mm 2mm 69mm 2mm, clip, width=0.445\textwidth]{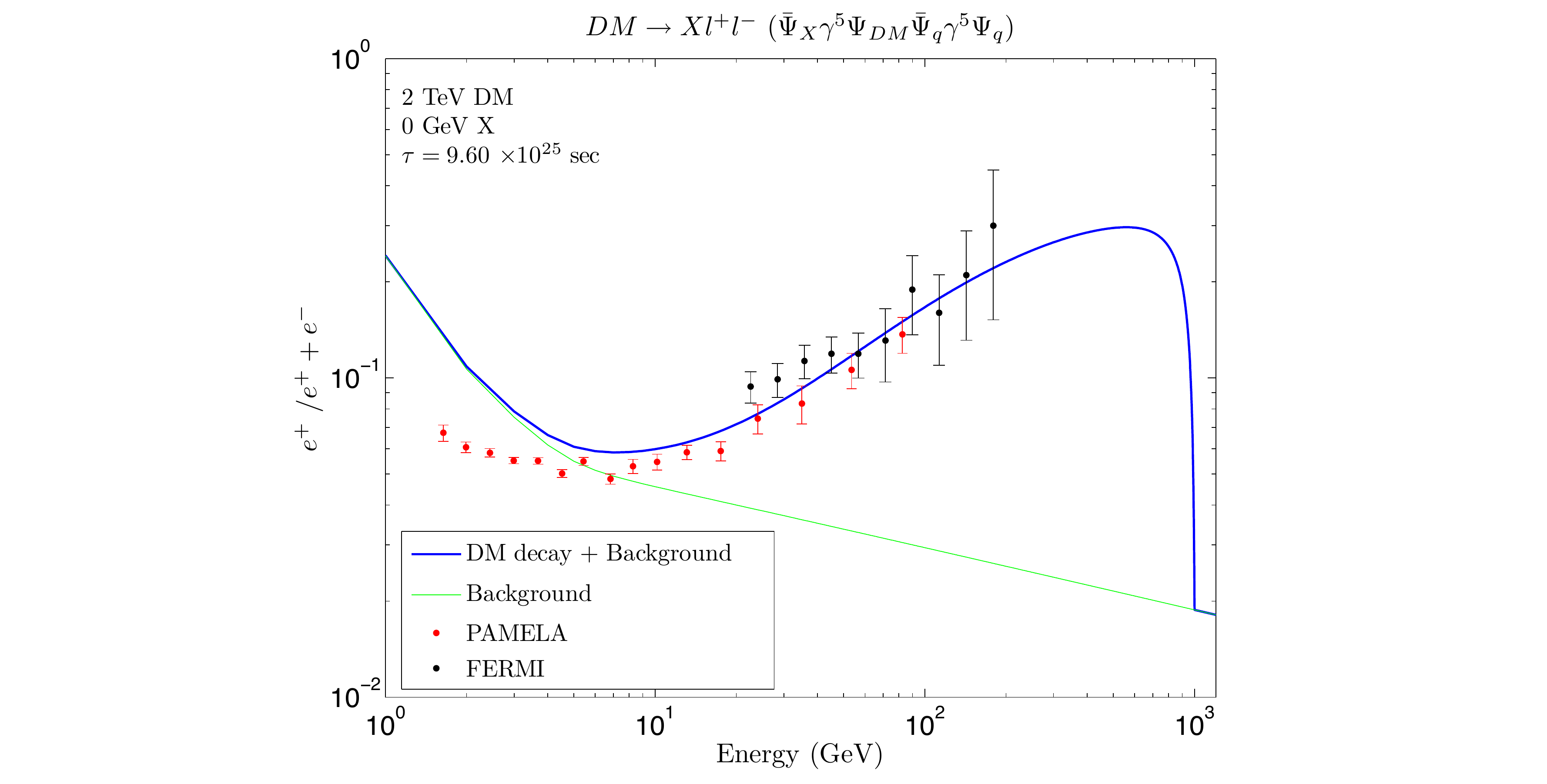}
\includegraphics[trim =55mm 2mm 69mm 2mm, clip, width=0.445\textwidth]{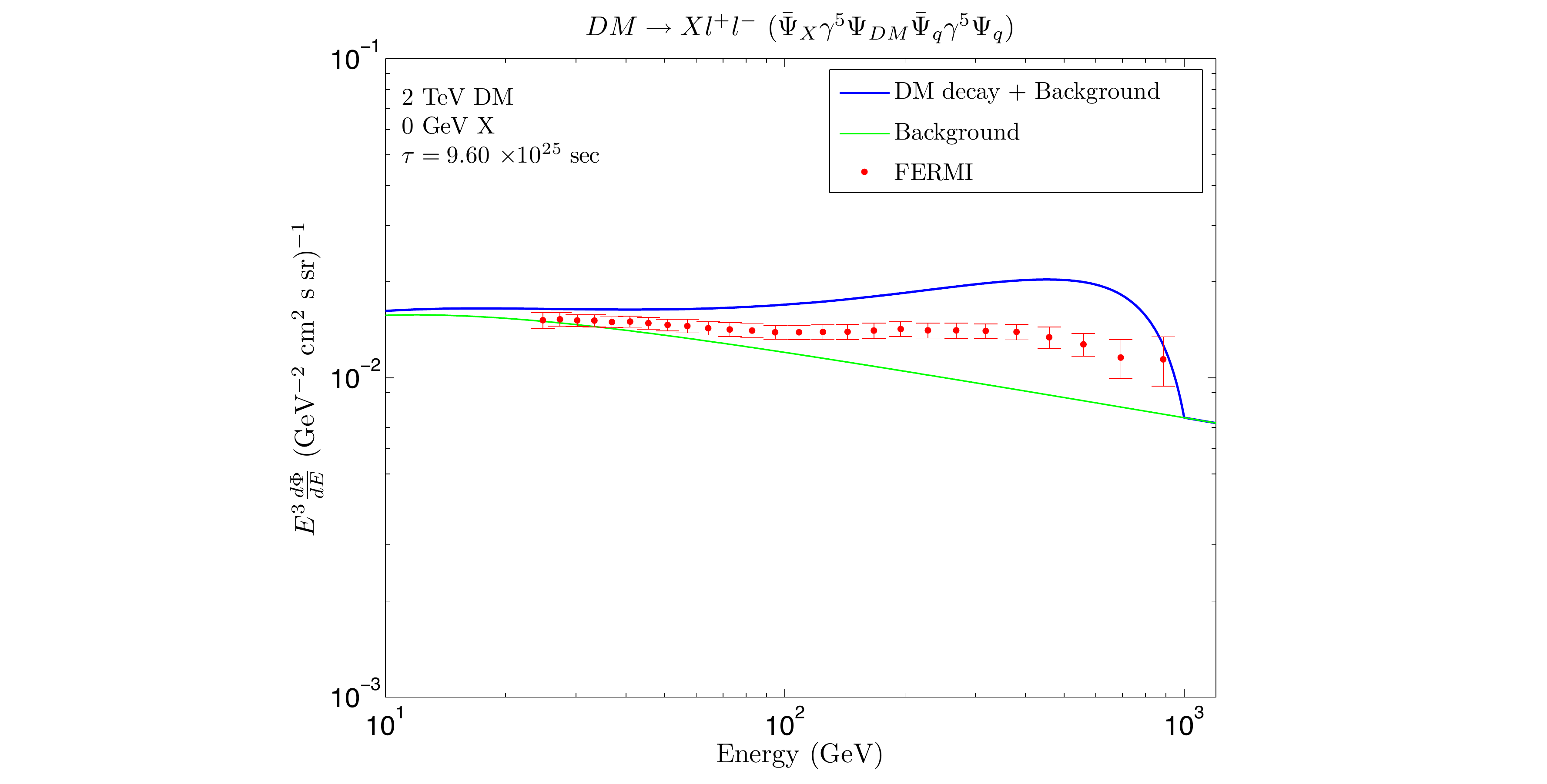}
\caption{\label{fig:D6psmass}\em  Best fits for the pseudo-scalar operator to PAMELA and Fermi-LAT combined, assuming couplings proportional to the fermion mass. The PAMELA(Fermi-LAT) excess can be accounted for properly
at the cost of a longer lifetime and $m_X=0$, which yields an unwanted bump in the Fermi-LAT $e^- +e^+$ data.}
\end{figure}

So far we have assumed universal couplings to all three lepton flavors. However,  the (pseudo-)scalar operators break the chiral symmetry and one may expect that the coefficients are proportional to the fermion masses, which implies that
the decay to $\tau$ leptons will be dominant. In this case, these two operators can not explain both PAMELA and Fermi-LAT data.
The  spectrum of $e^+(e^-)$ from $\tau$ decays is quite soft and would populate mostly the low-energy region. In order to explain the excess in PAMELA data, the lifetime has to be increased significantly and the missing particle mass $m_{X}$ has to be reduced, which results in a much harder spectrum than the Fermi-LAT total $e^- + e^+$ flux data as shown in Fig.~\ref{fig:D6psmass}.

\section{Astrophysical Constraints}
\label{sect:constraints}

In this section we consider astrophysical constraints on three-body decaying dark matter. There are three main categories: 1) the diffuse gamma-ray due to inverse Compton scatterings (ICS) and final state radiation (FSR), 2) prompt photons that are direct decay product of the dark matter, and 3) anti-proton flux measurements which constrain decays into hadronic final states.

\subsection{Diffuse $\gamma$}

It is known that the leptonic final states from dark matter decays yield photons via FSR as well as ICS when the produced leptons interact with background photons.  Both are subject to the constraints from the Fermi-LAT gamma ray data. In this subsection, we study such constraints  by assuming the $e^+/e^-$ excess in PAMELA and Fermi-LAT is the consequence of dark matter decays. Since we have shown that different three-body decay operators have similar best-fits to $e^+/e^-$ data for $2$ TeV dark matter mass, it is sufficient to focus on the goldstino case only.

We make use of the Fermi-LAT gamma ray data in Ref.~\cite{Abdo:2010nz}, which provides information of diffuse Galactic emission (DGE) and extragalactic gamma ray background (EGB).  For DGE we fit to the ``Galactic diffuse (fit)'' data from Table I in Ref.~\cite{Abdo:2010nz}, which is DGE averaged over the Galactic latitude range $|b| \geq 10^{\circ}$ as measured by Fermi-LAT. The DM signal results from the sum of ICS and FSR produced by charged leptons, the products of the goldstino decay. The averaged photon flux from FSR over a solid angle of interest from Table 3 in Ref.~\cite{Cirelli:2010xx} and a numerical code for ICS in Ref.~\cite{cookbook website} are utilized for calculating the photon flux from the Galactic dark matter decay. For the fitting procedure, we vary the normalization of the background, whose shape is taken from the ``Galactic diffuse (model)'' numbers in Table 1 of Ref.~\cite{Abdo:2010nz}, to minimize the $\chi^2$ for various goldstino decay widths.  In addition, for the goldstino decay, the gravitino mass is extracted from the best fits to the $e^+/e^-$ and $e^+ + e^-$ data.

Before presenting the results, we would like to comment on the effect of different electron propagation models, dubbed MIN, MED, and MAX~\cite{Delahaye:2007fr}~\cite{Donato:2003xg}.
For FSR, secondary photons are produced near where primary leptons are produced and once
produced, photons propagate directly toward the Earth; therefore, they are
independent of the propagation models of electrons through the galaxy.
On the other hand, ICS photons are produced from electrons scattering off background photons while the electrons propagate and lose
energy. Hence, it has dependence on the propagation models of electrons.
For the MIN model, the low/intermediate energy electron ($\lesssim 200$ GeV) will be more easily stopped and fail to produce ICS photons. Fermi-LAT gamma ray data ranging from $0.2$ to $100$ GeV are influenced by this effect.
Therefore, for both two- and three-body decay, decay scenarios adopting the MIN propagation model are less constrained
compared to MAX and MED. MAX has an opposite feature, i.e., producing more low/intermediate ICS photons but the effect is less dramatic than MIN. In other words, MAX is similar to MED.

\begin{figure}[t]
 \vspace{-20pt}
\begin{center}
\includegraphics[trim =55mm 2mm 69mm 2mm, clip, width=0.45\textwidth]{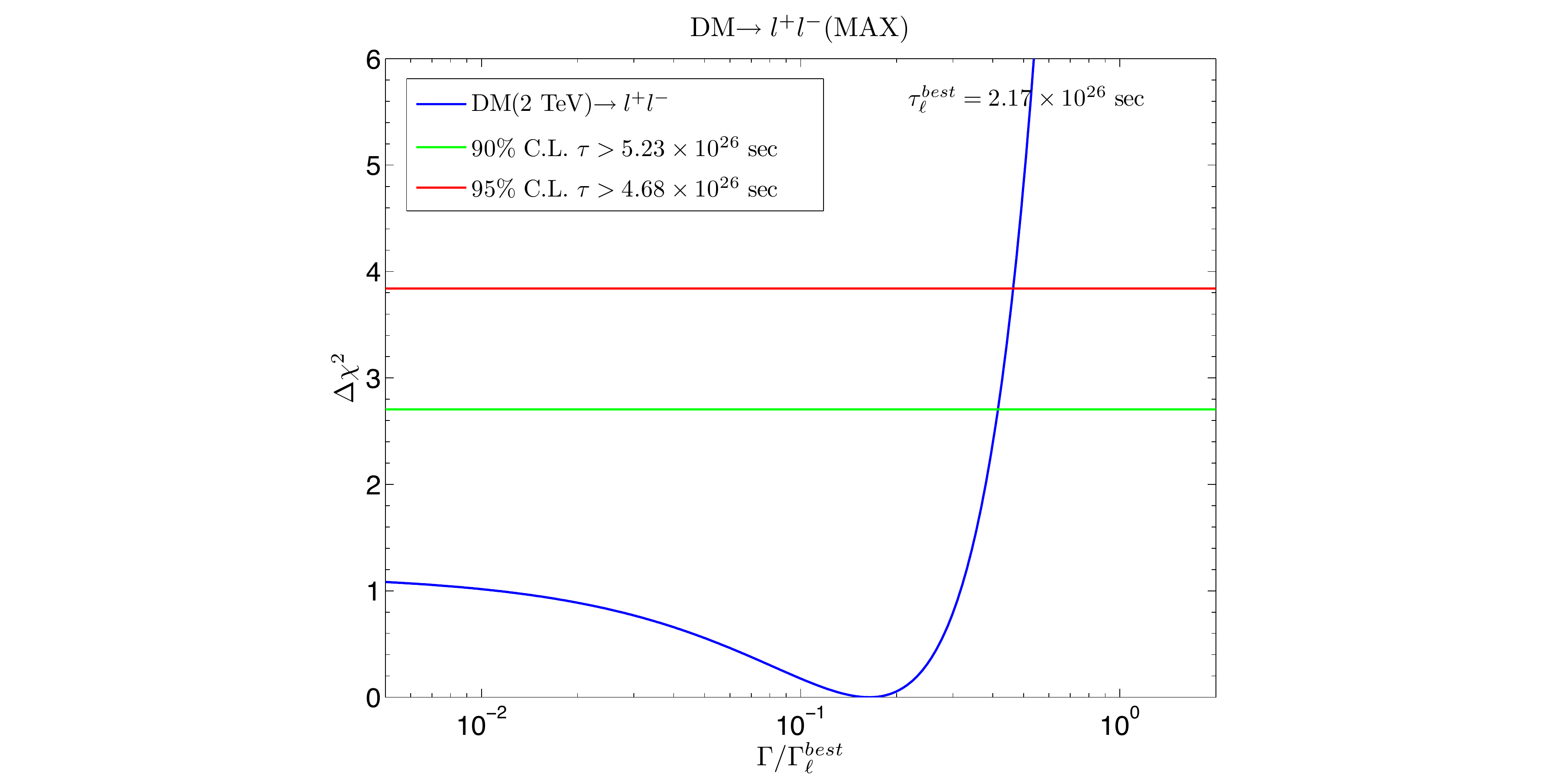}
\includegraphics[trim =55mm 2mm 69mm 2mm, clip, width=0.45\textwidth]{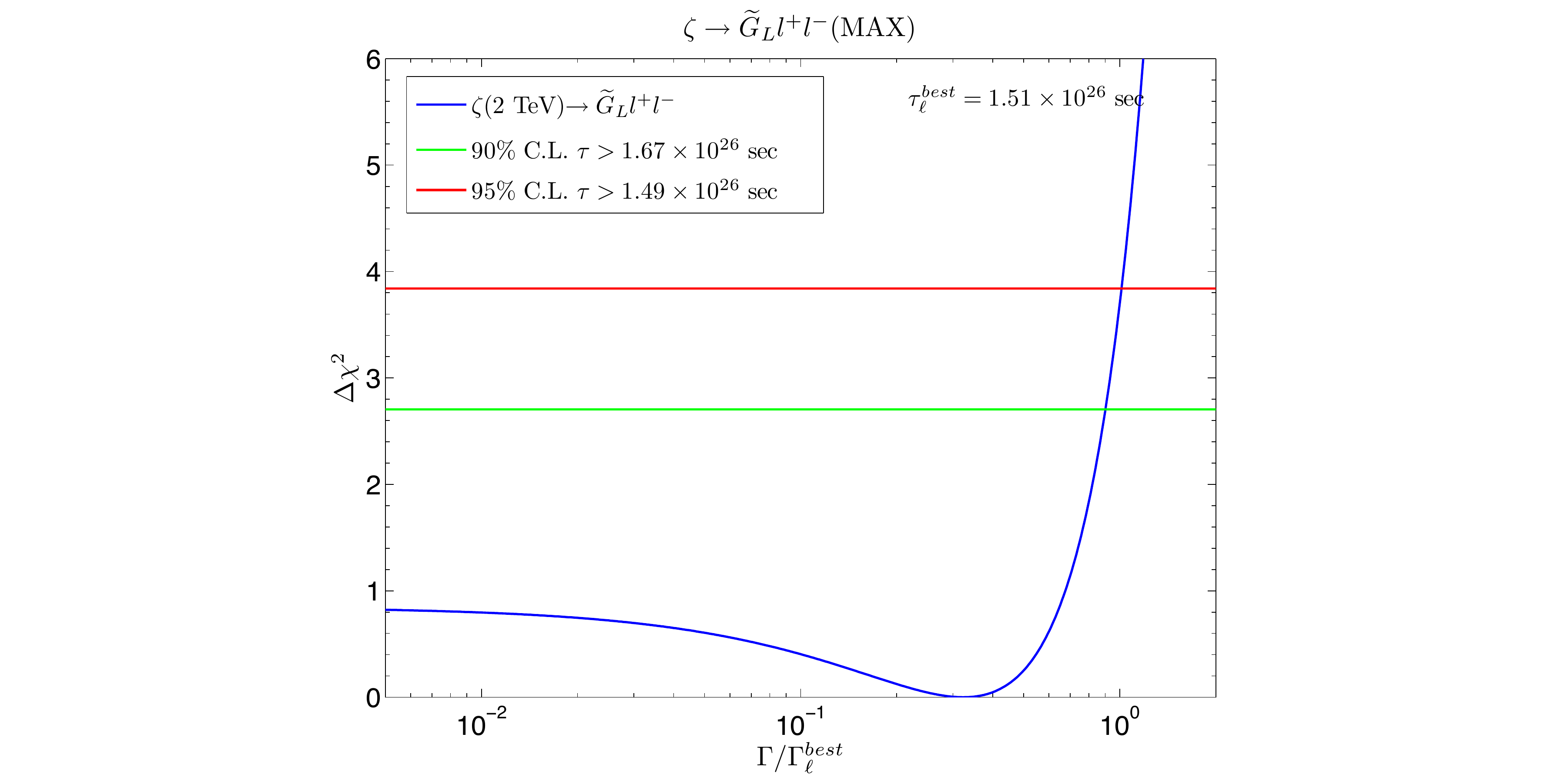}\\
\includegraphics[trim =55mm 2mm 69mm 2mm, clip, width=0.45\textwidth]{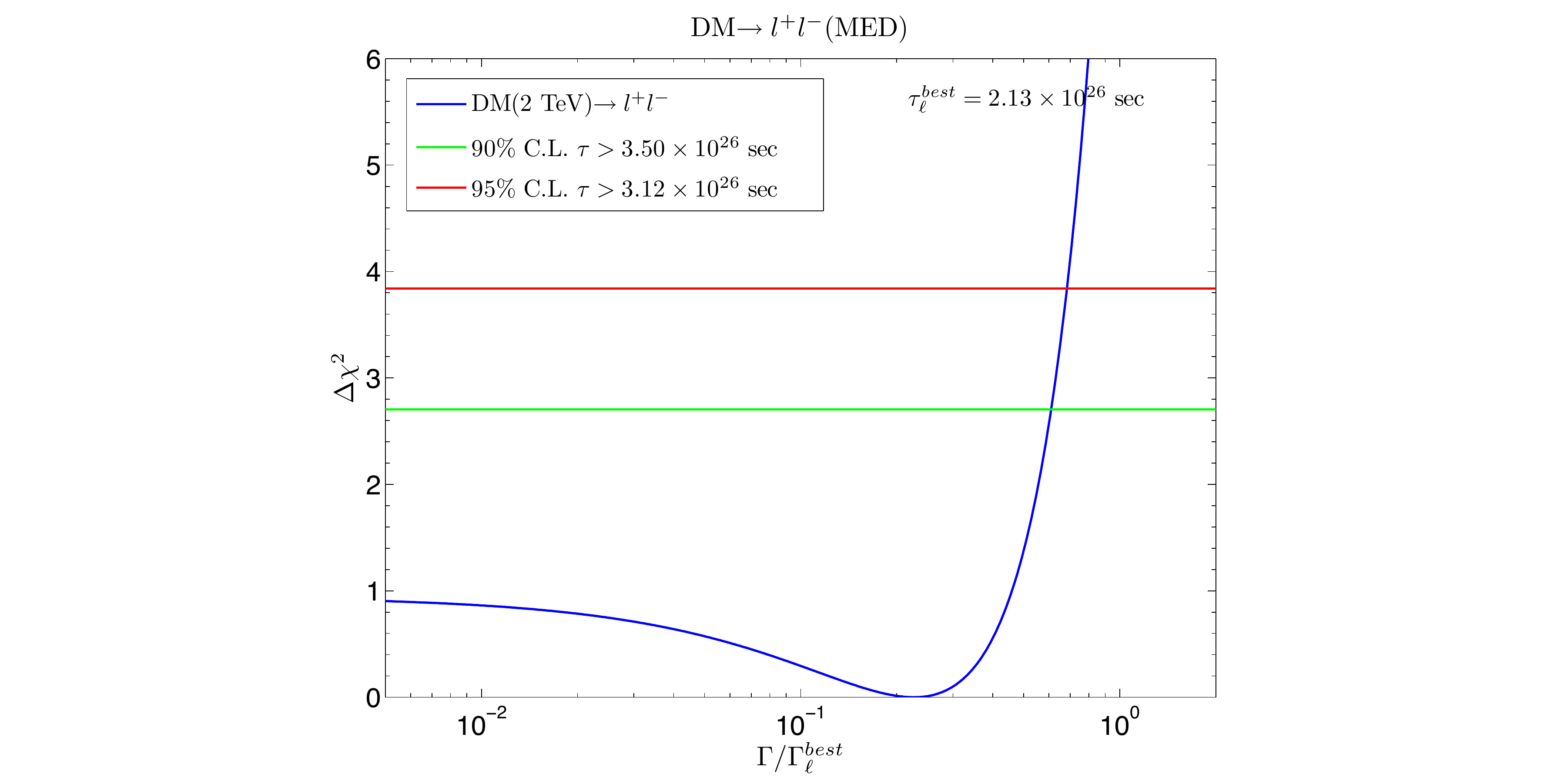}
\includegraphics[trim =55mm 2mm 69mm 2mm, clip, width=0.45\textwidth]{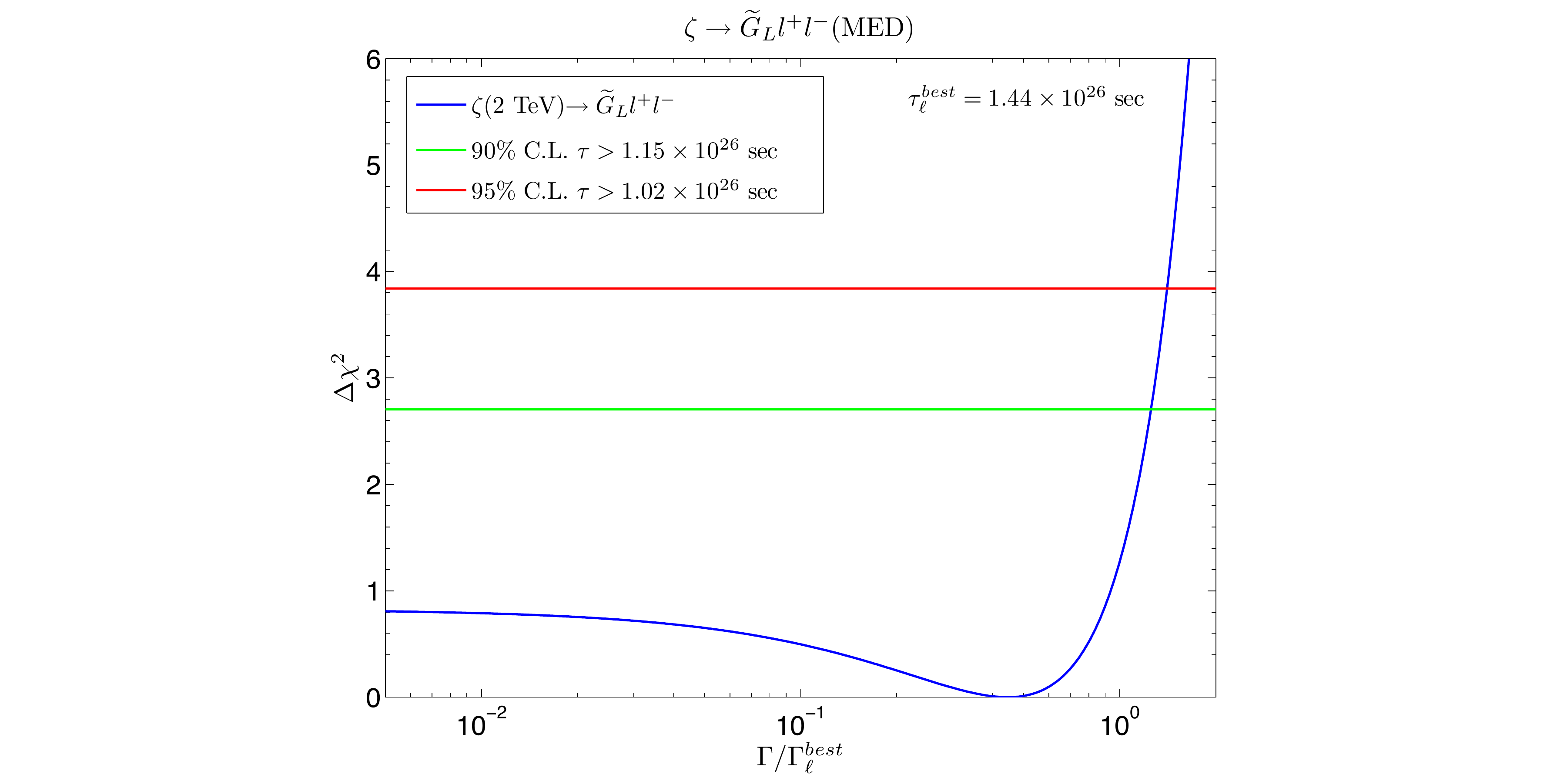}\\
\includegraphics[trim =55mm 2mm 69mm 2mm, clip, width=0.45\textwidth]{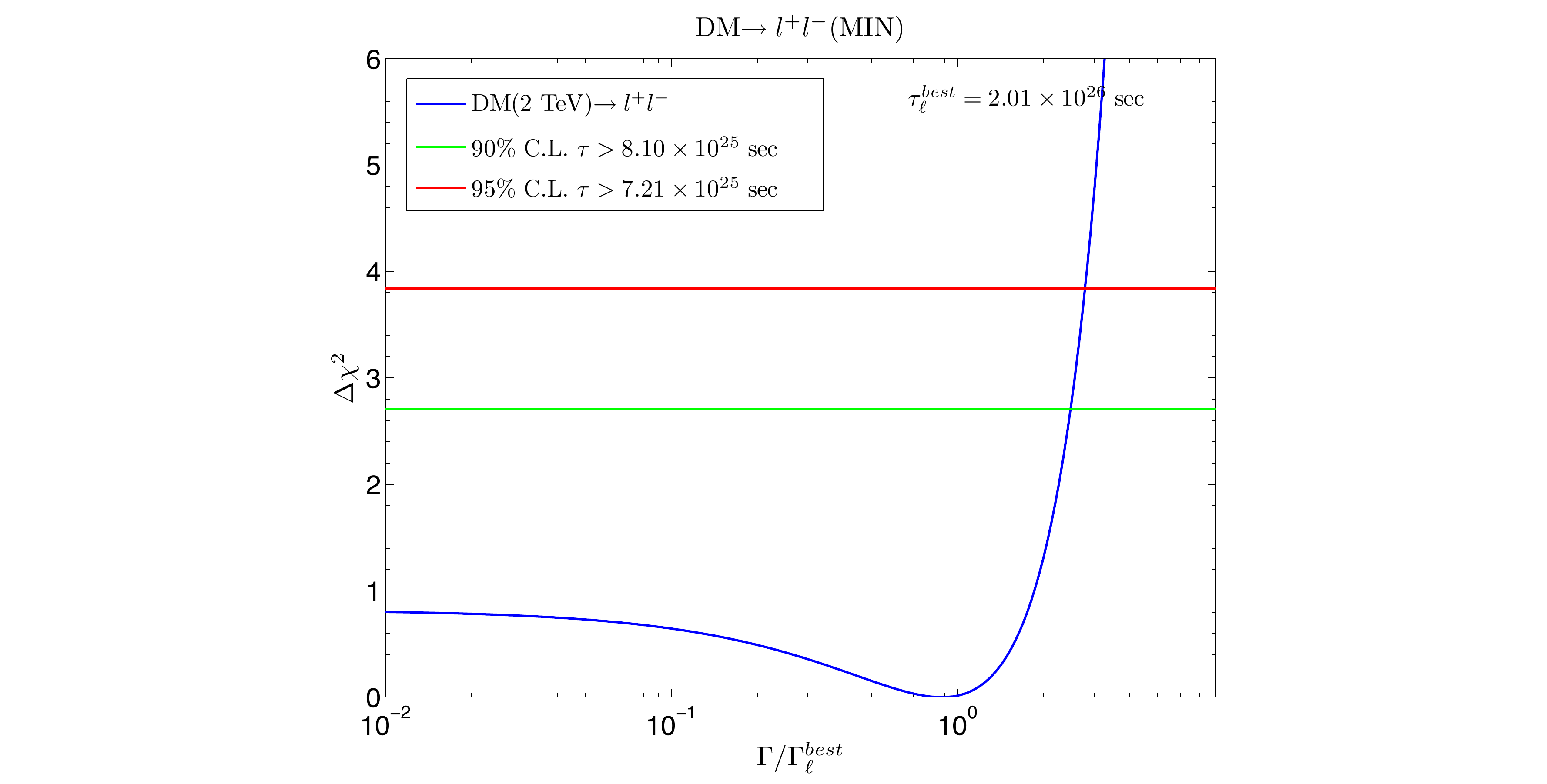}
\includegraphics[trim =55mm 2mm 69mm 2mm, clip, width=0.45\textwidth]{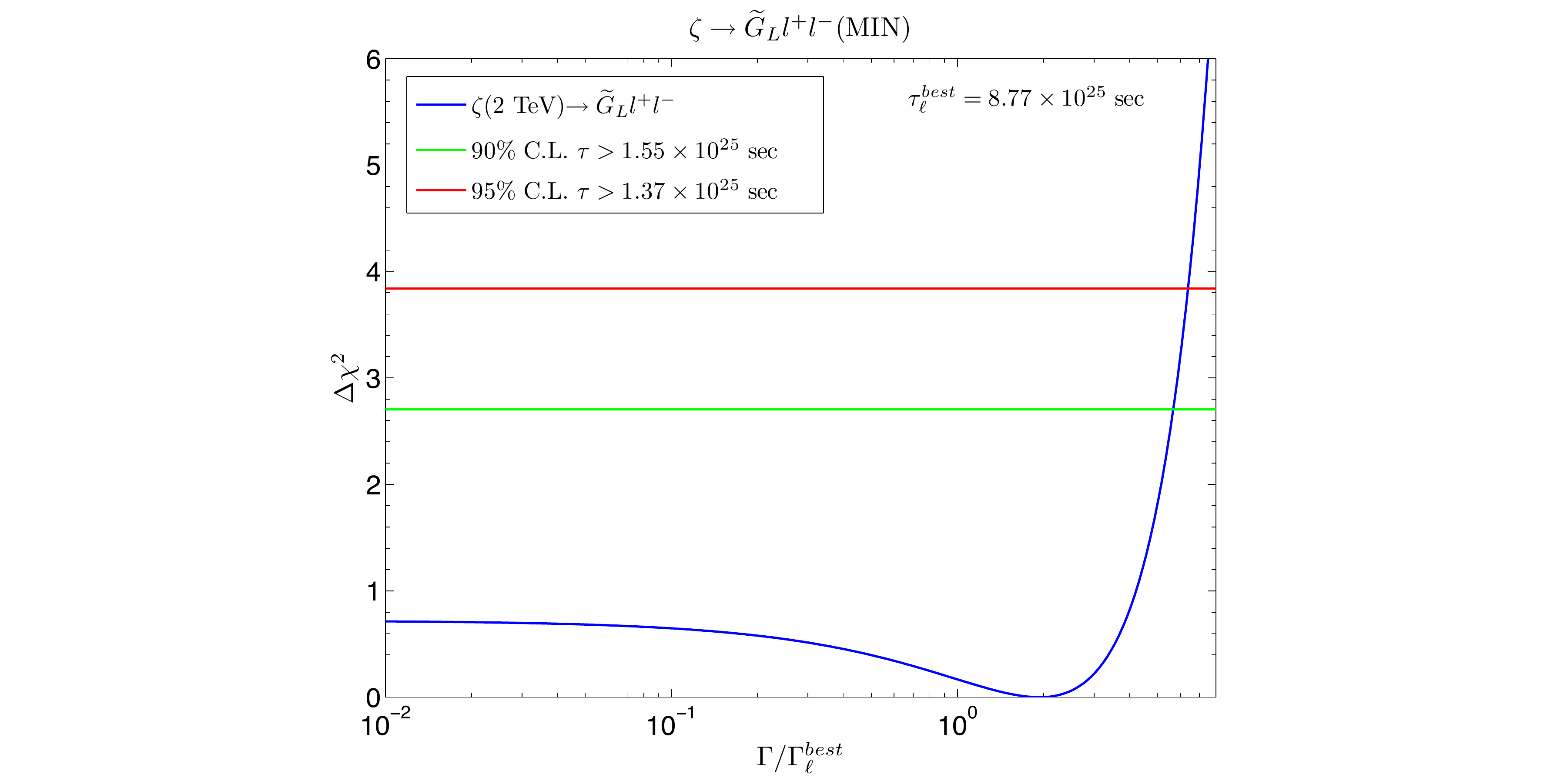}\\
 \vspace{-20pt}
\caption{\em The goldstino decay can satisfy the Fermi-LAT DGE gamma ray constraints. In contrast, for the two-body decay, only MIN is allowed by virtue of the reason mentioned in the text. Note that for the certain range of the DM lifetime, the existence of the DM signal compensates the difference between the data and background and in turn results in dips in $\chi^2$ fits.}
 \vspace{-20pt}
\label{ICSFSR two-three}
\end{center}
\end{figure}

We present results in Fig.~\ref{ICSFSR two-three} corresponding to the 90\% and 95\% C.L. upper (lower) limit on the decay width (or equivalently lifetime) of the goldstino, normalized to $\Gamma^{best}_{\ell}$ from best fits to $e^+$ and $e^-$ data, based on the Fermi-LAT DGE data. Again, the photon flux resulted from the goldstino decay consists of both ICS and FSR contributions. The goldstino decay can satisfy the diffuse gamma ray constraints. In contrast, for two-body decays with the MAX or MED propagation model, the decaying dark matter is ruled out as an explanation to the excess of $e^+$ and $e^-$. In the case of MIN, both two- and three-body decay can avoid the gamma ray constraint as explained before. Note that FSR contribution is numerically subdominant compared to those of ICS on both two- and three-body cases. The reason is that the $\tau$ decay channel, which produces relatively more photons via showering than $e$ and $\mu$ channel, gets diluted by the assumption of universal leptonic couplings. Therefore, if the dark matter decays into $\tau$ only, FSR would be far more stringent than ICS as demonstrated in Refs.~\cite{Cirelli:2009dv, Papucci:2009gd}.

\begin{figure}[t]
  \vspace{-20pt}
\begin{center}
\includegraphics[trim =55mm 2mm 69mm 2mm, clip, width=0.45\textwidth]{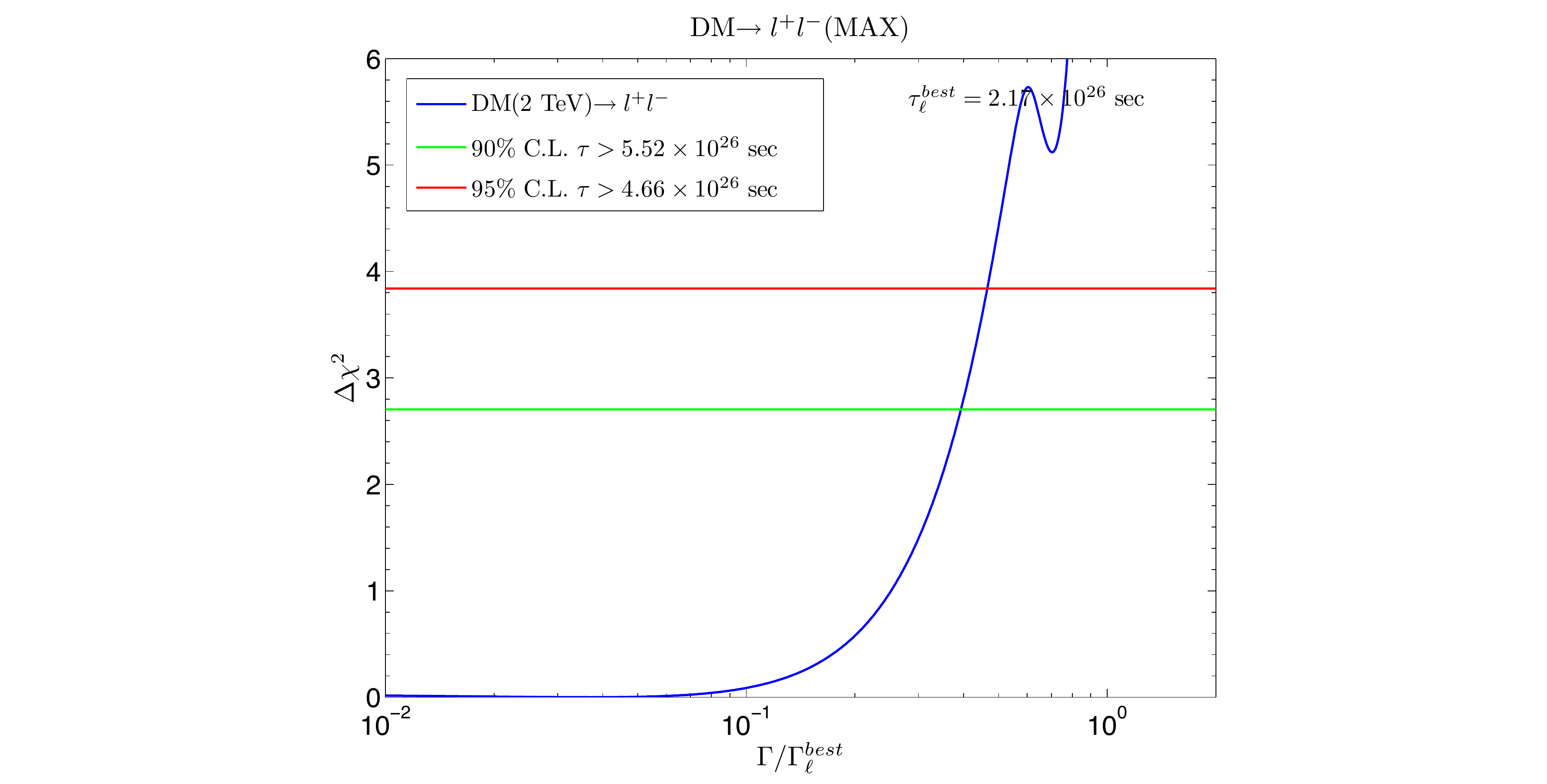}
\includegraphics[trim =55mm 2mm 69mm 2mm, clip, width=0.45\textwidth]{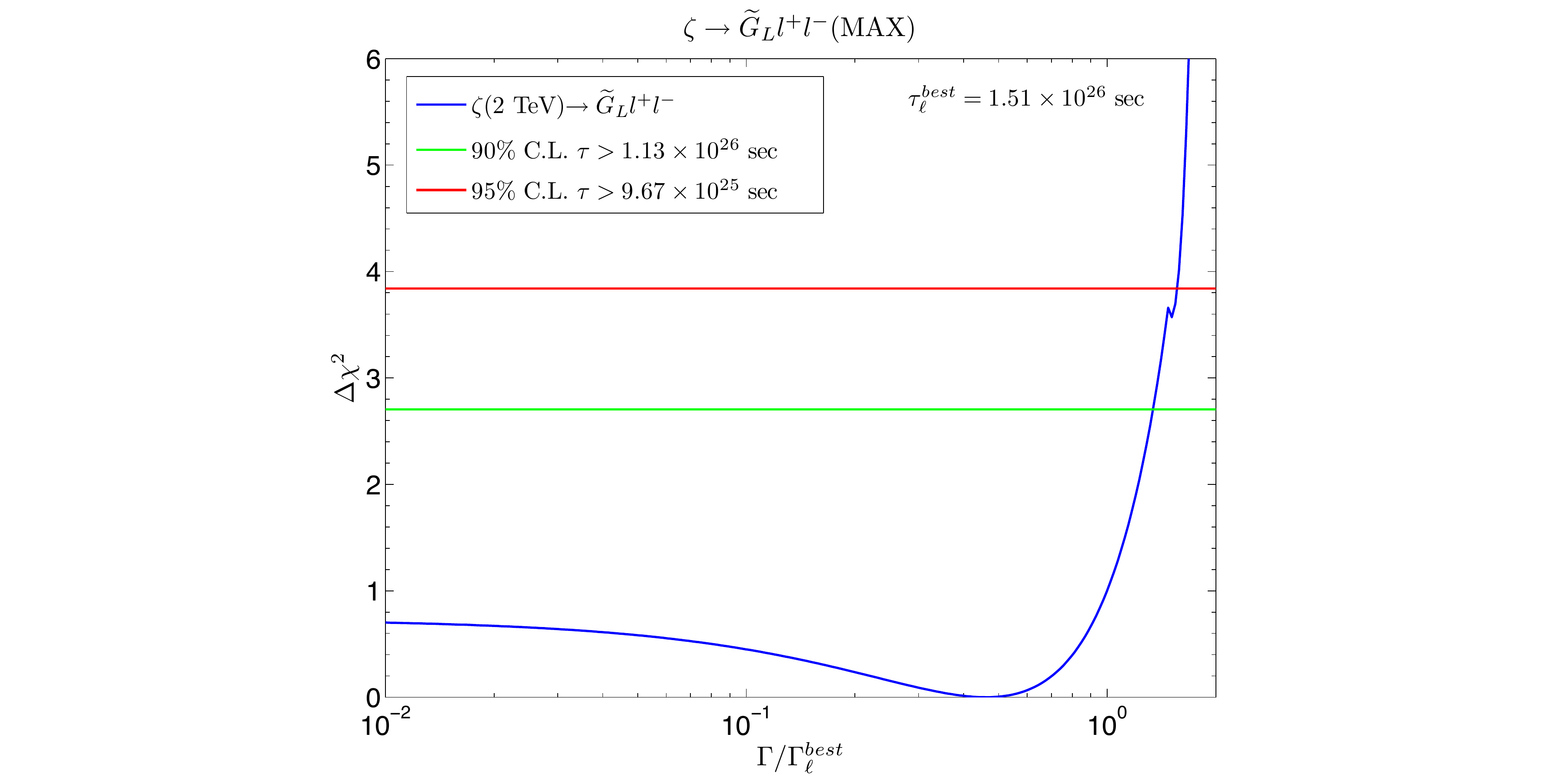}
\includegraphics[trim =55mm 2mm 69mm 2mm, clip, width=0.45\textwidth]{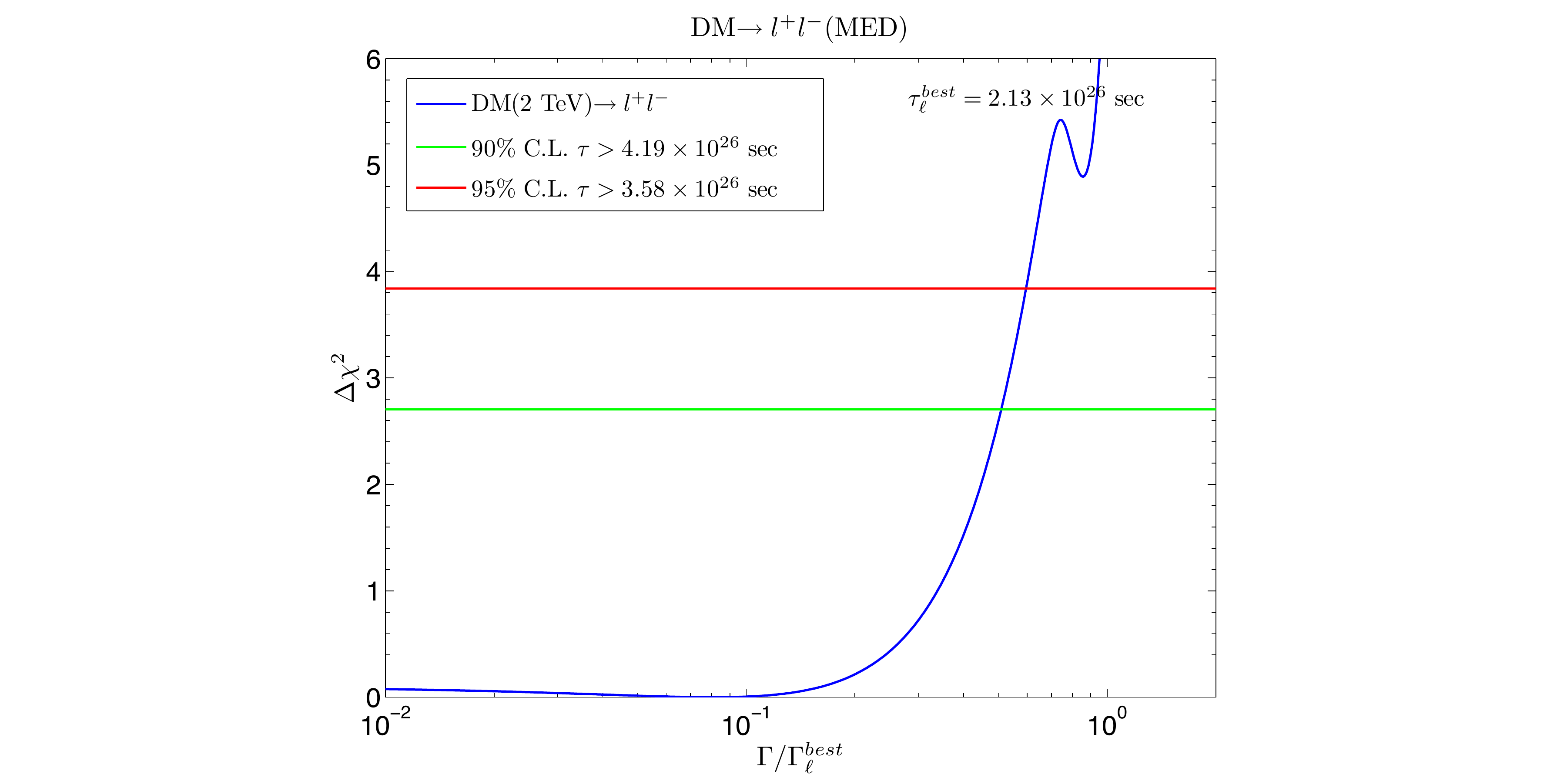}
\includegraphics[trim =55mm 2mm 69mm 2mm, clip, width=0.45\textwidth]{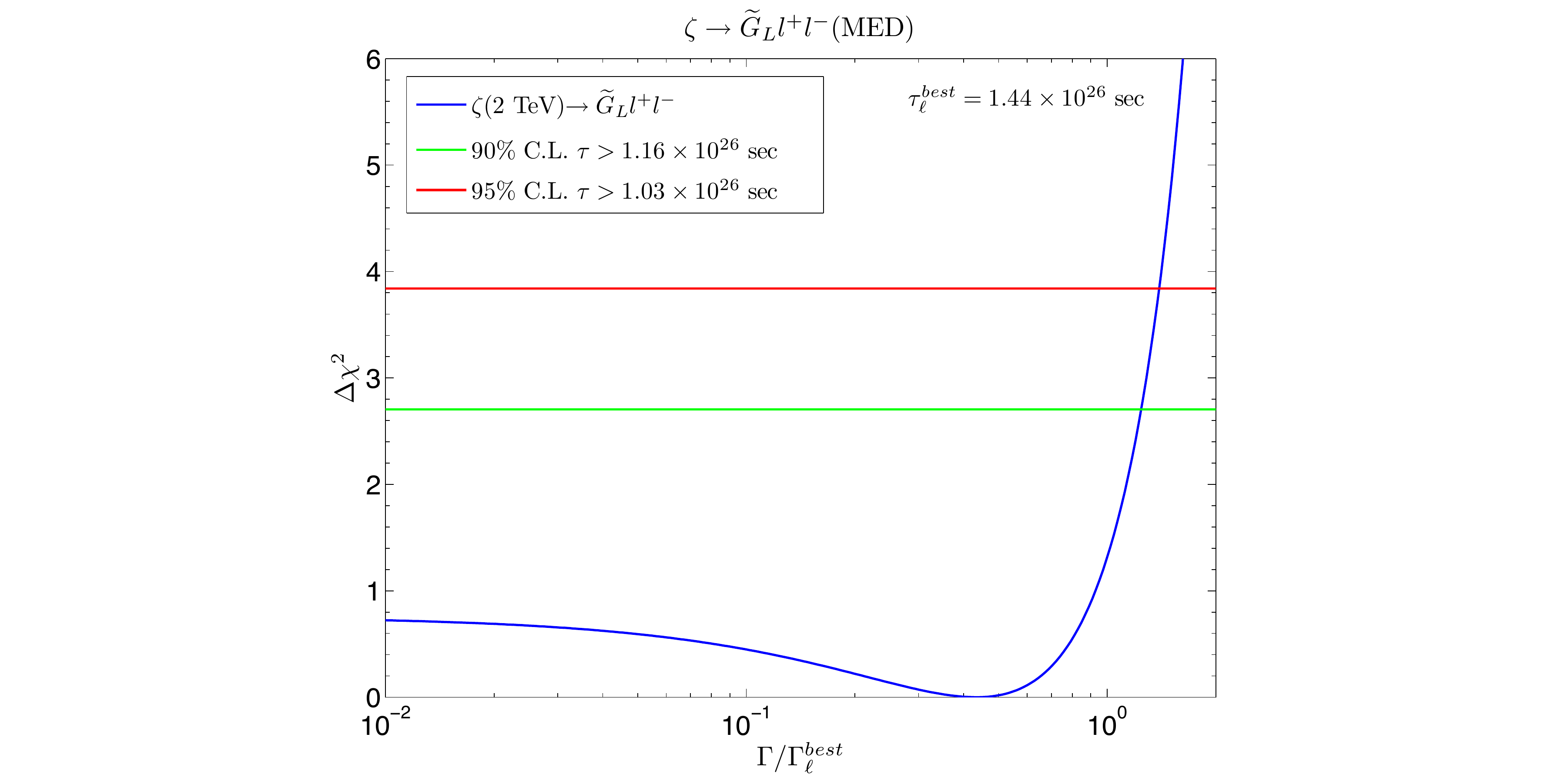}
\includegraphics[trim =55mm 2mm 69mm 2mm, clip, width=0.45\textwidth]{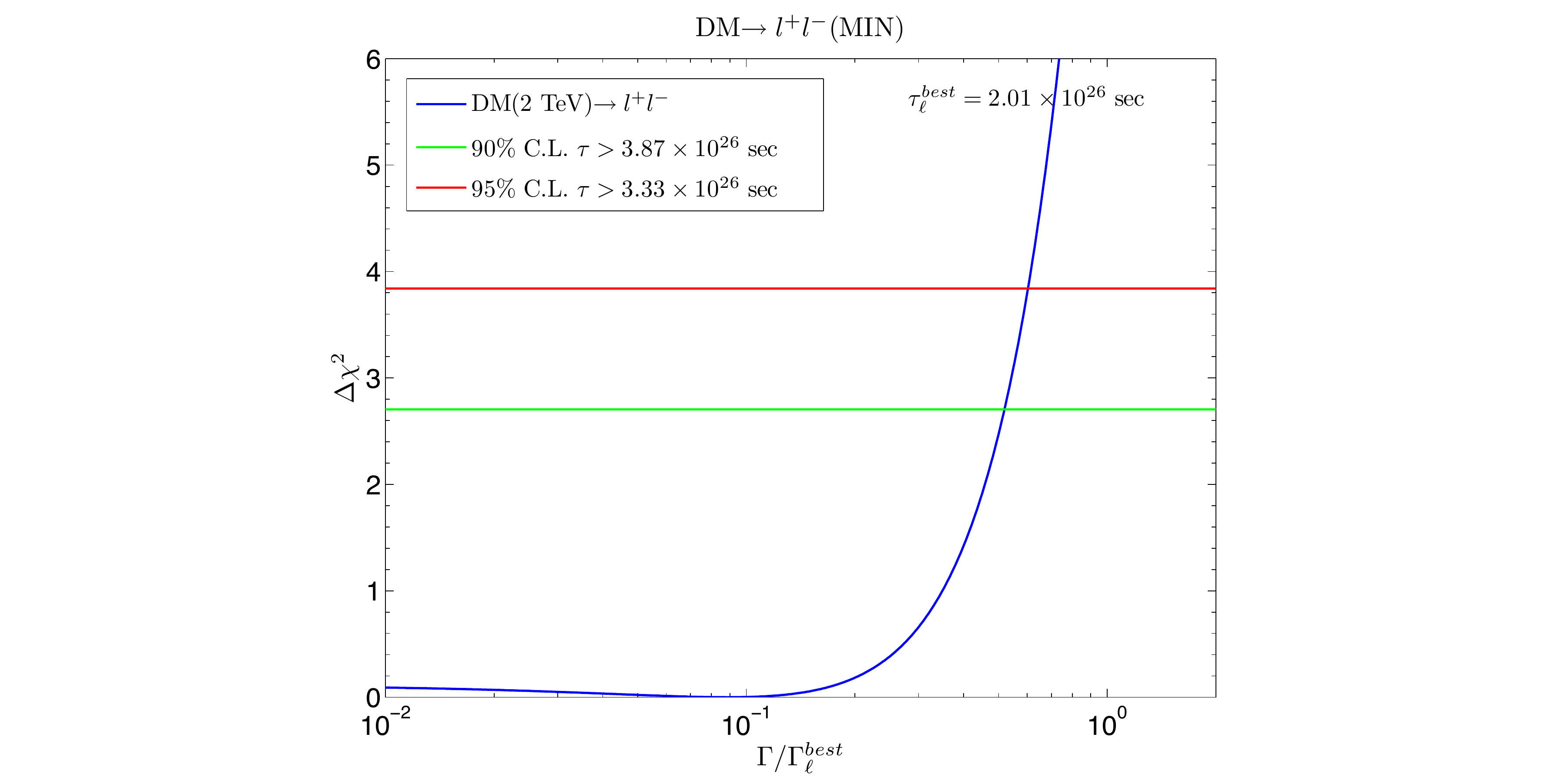}
\includegraphics[trim =55mm 2mm 69mm 2mm, clip, width=0.45\textwidth]{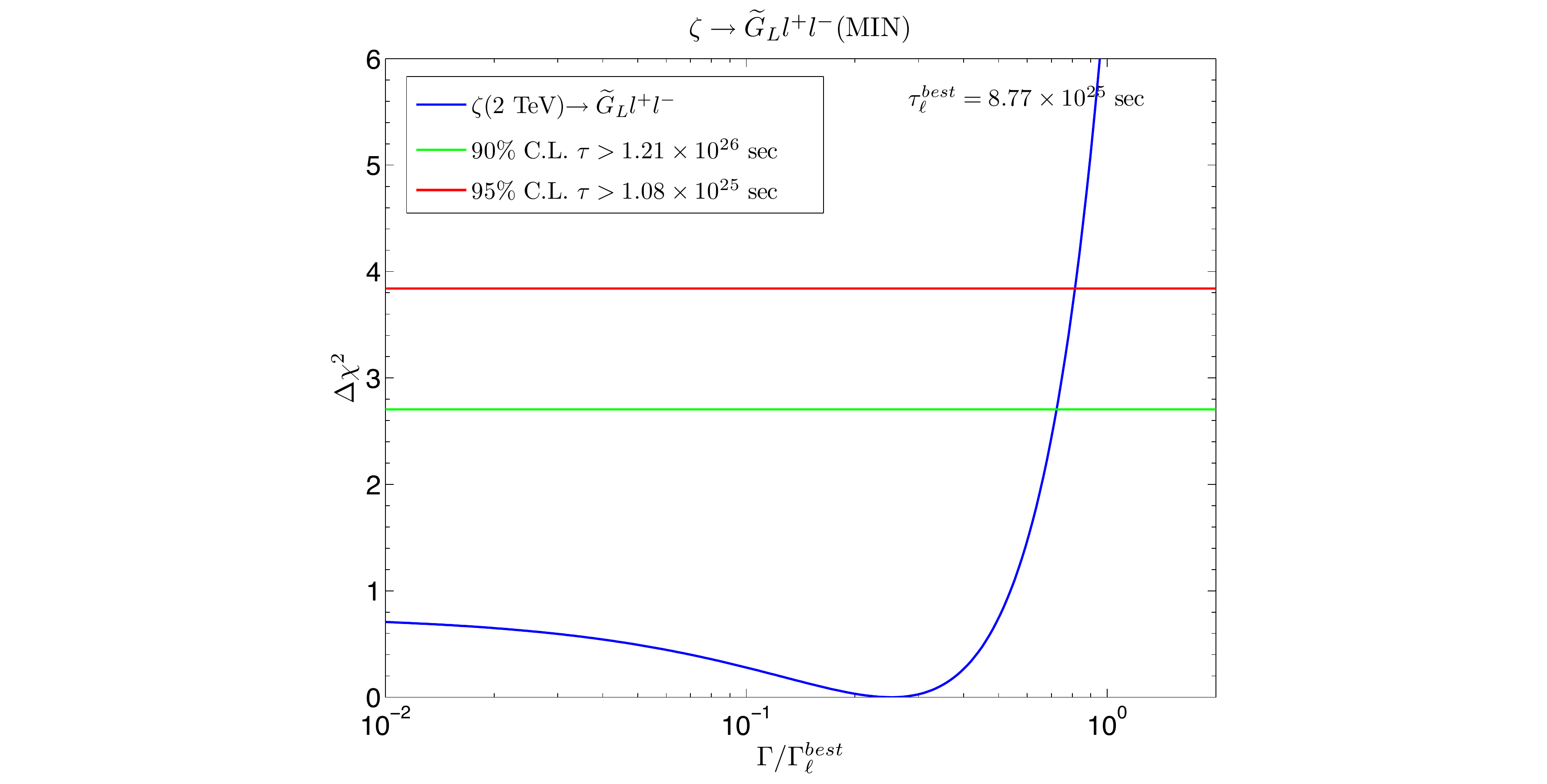}
  \vspace{-20pt}
\caption{\em The constraints from the Fermi-LAT EGB data. Only the goldstino decay with MAX and MED pass the test of the isotropic diffuse gamma ray flux. Like in Fig.~\ref{ICSFSR two-three}, we see the dip appears especially for the goldstino decay, but here the power-law index of the background is being varied and therefore the shape of the background changes for the different DM lifetime. In addition, the change on the power-law index is also the underlying reason for the local minimum around $\Delta \chi^2 \sim 5$ in the case of the two-body decay in MAX and MED.}
  \vspace{-20pt}
\label{EGB two-three}
\end{center}
\end{figure}

On the other hand, the Fermi-LAT EBG data provide constraints on the isotropic diffuse gamma ray. From Ref.~\cite{Cirelli:2012ut}, the DM decay contributes to the isotropic flux in two ways, which can be expressed in terms of the differential flux,
\beq
\frac{d\Phi_{\rm{Isotropic}}}{dE_{\gamma}}= \frac{d\Phi_{\rm{ExGal}}}{dE_{\gamma}}+\left. 4\pi\frac{d\Phi_{\rm{Gal}}}{dE_{\gamma}d\Omega}\right|_{\rm{minimum}},
\label{eq:cosmo and mini}
\eeq
 where the first term is an isotropic extragalactic cosmological flux from the decays at all past redshifts and the second term is the residual contribution from the Galactic DM halo. Note that the latter is of course not isotropic but the minimum will be the irreducible component of the isotropic gamma ray flux.   We employ the same assumption as in Ref.~\cite{Cirelli:2012ut} that the minimum of Galactic contribution is located at the anti-Galactic Center. With the help of the code, {\tt EGgammaFluxDec} in Ref.~\cite{cookbook website}, which is based on the two-body decay and includes both FSR and ICS  from the primary charge leptons \cite{marco_private}, we properly convolute it with our three-body injection spectra to obtain the isotropic cosmological flux. For the fitting procedure, we again use the EGB data from Table I of Ref.~\cite{Abdo:2010nz}, which can be described very well by a featureless power law with index $\gamma= 2.41 \pm 0.05$, and hence a power law background. The normalization of the background and the index of the power law are being varied to find the minimum of $\chi^2$ for different goldstino lifetimes. The results are presented in Fig.~\ref{EGB two-three}. Based on the numerical results, the cosmological flux, which is independent of the propagation model, is dominant over that of the DGE minimum and clearly the bounds on the goldstino lifetime are similar for different propagation models. Therefore, the EGB data impose more stringent constraints than those of DGE, i.e., Fig.~\ref{ICSFSR two-three}, subject to uncertainties on the propagation models. We can see that all of two-body decay are excluded and even the goldstino decay in the case of MIN is also disfavored, since the lifetime (decay width) of the goldstino is shorter (larger) in MIN to account for fewer electrons and positrons reaching the Earth.

\subsection{Prompt $\gamma$}
\begin{figure}[t]
\begin{center}
\includegraphics[trim =55mm 2mm 69mm 2mm, clip, width=0.445\textwidth]{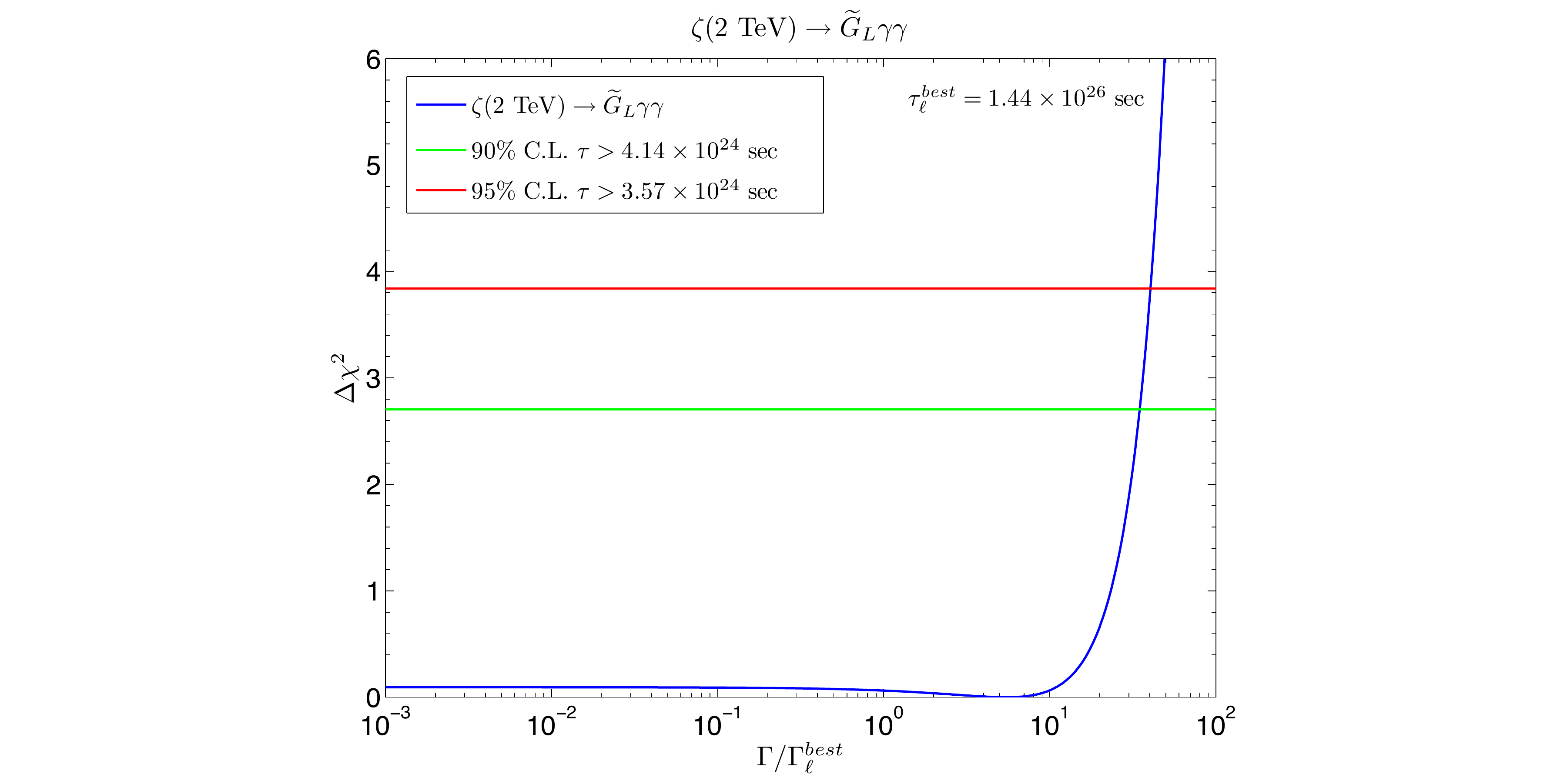}
\includegraphics[trim =55mm 2mm 69mm 2mm, clip, width=0.445\textwidth]{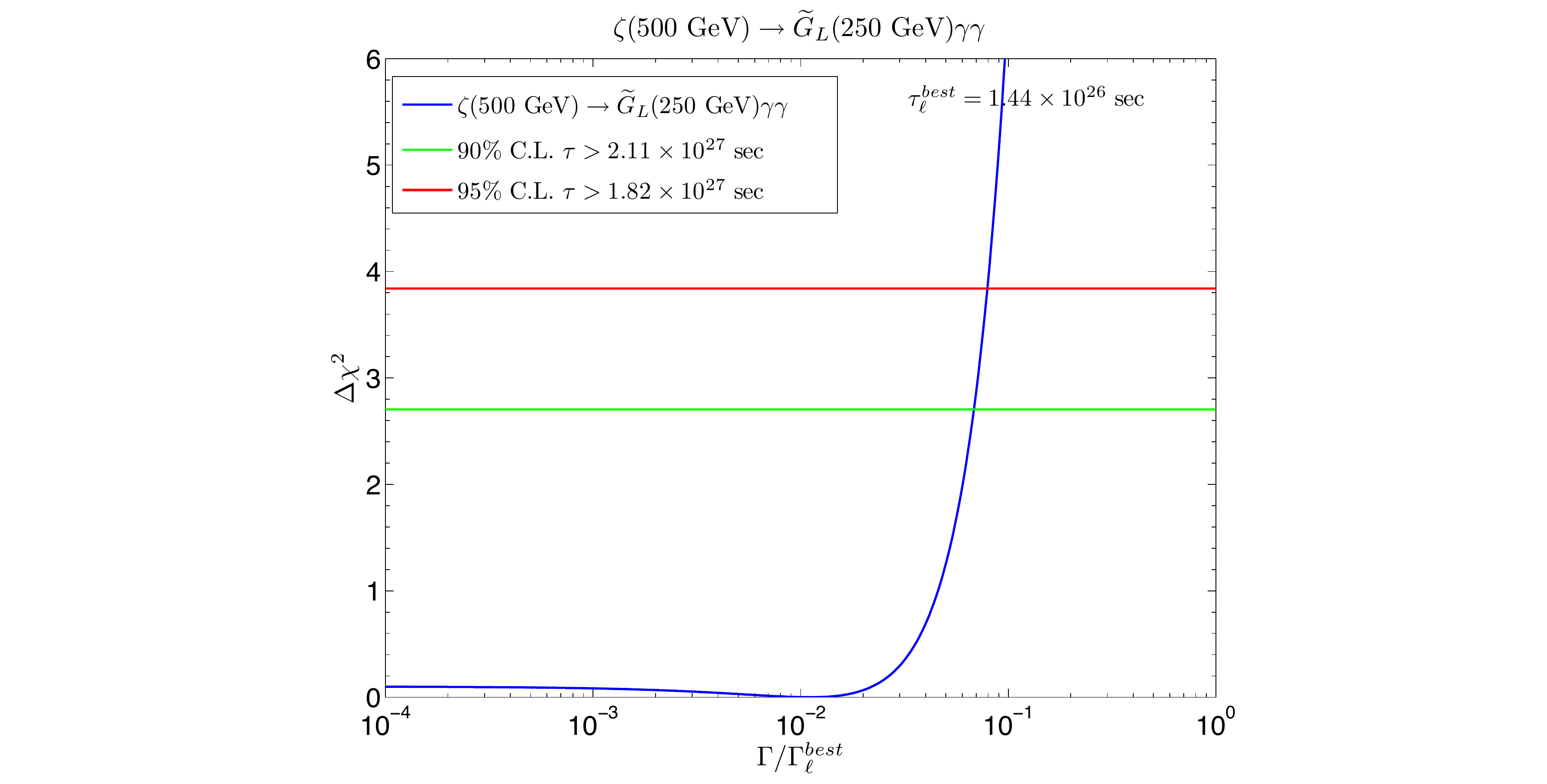}\\
\includegraphics[trim =55mm 2mm 69mm 2mm, clip, width=0.445\textwidth]{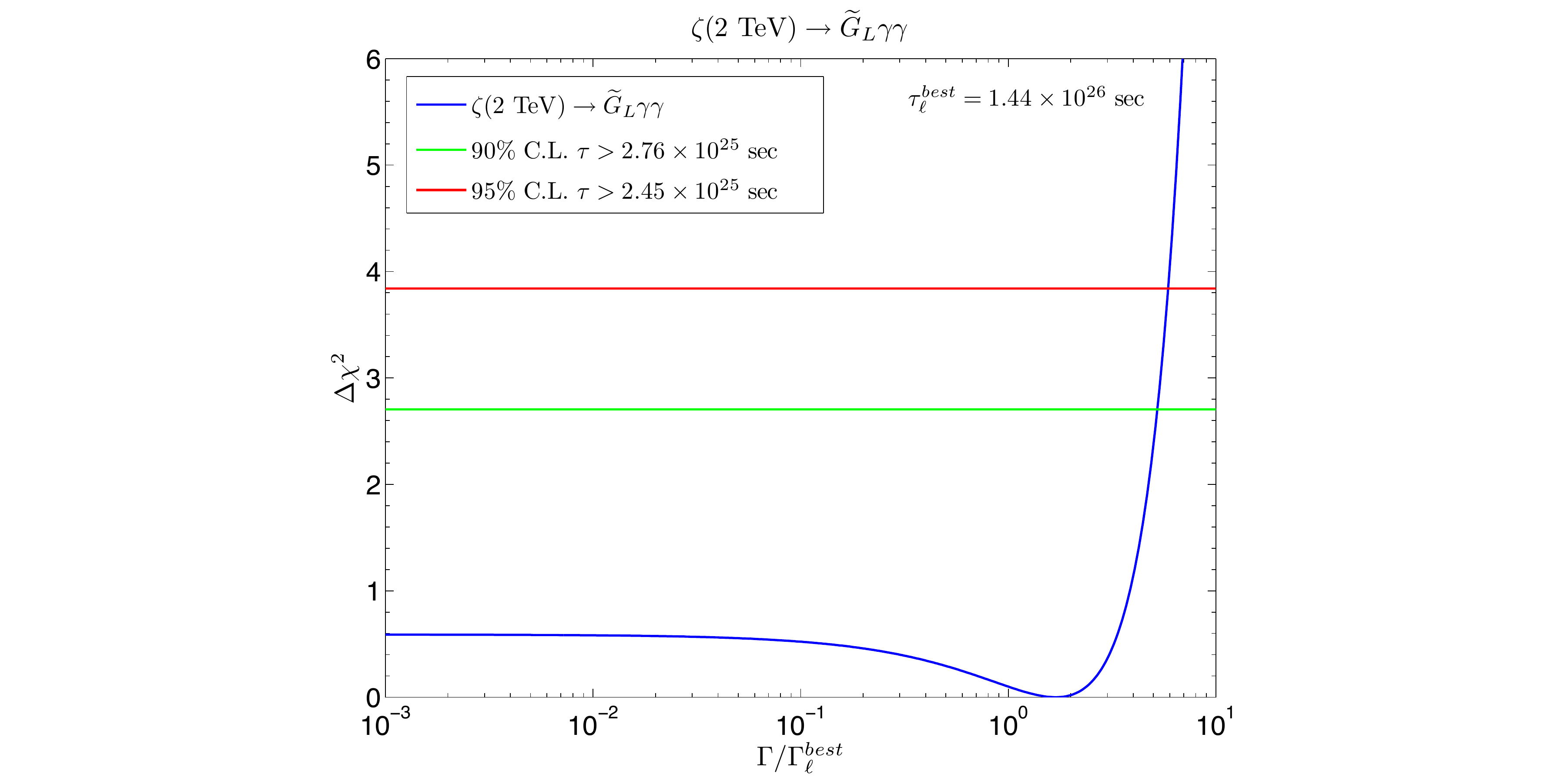}
\includegraphics[trim =55mm 2mm 69mm 2mm, clip, width=0.445\textwidth]{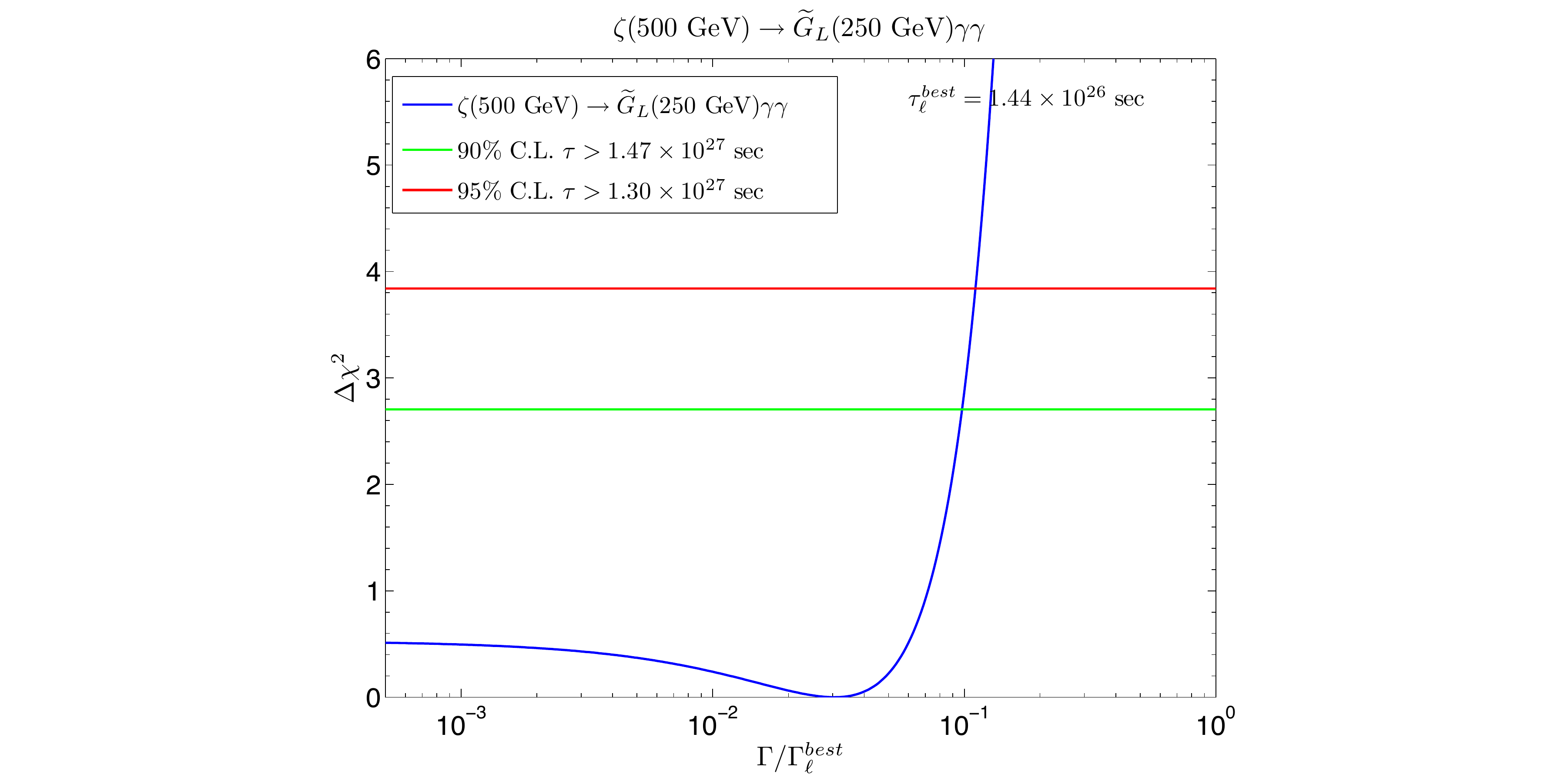}\\
\caption{\em  The Fermi-LAT DGE constraints (upper panels) and EGB constraints (lower panels). The 2 TeV goldstino decay to photons could easily pass the test of the Fermi-LAT gamma ray data because the injection spectrum of photons peaks well beyond 100 GeV. On the other hand, a 500 GeV goldstino is severely constrained by Fermi gamma ray data.}
\label{prompt three}
\end{center}
\end{figure}

In this subsection, we explore the situation where photons are direct products of three-body dark matter decays, i.e., primary photons.\footnote{Note that the signal of the two-body prompt decay is situated beyond the reach of the Fermi-LAT gamma ray data for 2 TeV dark matter~\cite{Abdo:2010nz}.} In this situation, we have two independent parameters: the decay widths of goldstino into leptons and into photons. Moreover, the results are obviously independent of the electron propagation models like FSR and the cosmological diffuse flux, and the constraints would be model-dependent. As the injection spectra of different operators from~\ref{sect:kinematics} are similar, we expect those  four operators to have the similar results. Here, we also involve both DGE and EBG data with the same fitting procedure described before. From Fig.~\ref{prompt three}, the constraint mainly comes from EGB data and the goldstino decay can not be ruled out even if the goldstino decays $80\%$ into photons and $20\%$ into leptons, which accounts for the $e^+/e^-$ excess observed by PAMELA and Fermi-LAT.
The underlying reason is, for 2 TeV goldstino, the prompt photons are mainly located beyond 100 GeV and consequently the DGE data yield a very loose bound. In these cases, Air Cherenkov Telescopes such as Veritas are well poised for DM decay to photons of ${\cal O}$(TeV) energy. On the other hand, the cosmological diffuse photons will get redshifted and constrained by the EGB data. For comparison, 500 GeV goldstino would be severely constrained by both of DGE and EGB data as shown in Fig.~\ref{prompt three}.


\subsection{$\bar p$}
\label{sect:pbardata}
In addition to measurements of the $e^+$ spectra made by PAMELA and Fermi-LAT, the $\bar p$ spectra has also been well measured by PAMELA.  While the $e^+$ species show an excess at high energies, the $\bar p$ measurement shows a spectra consistent with astrophysical sources~\cite{Adriani:2008zq}.  The most recent PAMELA anti-proton measurement from 60 MeV to 180 GeV has further confirmed this result~\cite{Adriani:2010rc}.  This agreement with the expected background can serve to limit the total annihilation rate of dark matter to hadronic final states~\cite{Cirelli:2008pk} as well as hadronic interactions of the dark matter, which could have important implications for direct detection experiments \cite{Cao:2009uv}.

\begin{figure}[t]
\begin{center}
\includegraphics[scale=0.40]{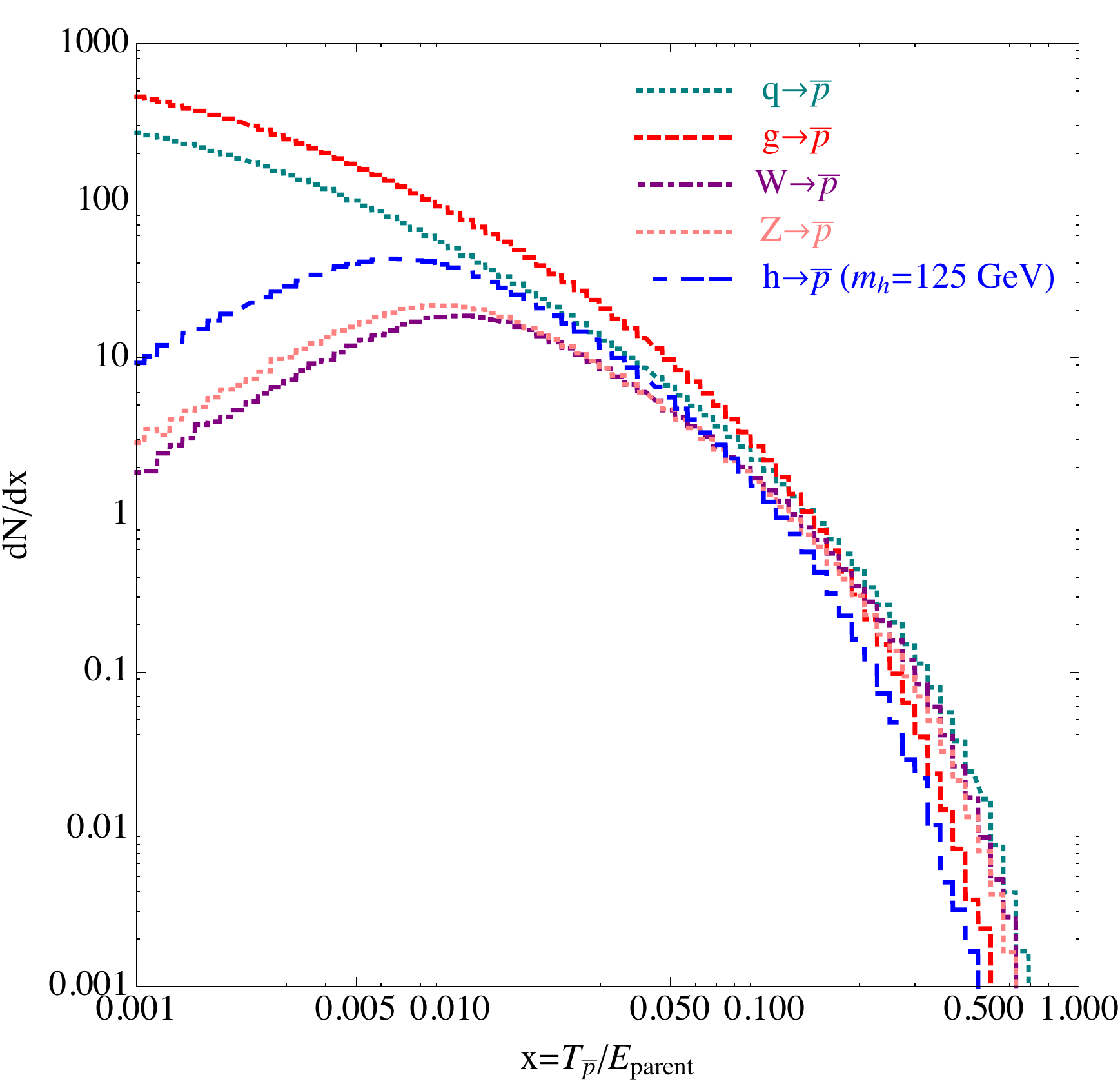}
\includegraphics[scale=0.40]{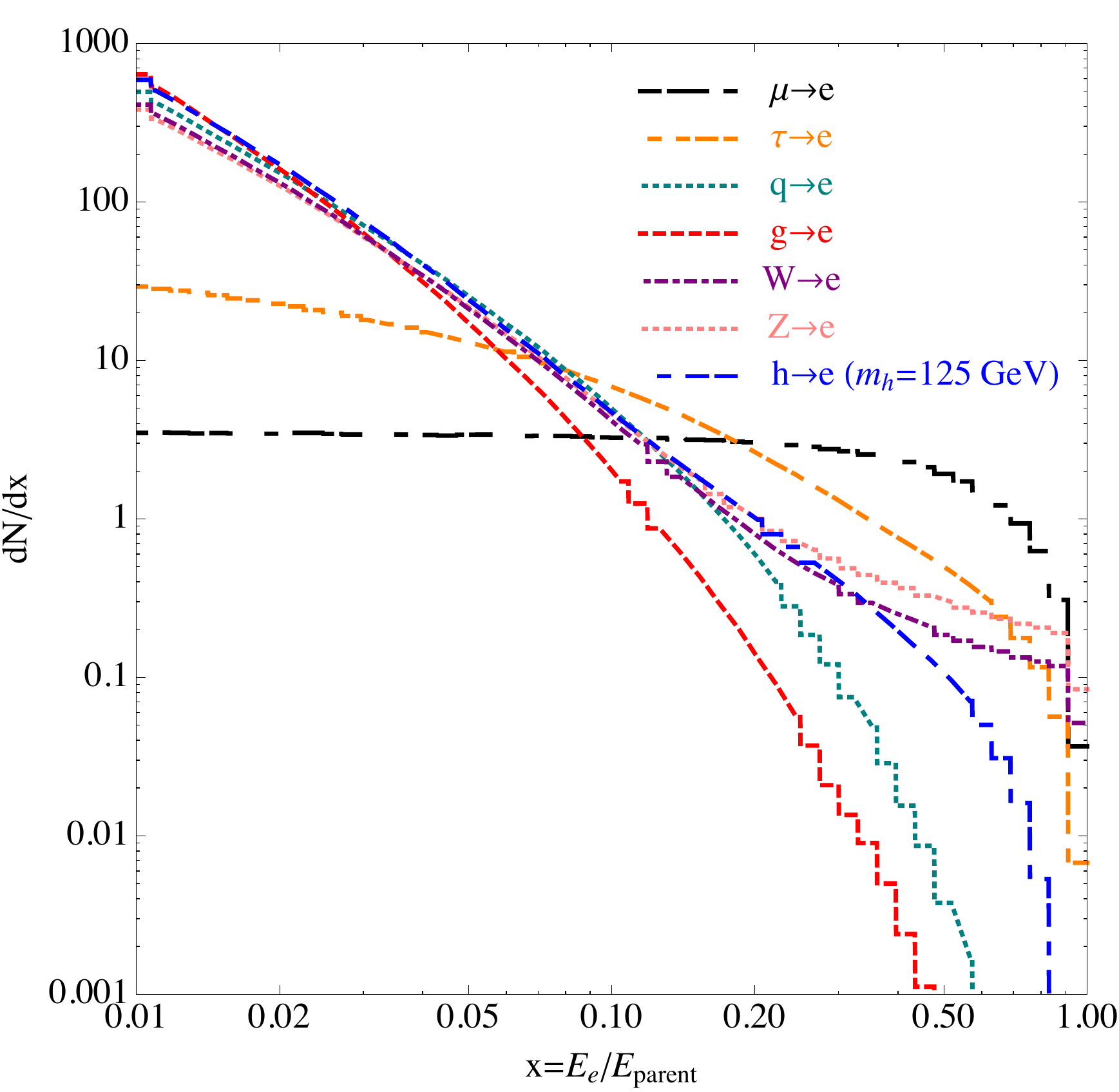}
\caption{\em Antiproton and electron energy distribution functions of $\mu, \tau$-leptons, $u$-quarks, gluons, $W, Z$ and Higgs boson decay.  The hardness of the electron in the lepton and $W/Z$ decays result from prompt production, whereas the quark/gluon decays are very soft, producing electrons only after QCD showering and hadronization.  Similarly, the antiproton energies are very soft and are similar for quarks and $W$ bosons, with a slightly harder component for $W$ decays.  For two body decay, the resulting positron and antiproton spectra are similar to these distributions.  The corresponding three body decay convolves the intermediate state energy with these distributions.}
\label{fig:frag}
\end{center}
\end{figure}

We study the restrictions of hadronic decay modes of the dark matter by the potential contribution from the decay to $q\bar q$, $gg$, $W^+W^-$, $ZZ$ and $hh$ modes, with $m_h=125$ GeV.  In the left panel of Fig.~\ref{fig:frag} we show the fragmentation function of the above final states into an anti-proton. Since these final states may also produce electrons upon further decays, it is important to study whether the partial width allowed by the anti-proton spectra could result in additional noticeable contributions in the positron measurements that were not included in our fits in the previous section. Thus in the right panel of Fig.~\ref{fig:frag} we also show the fragmentation function of all possible final states into the positron. If the dark matter decays through two-body kinematics, then Fig.~\ref{fig:frag} gives precisely the  injection energy spectra of the anti-proton and the positron.  For three body decay, these energy distributions are convolved with the intermediate state's energy distribution.  It is clear that due to the nature of fragmentation and hadronization, all non-leptonic modes provide a soft anti-proton spectra.

Once produced, the anti-protons propagate through the galaxy under the effects of diffusion due to the galactic magnetic field, the convective wind away from the plane of the galaxy and annihilations with interstellar protons; we model the anti-proton propagation according to Ref.~\cite{Cirelli:2008id}.  As with the modeling of the $e^+$ propagation, we primarily adopt the MED model and assume the dark matter is distributed according to the Moore profile.  In addition, once the anti-protons approach Earth, solar modulation effects alter their low energy spectra.  We include this effect for anti-protons since the energy range for solar modulation is well positioned within the PAMELA data.  To this end, we adopt a Fisk potential of $\phi=500$ MV~\cite{Gleeson:1968zz}.

To illustrate how consistent the two-body and three-body decay scenarios are with the anti-proton data, we fit the prediction of each model to the PAMELA $\bar p / p$ data~\cite{Adriani:2010rc}.  We take the best-fits to the electron data in Sect.~\ref{sect:edata} and simultaneously vary the dark matter signal normalization and total $\bar p$ background normalization.  We show in detail the fits for the $q\bar q$ and $W^+W^-$ mode in the MED propagation model.  We later summarize the results for other decay modes and propagation models.

\begin{figure}[htbp]
\begin{center}
\includegraphics[scale=0.48]{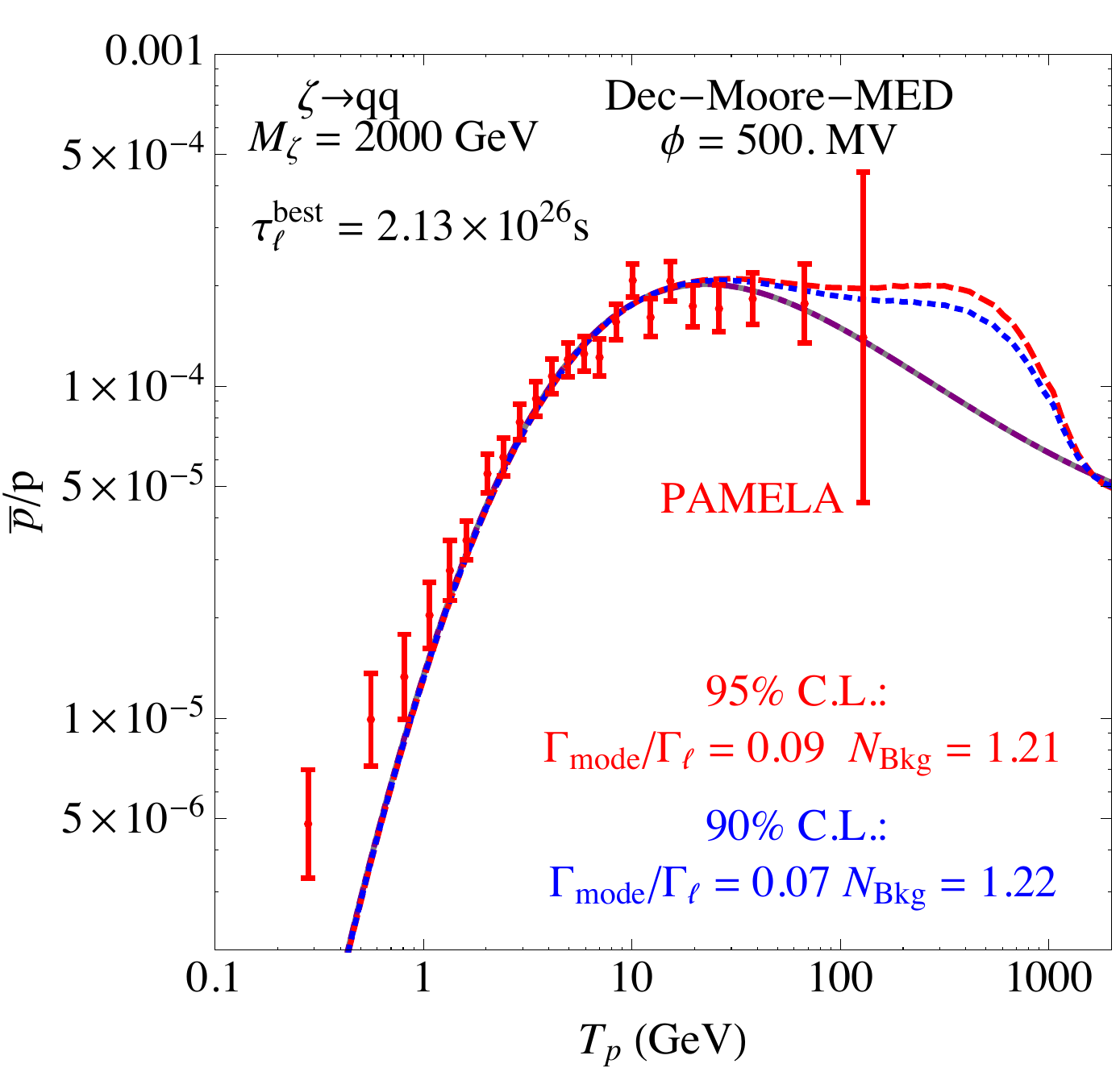}
\includegraphics[scale=0.48]{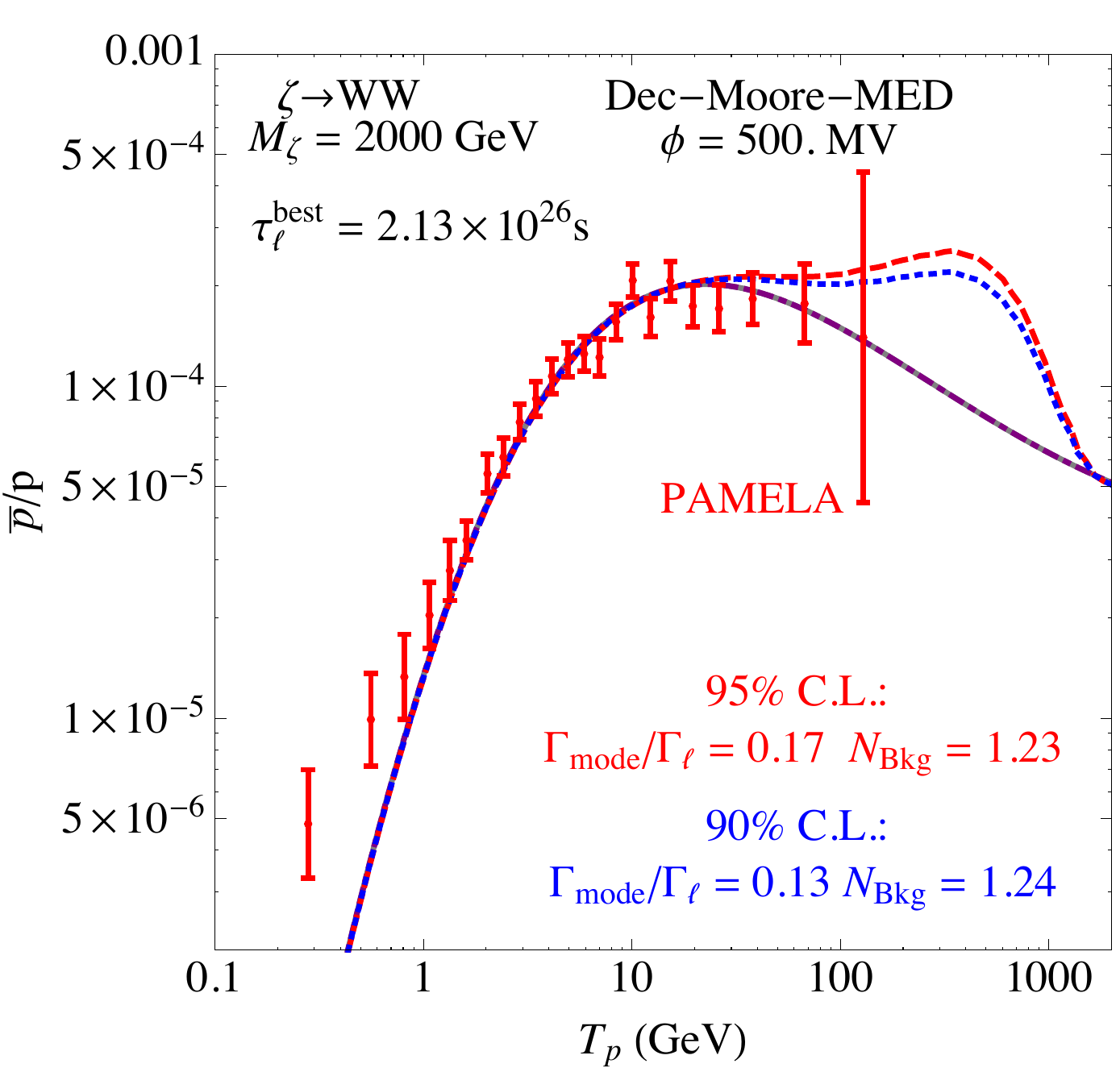}\\
\hspace{.2in}\includegraphics[scale=0.45]{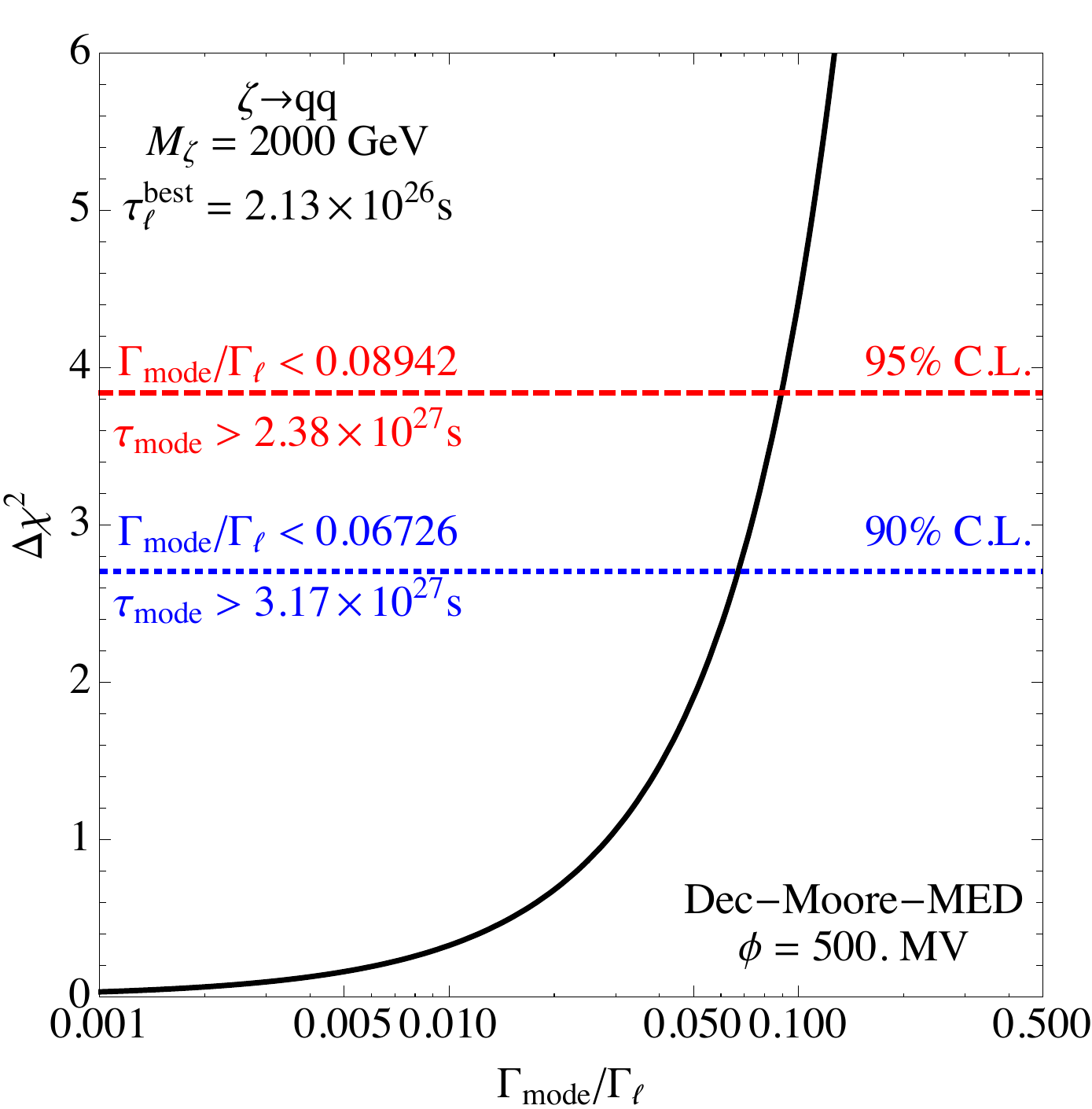}\hspace{.2in}
\includegraphics[scale=0.45]{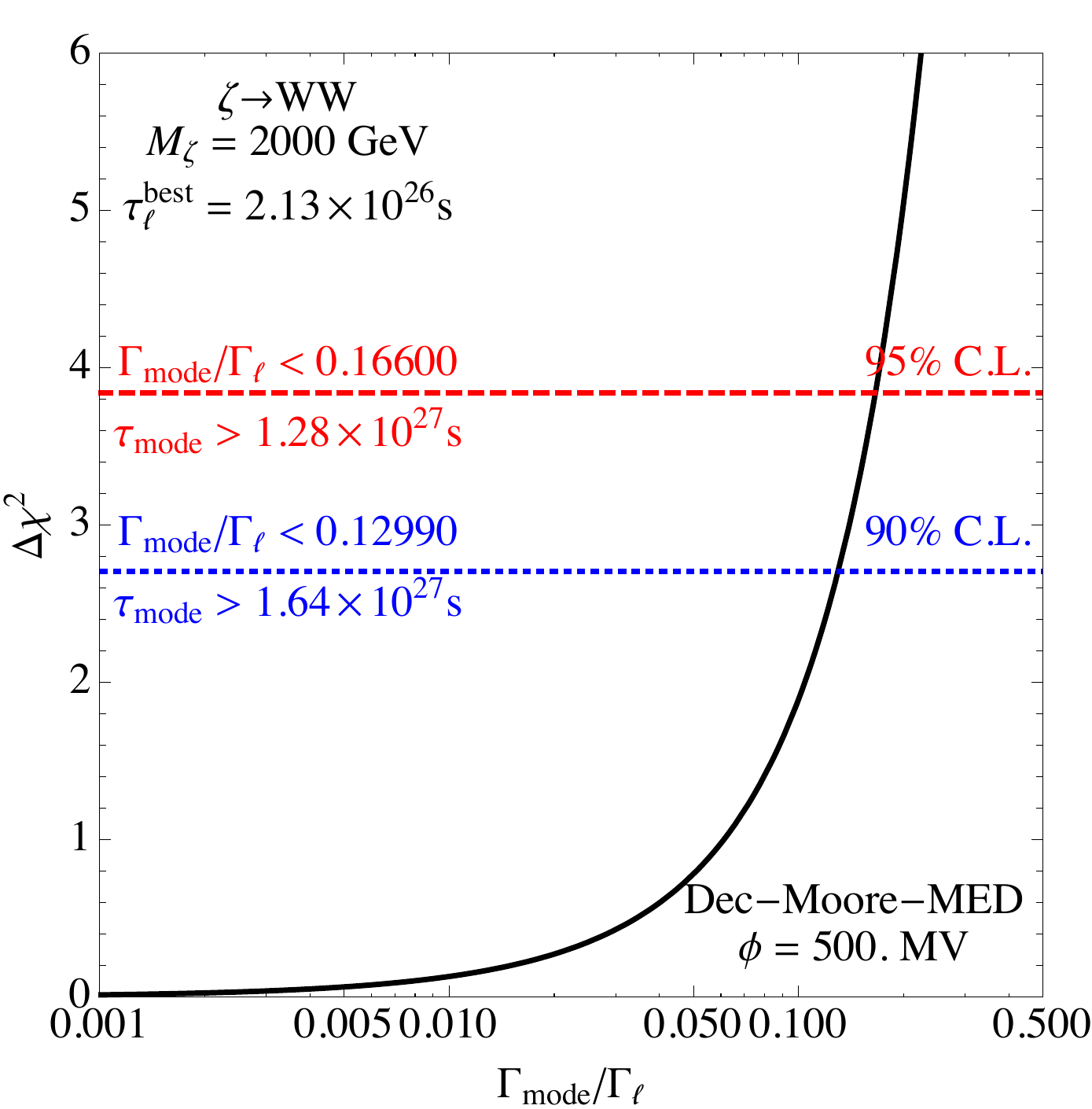}\\
\includegraphics[scale=0.48]{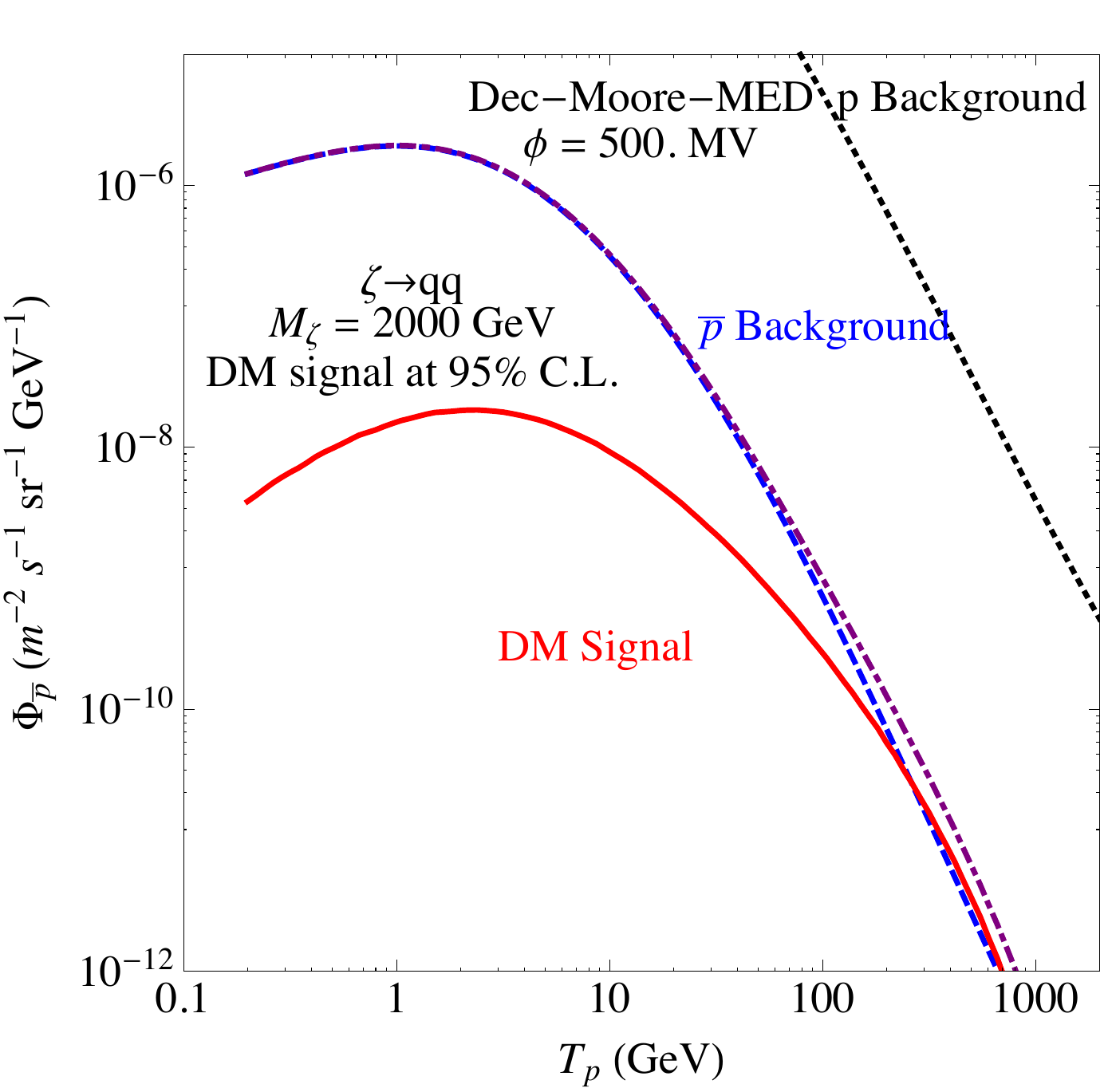}
\includegraphics[scale=0.48]{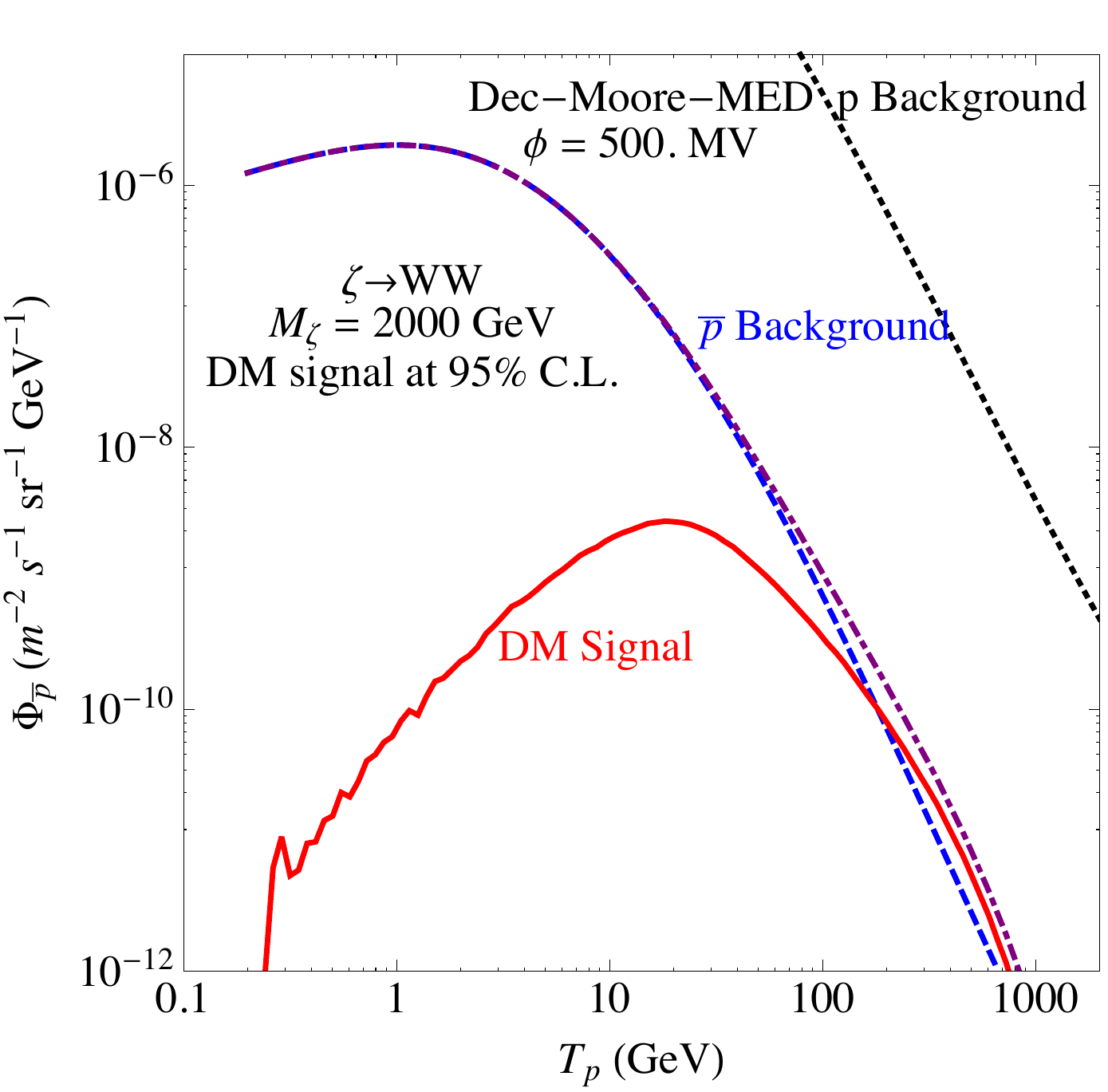}
\caption{\em Contribution to the $\bar p / p$ ratio from dark matter 2-body decay for the $W$ and $q$ decay modes.  Since the $W$-boson tends to produce fewer $\bar p$  than the $q$-mode, the decay rate allowed by the PAMELA $\bar p$ data is larger by nearly a factor of two. Dark matter decay rates for $W$-bosons ($q$) as high as $0.17\, (0.09) \times \Gamma_{DM\to \ell^+\ell^-}$ can be accommodated by the PAMELA $\bar p$ data.}
\label{fig:pbar2bdy}
\end{center}
\end{figure}

The two-body decay of dark matter  yields two particles with well defined energy.  If they are massless, the resulting anti-proton spectra appears identical to Fig.~\ref{fig:frag} with $E_{parent} = m_{DM}/2$.  The energy of injected anti-protons from massive parents are altered with two-body kinematics.  In Fig.~\ref{fig:pbar2bdy}, we show the two-body dark matter decay to quark modes (left panels) and $W$ boson modes (right panels) for $m_{DM}=2$ TeV.  The comparison with the PAMELA anti-proton data (top panels) shows that a considerable anti-proton flux is possible in energies beyond the present data.  The $\Delta \chi^2$ fits (middle panels) show no preferred value for the partial width, as expected since no excess is observed in this data.  Had there been a preferred decay rate, the position of the  minimum $\chi^2$ would not be asymptotically approaching zero.  Finally, we show in the bottom panels the components of the proton and anti-proton flux based on the 95\% C.L. fit.  Overall, we see that decay rate to the quark mode at 95\% C.L. must be smaller than $\approx 1/(2.4\times 10^{27}$~s) and the rate for the $W$-boson mode must be smaller than $\approx 1/(1.3\times 10^{27}$~s).  One could also compare these bounds on the hadronic decay width with the best fit decay widths into the lepton modes from fitting the electron/positron data. If we take the best-fit lifetime from fitting two-body decays to both the positron fraction and the total $e^++e^-$ flux, which is shown in the upper panels in Fig.~\ref{e+e- twobody}, the hadronic partial widths fall around ${\cal O}(10\%)$  of the lepton modes. The upper limit on the two body decay of dark matter in the $W$-boson mode is roughly $\Gamma_{DM\to W^+W^-} < 0.17\times \Gamma_{DM\to \ell^+\ell^-}$ from the PAMELA anti-proton data.  Likewise the limit on the $q$ mode is $\Gamma_{DM\to q\bar q} < 0.09\times \Gamma_{DM\to \ell^+\ell^-}$. However, these ratios are a factor of $2 - 3$ smaller if one only includes the positron fraction in the fit for the leptonic widths, which was demonstrated in the lower panels of Fig.~\ref{e+e- twobody}.

The quark mode is more constrained by roughly a factor of two relative to the $WW$ mode since the anti-protons originate from quark hadronization. In comparison, the $W$-boson can decay to $\ell\nu$, giving a reduction in the overall $\bar p$ rate.  Moreover, as the $W$-boson decays to pairs of quarks, the resulting anti-protons are softer compared with the dark matter decay to quarks.  However, below $T_{\bar p} = m_p E_{\rm inj}/ m_W $, the $\bar p$ spectra is cut off.  This can be understood since the $\bar p$'s are predominantly relativistic, meaning they have a kinetic energy of at least ${\cal O}(m_p)$ in the $W$ boson rest frame.  Boosting to the dark matter rest frame places the energy cutoff near $T_{\bar p} = m_p E_{\rm inj}/ m_W$.  Overall, the flux from lower energy anti-protons becomes buried in the increasing background $\bar p$ flux, thus providing a lower  $\bar p / p$ contribution.  This is clearly seen in the position of the peak of the DM signal in the lower panels.  The propagation model dependence is shown in Table~\ref{tab:2bdy}.  The decay lifetime fit to the lepton modes from the PAMELA data are shown in units of $10^{26}$~s.  The subsequent 95\% C.L. lower limits from the PAMELA $\bar p$ data on the decay lifetime of the two body decay modes to $q\bar q$, $gg$, $W^+W^-$, $ZZ$, and $hh$ are also summarized.  In parentheses are the respective ratios between the leptonic lifetime fit and the hadronic lifetime limit.  We see that obviously the most constrained modes are the purely hadronic $q\bar q$ and $gg$ modes.  Generally, within the MAX propagation model, more suppression into the decay of hadronic final states is required to maintain agreement with data.  Within the MIN propagation model, the hadronic states have to be suppressed no more than an order of magnitude.

\begin{table}[t]
\caption{\em
Dark matter two-body decay mode lifetime dependence on the propagation model for fits to the PAMELA/Fermi $e^+$, Fermi $e^+/(e^++e^-)$, and PAMELA $\bar p$ data with $m_{\zeta}=2$ TeV.  The decay  to the $\ell^+\ell^-$ mode is the best lifetime fit provided by the Fermi-LAT and PAMELA electron data, while the other decay modes are 95\% C.L. lower limits on the lifetime and the ratio with respect to the best fit lepton lifetime in parenthesis.}
\begin{center}
\begin{tabular}{|c|c|ccccc|}
\hline
${\tau \over 10^{26} \text{ $s$}}$& $\zeta\to\ell^+\ell^-$& $\zeta\to q\bar q$& $\zeta\to g\bar g$& $\zeta\to W^+W^-$& $\zeta\to ZZ$& $\zeta\to hh$\\
\hline
MAX & 2.17 & 49.7 (22.9)& 80.9 (37.3)& 29.6 (13.7) &  31.4 (14.5) &  47.5 (21.9) \\
MED& 2.13 & 23.8 (11.2) & 39.2 (18.4) & 12.8 (6.01) &  13.7 (6.42) &  21.0 (9.86) \\
MIN& 2.01 & 5.63 (2.80) & 9.30 (4.62)& 2.89 (1.44) &  3.09 (1.54) &  4.78 (2.38) \\
\hline
\end{tabular}
\end{center}
\label{tab:2bdy}
\end{table}%

\begin{figure}[htbp]
\begin{center}
\includegraphics[scale=0.48]{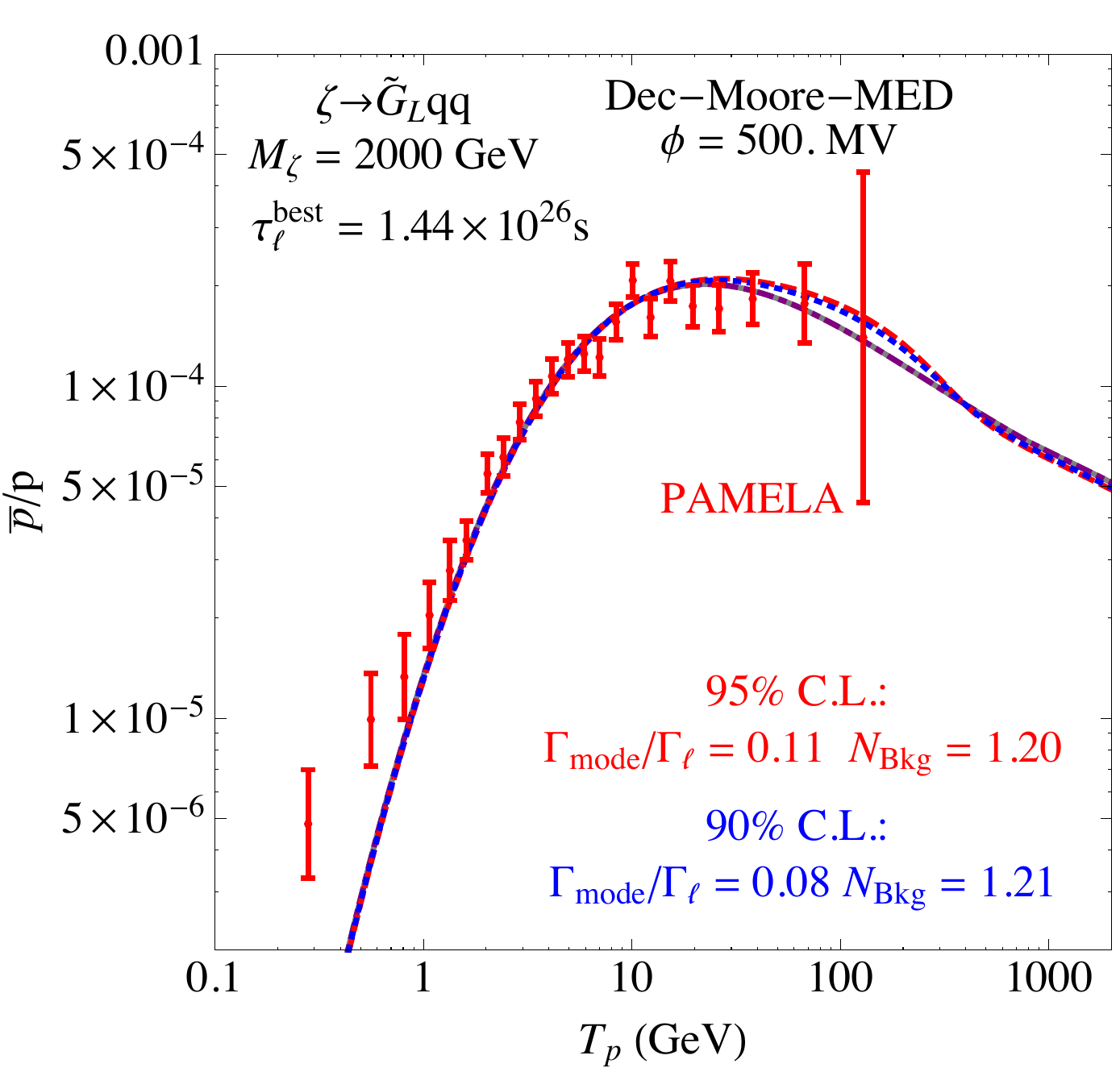}
\includegraphics[scale=0.48]{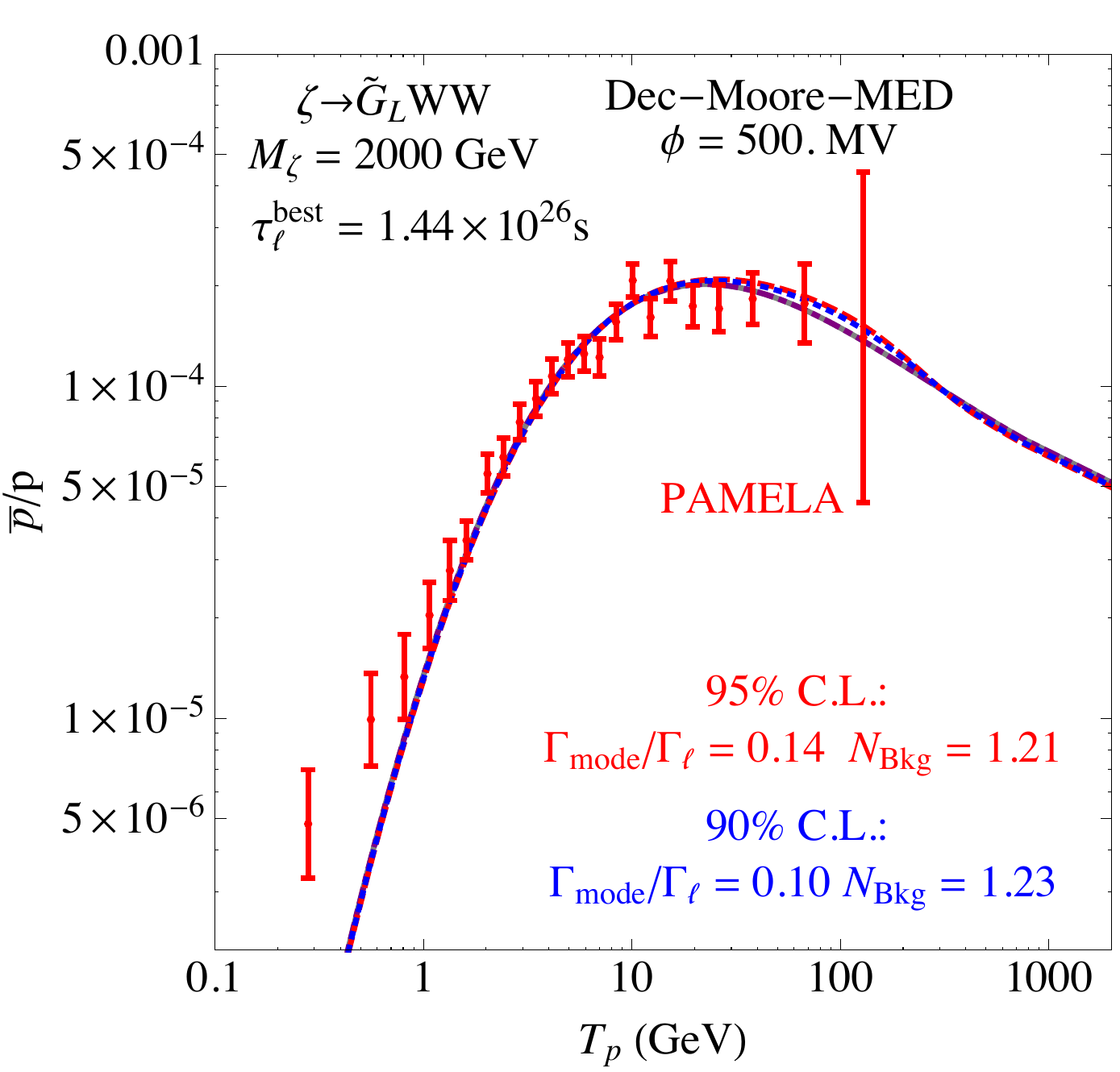}\\
\hspace{.2in}\includegraphics[scale=0.45]{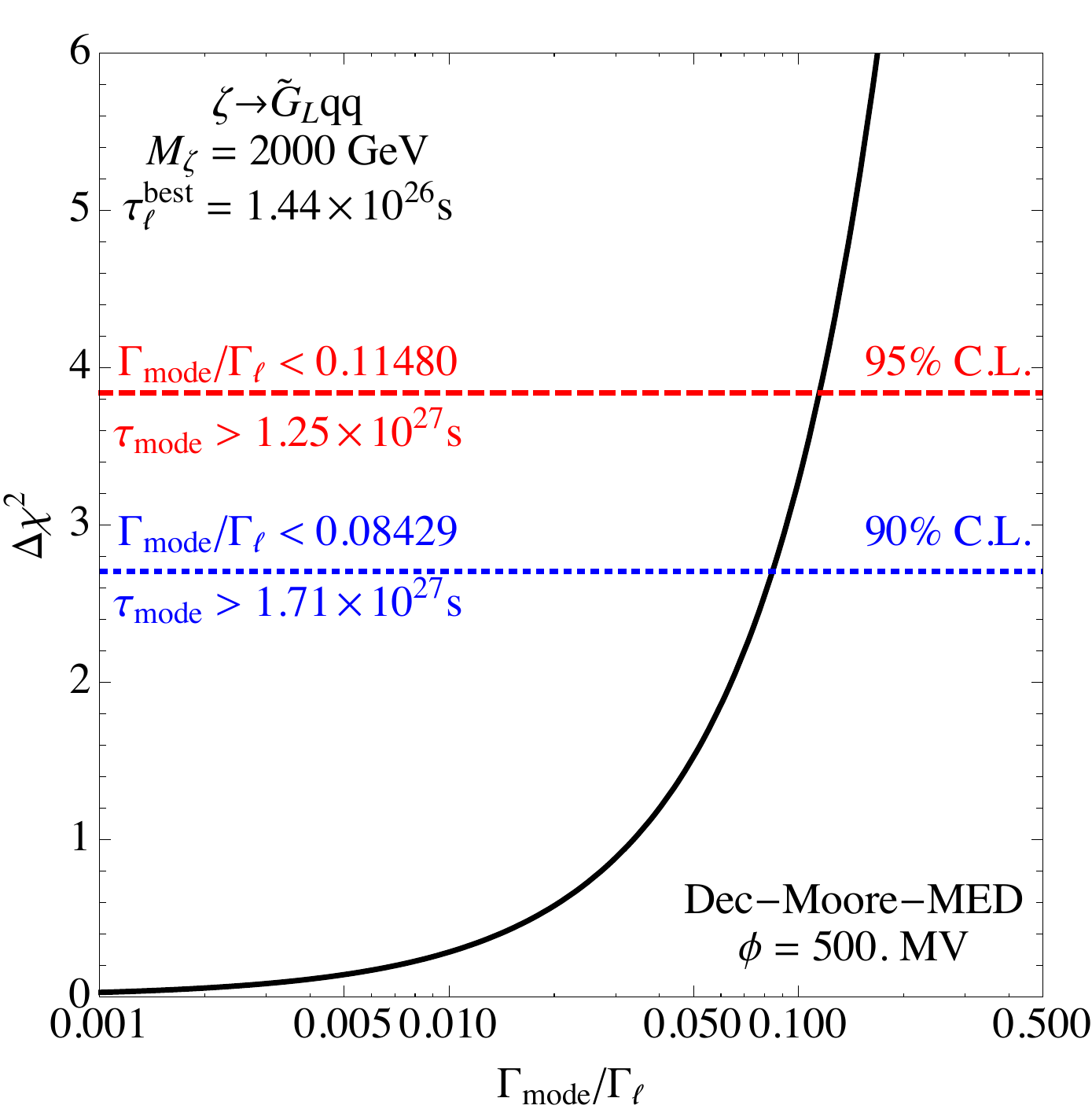}\hspace{.2in}
\includegraphics[scale=0.45]{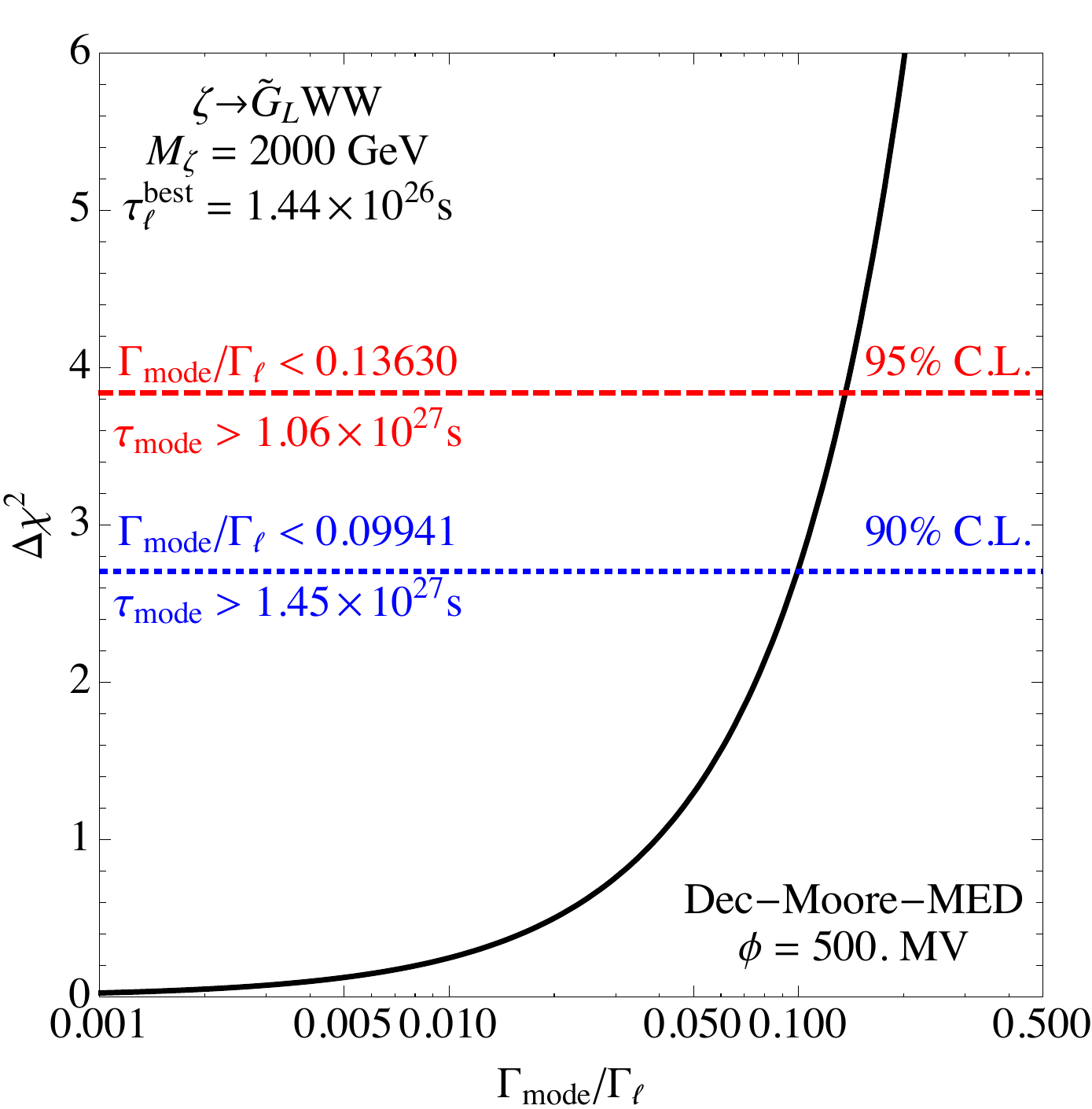}\\
\includegraphics[scale=0.48]{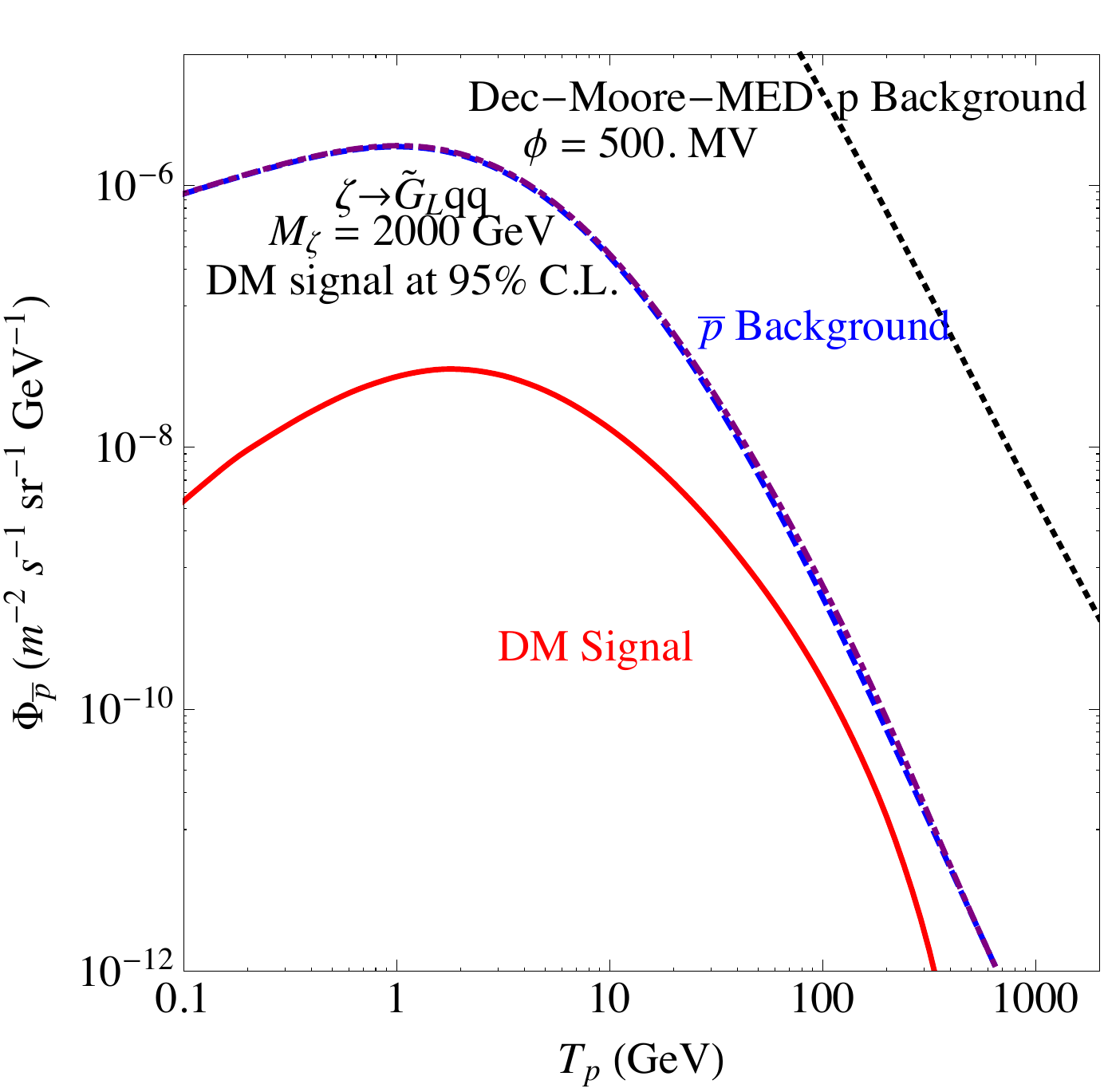}
\includegraphics[scale=0.48]{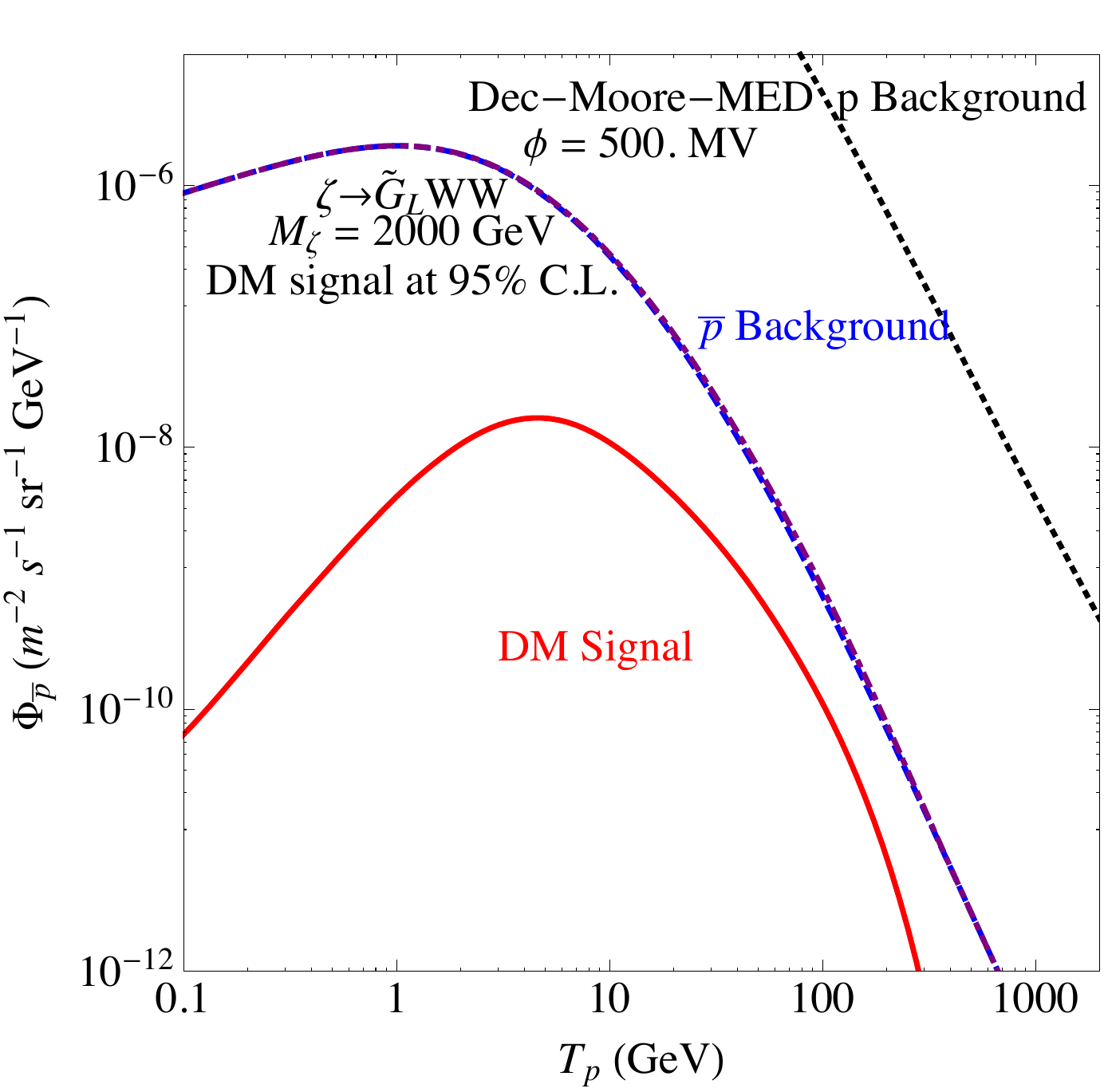}
\caption{\em Similar to Fig.~\ref{fig:pbar2bdy}, but for dark matter three-body decay via the Dim-8 goldstino decay operators.
Decay rate for $W$-bosons ($q$) as high as $0.14\, (0.11) \times \Gamma_{DM\to \widetilde G_L \ell^+\ell^-}$ can be accommodated by the PAMELA $\bar p$ data. }
\label{fig:pbar3bdy}
\end{center}
\end{figure}

Generically, the three-body decays through various operators of dark matter to quarks and $W$-bosons create similar $\bar p$ spectra.  We therefore concentrate on the analysis of the $\bar p$ flux in the goldstino decay scenario where it decays through the dimension-8 operators.  Since the injected $q$ and $W$ in the three-body decay is of lower energy, the resulting anti-protons are softer than the two-body case.  As a result, the constraint on the allowed decay width in every hadronic channel is weaker than the corresponding two-body case, which can be seen by comparing Table \ref{tab:3bdy} with Table \ref{tab:2bdy}. The decay rate to the quark mode at 95\% C.L. must be smaller than $\approx 1/(1.3\times 10^{27}$~s)  while the rate for the $W$-boson mode must be smaller than $\approx 1/(1.1\times 10^{27}$~s).  In terms of the ratio of the hadronic decay rates over the best-fit leptonic decay rates, the 95\% C.L. upper limit on the three-body decay of dark matter  in the $W$-boson mode is roughly $\Gamma_{DM\to W^+W^-} < 0.14\times \Gamma_{DM\to \ell^+\ell^-}$ from the PAMELA anti-proton data.  Likewise the limit on the $q$ mode is $\Gamma_{DM\to q\bar q} < 0.11\times \Gamma_{DM\to \ell^+\ell^-}$.  Similar to the Table~\ref{tab:2bdy} for the two-body decays, the propagation model dependence for three-body decays is shown in Table~\ref{tab:3bdy}.

\begin{table}[t]
\caption{\em
Goldstino decay mode lifetime dependence on the propagation model for fits to the PAMELA/Fermi $e^+$, Fermi $e^+/(e^++e^-)$, and PAMELA $\bar p$ data with $m_{\zeta}=2$ TeV.  The decay  to the $\widetilde G_L \ell^+\ell^-$ mode is the best lifetime fit provided by the Fermi and PAMELA electron data, while the other decay modes are 95\% C.L. lower limits on the lifetime and the ratio with respect to the best fit lepton lifetime in parenthesis.}
\begin{center}
\begin{tabular}{|c|c|cccccc|}
\hline
${\tau \over 10^{26} \text{ s}}$& $\zeta\to \widetilde G_L \ell^+\ell^-$& $\zeta\to \widetilde G_L q\bar q$& $\zeta\to \widetilde G_L g\bar g$& $\zeta\to \widetilde G_L W^+W^-$& $\zeta\to \widetilde G_L ZZ$& $\zeta\to \widetilde G_L hh$& $\zeta\to \widetilde G_L h$\\
\hline
MAX & 1.51 & 23.2 (15.3)& 23.6 (15.6)& 22.8 (15.1) &  21.8 (14.4) &  22.0 (14.5) & 23.8 (15.7) \\
MED& 1.44 & 12.5 (8.68) & 14.7 (10.2) & 10.6 (7.36) &  10.2 (7.08) &  10.0 (6.94) & 10.5 (7.29) \\
MIN& 0.89 & 2.48 (2.83) & 2.89 (3.29)& 2.10 (2.39) &  1.99 (2.27) &  1.99 (2.27) & 2.39 (2.73) \\
\hline
\end{tabular}
\end{center}
\label{tab:3bdy}
\end{table}%

\begin{figure}[t]
\begin{center}
\includegraphics[trim =55mm 2mm 69mm 2mm, clip, width=0.445\textwidth]{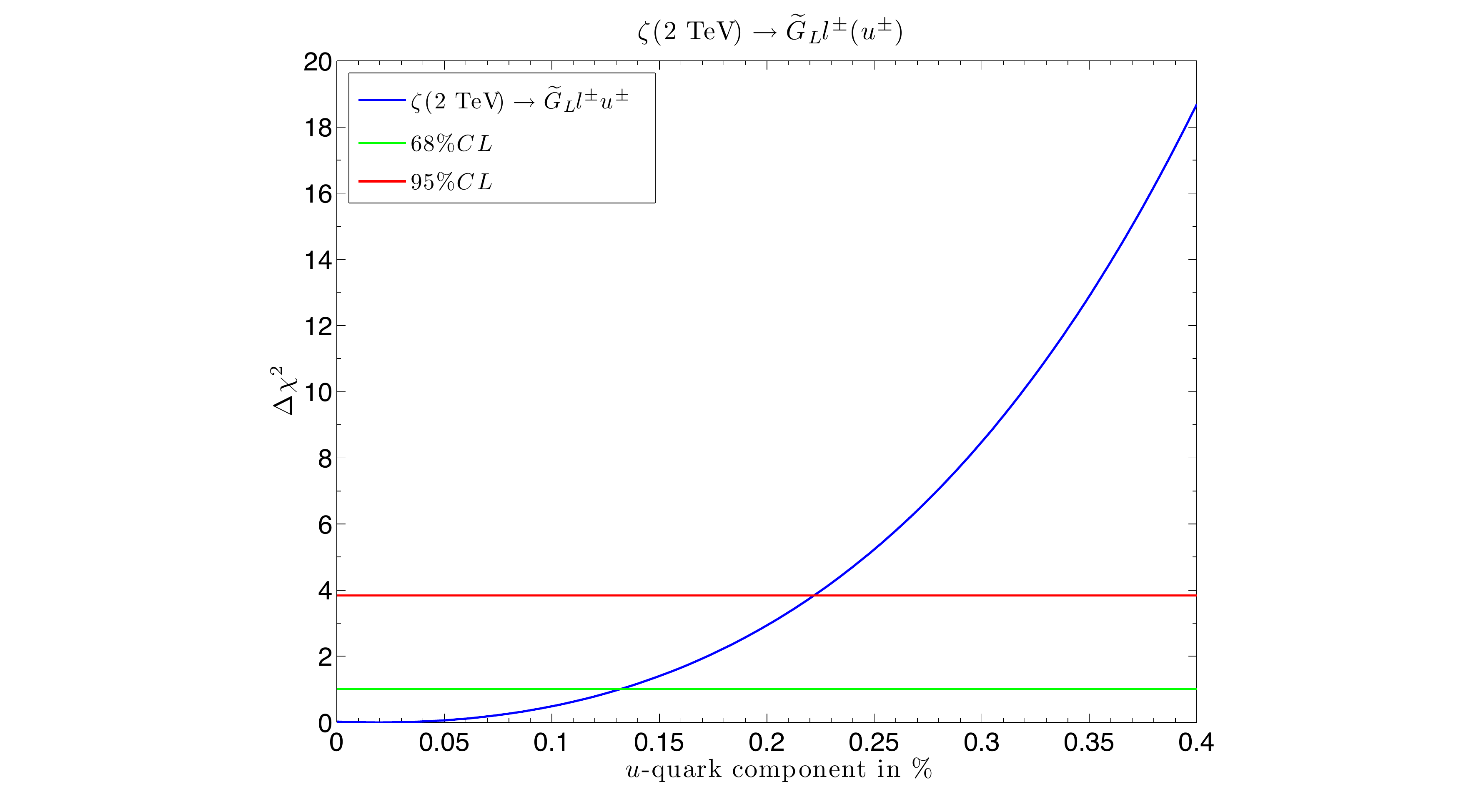}
\includegraphics[trim =55mm 2mm 69mm 2mm, clip, width=0.445\textwidth]{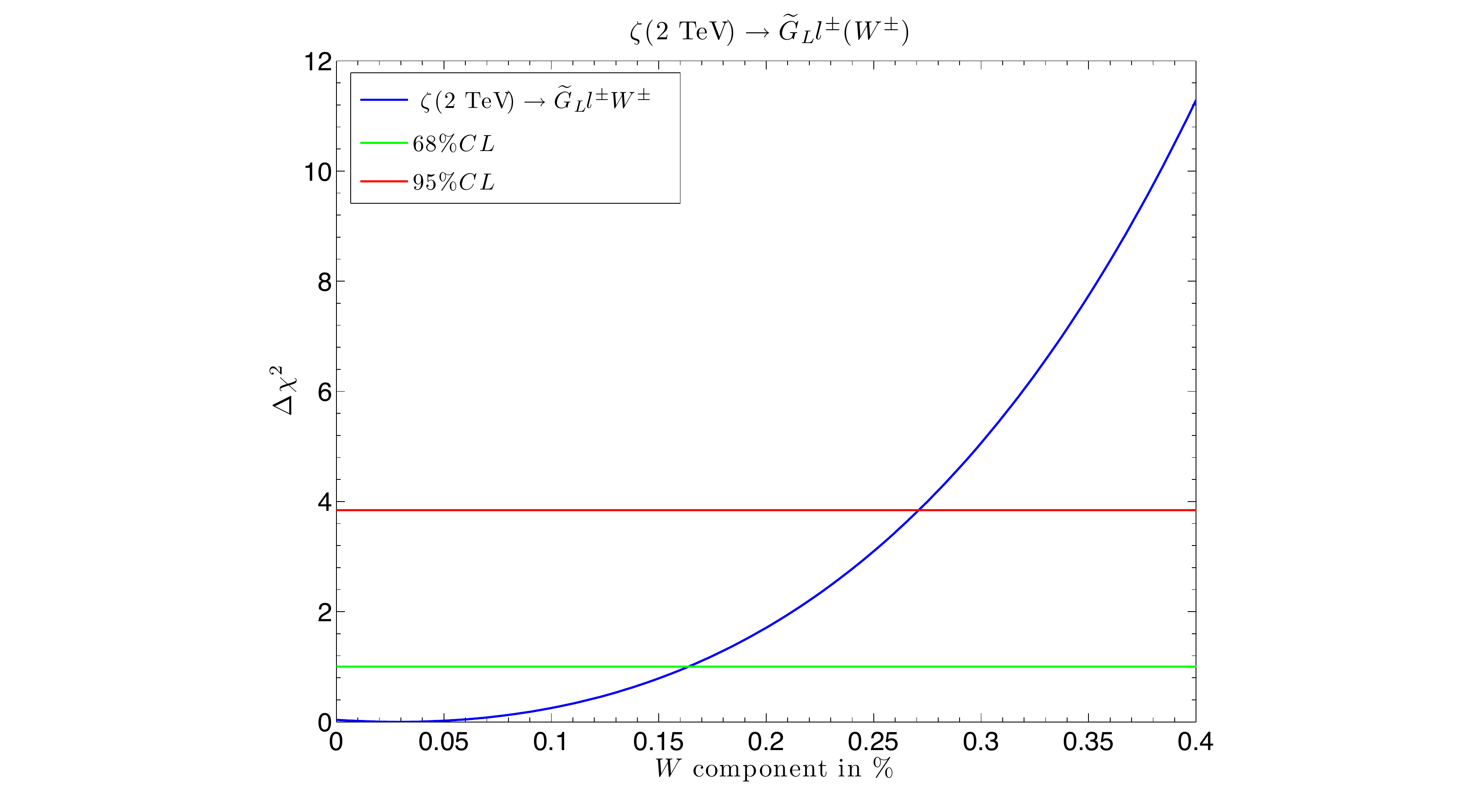}
\caption{\em The effect on fits to PAMELA and Fermi-LAT from inclusion of hardonic final states. The bounds is much looser than ones from antiproton data which justify our fitting procedure.}
\label{fig:uw-con}
\end{center}
\end{figure}

It is worthwhile to mention that, for the three-body decay, in principle, we should simultaneously fit both of the positron (electron) and the antiproton data by including all kinds of decay products such as leptons, quarks and gauge bosons, etc. Instead, we fit positron data first with universal couplings to leptons only and obtain $m_{\widetilde{G}_L}$ or $m_X$, and then fit antiproton data with the extracted $m_{\widetilde{G}_L}$(or $m_X$) to see how much the hadronic final state is allowed.
In this procedure, first it is easier to pin down the sufficient conditions for good fits to positron data and second it is numerically faster on searching for minima.
Nevertheless, one can argue that including other final states might change best fits to positron data significantly since these final states yield positrons (electrons) as well, and therefore modify the whole picture.
Fig.~\ref{fig:uw-con} clearly shows that inclusion of hadron final states will have a modest effect in terms of $\chi^2$ on fits to PAMELA and Fermi-LAT positron data even if the quarks or $W$-bosons comprise more than $25\%$ of decay products, which is much higher than the upper bounds from the antiproton data consideration. Similar conclusions hold for the effect on the gravitino mass and the goldstino lifetime. As a result, our way of fitting is properly justified.

\section{Conclusions}
\label{sect:concl}

In this work we have attempted to make a strong case for the three-body decaying dark matter, by studying how the three-body kinematics could allow simultaneous fits to both the PAMELA and Fermi-LAT positron excesses without being excluded by other astrophysical measurements such as the gamma-ray and anti-proton data. As a contrast, conventional decaying dark matter models have trouble achieving the above goal due to the restrictive two-body decay kinematics.

Using the goldstino as a prime example of the three-body decaying DM, which arises in a certain class of supersymmetric theories where SUSY is broken by multiple sectors,  we found that the goldstino decay with universal leptonic couplings could fit PAMELA and Fermi-LAT positron data well, if the goldstino mass is at around 2 TeV and the gravitino, which escapes detection and results in missing energy, has a mass around 1 TeV.  The mass difference is driven by the hardening feature around $300\sim500$ GeV in the Fermi-LAT $e^+ + e^-$ data. A slightly smaller (larger) mass difference than $1$ TeV can be counterbalanced with a shorter (longer) lifetime.

We further demonstrated that these features of goldstino decays persist in other types of three-body decay mechanisms, by studying the injection energy spectra of several  different types of four-fermi interactions with universal leptonic couplings, as well as scalar dark matter decays. We found that these other mechanisms all have softer injection spectra and consequently smaller missing particle mass for best fits to the positron data. However, the best fit $\chi^2$ are all similar to the goldstino case. The only exception is when certain operators are associated with chiral symmetry breaking and  might have couplings proportional to the masses of SM fermions. Then the $\tau$ decay channel would be the dominant one and in this situation the resulting soft injection spectrum can not fit $e^+/(e^- + e^+)$ and $e^- + e^+$ at the same time, while satisfying the gamma ray constraints. Interestingly, the scalar dark matter have the very similar injection spectrum for the mass region of interest and in turn the similar best fits. These observations suggest that it would be difficult to distinguish different decay mechanisms using data considered in this work.

One important advantage of three-body decaying dark matter over other conventional models is the ability to avoid null searches in cosmic gamma-ray and anti-proton data for dark matter. Due to the softer injection energy spectra of the three-body kinematics, we showed that the diffuse gamma-ray measurements are compatible with the three-body decays, while at the same time maintaining the fits to the PAMELA and Fermi-LAT positron data. The allowed hadronic decay widths from the anti-proton data are also larger in the three-body decay scenario than in the two-body case, due to the softness of the decay energy spectra.

In the end, we hope it is clear that there is a strong case for three-body decaying dark matter, if the  excesses in the positron measurements by PAMELA and Fermi-LAT are believed to be due to dark matter. It would therefore be important to explore ways to definitively determine the decay mechanism of the dark matter in other types of measurements, which will be a subject for future studies.

\begin{acknowledgments}
This work was supported in part by the U.S. Department of Energy under
contracts No. DE-AC02-06CH11357, No. DE-FG02-91ER40684 and No. DE-FG02-95ER40896. We thank Marco Cirelli for pointing out the Fermi-LAT EGB data, useful discussions on diffuse gamma rays and help with {\sc Mathematica}$^{\tiny{\textregistered}}$ codes in~\cite{cookbook website}. H.-C.~C. would like to thank the hospitality of Fermilab Theory Group and Academia Sinica in Taiwan, and W.-C.~H. would like to thank the hospitality of Physics Division of National Center for Theoretical Sciences (NCTS) in Taiwan, where part of this work was performed.
\end{acknowledgments}

\end{document}